\documentclass[a4paper,11pt]{article}
\pdfoutput=1
\usepackage{jinstpub} % for details on the use of the package, please
\usepackage{amssymb}
\usepackage{lineno}

\usepackage{caption}
\usepackage{subcaption}

\usepackage{hyperref}

\usepackage{amsmath}
\usepackage{amssymb}
\usepackage{amsthm} % Theorem formatting. http://en.wikibooks.org/wiki/LaTeX/Theorems
\usepackage{bm} % bold math
\usepackage{xspace}
\usepackage{upgreek}
\usepackage{float}  % for figure positioning

\usepackage{booktabs}  % for \toprule, \midrule

\usepackage[svgnames]{xcolor}
\newcommand{\be}{\begin{equation}}
\newcommand{\ee}{\end{equation}}
\newcommand{\bea}{\begin{eqnarray}}
\newcommand{\eea}{\end{eqnarray}}

\numberwithin{lemma}{theorem}

\usepackage{moreverb} 

\newcommand{\Rutgers}{Rutgers, The State University of New Jersey, Piscataway, New Jersey 08854, USA}

\newcommand{\Florida}{University of Florida, Department of Physics, Gainesville, FL 32611}

\newcommand{\CBPF}{Centro Brasileiro de Pesquisas F\'{i}sicas, Rua Dr. Xavier Sigaud 150, Urca, Rio de Janeiro, Rio de Janeiro, 22290-180, Brazil}
\newcommand{\PUCP}{Secci\'{o}n F\'{i}sica, Departamento de Ciencias, Pontificia Universidad Cat\'{o}lica del Per\'{u}, Apartado 1761, Lima, Per\'{u}}

\newcommand{\Pittsburgh}{Department of Physics and Astronomy, University of Pittsburgh, Pittsburgh, Pennsylvania 15260, USA}
\newcommand{\Guanajuato}{Campus Le\'{o}n y Campus Guanajuato, Universidad de Guanajuato, Lascurain de Retana No. 5, Colonia Centro, Guanajuato 36000, Guanajuato M\'{e}xico.}

\newcommand{\Tufts}{Physics Department, Tufts University, Medford, Massachusetts 02155, USA}
\newcommand{\WM}{Department of Physics, College of William \& Mary, Williamsburg, Virginia 23187, USA}
\newcommand{\FNAL}{Fermi National Accelerator Laboratory, Batavia, Illinois 60510, USA}

\newcommand{\MCLA}{Massachusetts College of Liberal Arts, 375 Church Street, North Adams, MA 01247}
\newcommand{\UMD}{Department of Physics, University of Minnesota -- Duluth, Duluth, Minnesota 55812, USA}
\newcommand{\Northwestern}{Northwestern University, Evanston, Illinois 60208}
\newcommand{\UNI}{Universidad Nacional de Ingenier\'{i}a, Apartado 31139, Lima, Per\'{u}}
\newcommand{\Rochester}{University of Rochester, Rochester, New York 14627 USA}

\newcommand{\USM}{Departamento de F\'{i}sica, Universidad T\'{e}cnica Federico Santa Mar\'{i}a, Avenida Espa\~{n}a 1680 Casilla 110-V, Valpara\'{i}so, Chile}

\newcommand{\OregonState}{Department of Physics, Oregon State University, Corvallis, Oregon 97331, USA}

\newcommand{\upenn}{Department of Physics and Astronomy, University of Pennsylvania, Philadelphia, PA 19104}
\newcommand{\AMU}{AMU Campus, Aligarh, Uttar Pradesh 202001, India}
\newcommand{\wroclaw}{University of Wroclaw, plac Uniwersytecki 1, 50-137 Wroc\l{}aw, Poland}

\newcommand{\chrismarshallThanks}{now at Lawrence Berkeley National Laboratory, Berkeley, CA 94720, USA}
\newcommand{\jwolcottThanks}{now at Tufts University, Medford, MA 02155, USA}

\newcommand{\oxford}{Department of Physics, University of Oxford, Oxford OX1 3PU, UK}

\newcommand{\ChicagoStat}{University of Chicago, Department of Statistics, Chicago, Illinois 60637, USA}
\newcommand{\IBM}{IBM Research, Cambridge, Massachusetts 02141, USA}
\newcommand{\ORNL}{Oak Ridge National Laboratory, Oak Ridge, Tennessee 37831, USA}

\newcommand{\minerva}{MINER\ensuremath{\upnu}A\xspace}

\newcommand{\locclass}{localization-classification\xspace}

\title{Reducing model bias in a deep learning classifier using domain adversarial neural networks in the \minerva experiment 
}

\author[a,b]{G.N.~Perdue,}   % FNAL, Roch
\author[c,d]{A.~Ghosh,}      % USM, CBPF
\author[e]{M.~Wospakrik,}    % Florida
\author[f]{F.~Akbar,}        % AMU
\author[g]{D.A.~Andrade,}    % Guanajuato
\author[h]{M.~Ascencio,}     % PUCP
\author[a]{L.~Bellantoni,}   % FNAL
\author[b]{A.~Bercellie,}    % Roch
\author[a]{M.~Betancourt,}   % FNAL
\author[d]{G.~F.~R.~Caceres~Vera,}   % CBPF
\author[b]{T.~Cai,}          % ROCH
\author[i]{M.F.~Carneiro,}   % OSU
\author[j]{J.~Chaves,}       % Penn
\author[k]{D.~Coplowe,}      % Oxford
\author[d]{H.~da~Motta,}     % CBPF
\author[b,h]{G.A.~D\'{i}az,} % Roch, PUCP
\author[g]{J.~Felix,}        % Guanajuato
\author[a,l]{L.~Fields,}     % FNAL, Northwestern
\author[b]{R.~Fine,}         % Roch
\author[h]{A.~M.~Gago,}      % PUCP
\author[c]{R.~Galindo,}      % USM
\author[m,b]{T.~Golan,}      % Wroclaw, Roch
\author[n]{R.~Gran,}         % UMD
\author[o]{J.~Y.~Han,}       % Pitt
\author[a]{D.~A.~Harris,}    % FNAL
\author[a]{D.~Jena,}         % FNAL
\author[b]{J.~Kleykamp,}     % Roch
\author[p]{M.~Kordosky,}     % WM
\author[k]{X.-G.~Lu,}        % Oxford
\author[q]{E.~Maher,}        % MCLA
\author[r]{W.~A.~Mann,}      % Tufts
\author[b,1]{C.~M.~Marshall, \note{~\chrismarshallThanks}}  % Roch
\author[b,a]{K.S.~McFarland,}   % Roch, FNAL
\author[b]{A.~M.~McGowan,}   % Roch
\author[o]{B.~Messerly,}     % Pitt
\author[c]{J.~Miller,}       % USM
\author[p]{J.~K.~Nelson,}    % WM
\author[e]{C.~Nguyen,}       % Florida
\author[p]{A.~Norrick,}      % WM
\author[s,c]{Nuruzzaman,}    % Rutgers, USM
\author[b]{A.~Olivier,}      % Roch
\author[t]{R.~Patton,}       % ORNL
\author[g]{M.A.~Ram\'{i}rez,}  % Guanajuato
\author[s]{R.~D.~Ransome,}   % Rutgers
\author[e]{H.~Ray,}          % Florida
\author[o]{L.~Ren,}          % Pitt
\author[e]{D.~Rimal,}        % Florida
\author[b]{D.~Ruterbories,}  % Roch
\author[i,l]{H.~Schellman,}  % OSU, Northwestern
\author[u]{C.~J.~Solano~Salinas,}   % UNI
\author[o]{H.~Su,}           % Pitt
\author[v,2]{S.~Upadhyay, \note{~now at \IBM}}   % Chicago
\author[p,g]{E.~Valencia,}   % WM, Guanajuato
\author[b,3]{J.~Wolcott, \note{~\jwolcottThanks}}  % Roch
\author[c]{B.~Yaeggy,}       % USM
\author[t]{S.~Young}         % ORNL

\affiliation[a]{\FNAL}
\affiliation[b]{\Rochester}
\affiliation[c]{\USM}
\affiliation[d]{\CBPF}
\affiliation[e]{\Florida}
\affiliation[f]{\AMU}
\affiliation[g]{\Guanajuato}
\affiliation[h]{\PUCP}
\affiliation[i]{\OregonState}
\affiliation[j]{\upenn}
\affiliation[k]{\oxford}
\affiliation[l]{\Northwestern}
\affiliation[m]{\wroclaw}
\affiliation[n]{\UMD}
\affiliation[o]{\Pittsburgh}
\affiliation[p]{\WM}
\affiliation[q]{\MCLA}
\affiliation[r]{\Tufts}
\affiliation[s]{\Rutgers}
\affiliation[t]{\ORNL}
\affiliation[u]{\UNI}
\affiliation[v]{\ChicagoStat}

\abstract{
We present a simulation-based study using deep convolutional neural networks (DCNNs) to identify neutrino interaction vertices in the \minerva passive targets region, and illustrate the application of domain adversarial neural networks (DANNs) in this context.
DANNs are designed to be trained in one domain (simulated data) but tested in a second domain (physics data) and utilize unlabeled data from the second domain so that during training only features which are unable to discriminate between the domains are promoted.
\minerva is a neutrino-nucleus scattering experiment using the NuMI beamline at Fermilab.
$A$-dependent cross sections are an important part of the physics program, and these measurements require vertex finding in complicated events.
To illustrate the impact of the DANN we used a modified set of simulation in place of physics data during the training of the DANN and then used the label of the modified simulation during the evaluation of the DANN.
We find that deep learning based methods offer significant advantages over our prior track-based reconstruction for the task of vertex finding, and that DANNs are able to improve the performance of deep networks by leveraging available unlabeled data and by mitigating network performance degradation rooted in biases in the physics models used for training.
}


\begin{document}

\maketitle
\flushbottom

%\linenumbers

\section{Introduction}
\label{sec:introduction}

In particle physics, machine learning algorithms trained with simulated data, but applied to physics detector data, face uncertainties associated with scientific modeling because the simulation is based on a physics model that is not identical to nature.
Therefore, algorithms must be vetted carefully for domain-dependent bias.
To control these effects in a deep learning classifier, we explore the use of domain adversarial neural networks (DANNs) and show they may be used to optimize networks trained in one domain (e.g., simulation) but applied in another (e.g., detector data), using neutrino scattering data from \minerva.

\minerva is a neutrino-nucleus scattering experiment at Fermilab in the NuMI neutrino beam \cite{Adamson:2015dkw}.
Precise determination of the interaction vertex is required to identify the target nucleus in \minerva, but vertex finding is challenging in a highly segmented, non-uniform detector.
Here we present a vertex-finding technique for use in the measurement of neutrino-nucleus cross sections in neutrino-nucleus deep inelastic scattering (DIS), based on deep convolutional neural networks (DCNNs) \cite{lenet}.
In simulation studies, we find significant advantages in signal purity and efficiency over track-based vertex finding methods for this analysis.

We study domain adversarial neural networks (DANNs) \cite{JMLR:v17:15-239} for control of un-modeled physics effects when training a neural network and find that DANNs improve cross-domain training and evaluation in a particle physics context.
Here a ``domain'' refers to a physics model (e.g., nature or a particular simulation model). 
When applying machine learning algorithms to physics problems, we often train using simulation but need to apply the models to detector data.
In this context, we refer to simulation as a \emph{source domain} and detector data as a \emph{target domain}.
If the physics model in our simulation were perfect, we would not have domain discrepancies, but because our models are not perfect, we must seek strategies for mitigating any biases in the algorithm that may come from training our models in one domain and applying them in another.

Ideally, we would benchmark the performance of the DANN algorithm using real detector data, but because detector data is unlabeled, we instead benchmark it by using simulation with a warped underlying physics model.
We neglect the physics labels in the warped sample during training, but we retain the domain labels.
Then, during evaluation, we can check the performance of the trained models on both domains (plain and warped simulation) to see if our strategy for mitigating the impacts of domain modeling was successful.
This is a semi-supervised approach.
Some of the training sample is fully-labeled and we use both physics labels and domain labels, and some of the training sample is partially labeled with domain information only. 
Our findings show that DCNNs can be a very effective reconstruction technique and that we may reduce biases inherent in a simulation-based training scheme by using the DANN algorithm to utilize unlabeled data from our target domain during training in a semi-supervised fashion.
These results suggest deep learning based event reconstruction schemes may outperform traditional methods in many circumstances.

\subsection{Traditional machine learning in particle physics analysis}
\label{sec:mloverview}

Modern physics experiments are large, with tens of thousands or millions of digital measurement channels, and it is often possible to build hundreds of physically-motivated features to describe an event.
Traditionally, creating and exploiting these features has been the predominant work of the physics analyzer.
%However, in practice it is impossible to exhaust all possible features that can be extracted and so the best features may be unused in a given analysis.
%The large datasets of modern particle and nuclear physics experiments as well as the sensitivity of the instrumentation have increased the desirability of utilizing machine learning algorithms (MLA) in both reconstruction and analysis.
To fully exploit the information in these features, machine learning algorithms (MLAs) have become a common tool \cite{hastieml,bishopml,murphyml}.
The most successful MLAs include boosted decision trees, support vector machines, and neural networks.
An advantage of MLAs over analyses based on collections of univariate cuts is that they easily incorporate non-linear correlations between the features in the event selection.
These have become part of the standard toolkit for particle physics analysis \cite{Brun:1997pa, Hocker:2007ht}. 

%Modern analysis procedure has been for the physicist to use procedural algorithms to extract features from data based study of the events and domain knowledge.
%%These features may then be used to form cuts for event selection directly or used as input for a MLA.
%For traditional analyses, a series of cuts are applied to these features, resulting in an event selection. 
%Analyses that utilize machine learning input a moderate to large number (generally from a few to dozens) of features into the MLA with training and testing sets delineated.
%The MLA then generates a classifier or regressor which is a complex, non-linear feature that may be adjusted to select signal at a given efficiency and purity.

In particle physics it is common to employ extensive and detailed simulations. 
These carefully tuned simulations include fundamental physics interactions and the description of the physics detector including material and electronics.
Particle physics experiments often produce very large simulated datasets to help analyze real detector data, and these simulations form good candidate training sets for MLAs.
However, these simulations are based on physics models that are only approximations to nature.
They include simplifications, parameterizations, and extrapolations.
For example, in neutrino scattering measurements, often we may directly calculate some aspects of the interactions we are modeling, but these calculations generally do not cover the full phase space of a reaction.
Therefore in the application of machine learning we must be very careful of the problem of \emph{domain adaption}.
We have labels (for example, particles produced in a scattering event) that are the same in both our simulation and detector data, but the underlying distributions can, perhaps inadvertently, be subtly different between the two domains.

\subsection{An introduction to DCNNs for physicists}
\label{sec:dnnintro}

In recent years, we have seen significant advances in machine learning.
``Deep learning,'' or the use of neural networks with many hidden layers, has revolutionized many fields including computer vision \cite{bengiodl,Razavian:2014:CFO:2679599.2679731,alexnet}.
It has made an impact at the LHC \cite{Guest:2018yhq} and in neutrino flavor classification \cite{Aurisano:2016jvx}. 

Neural networks are functions motivated by biological networks of neurons.
Inputs are mapped to outputs by weighted connections that may connect all inputs to all outputs or to a subset of the outputs.
Outputs ``fire'' as function of the weighted inputs and often feature threshold effects and non-linear responses.
Neural networks become ``deep'' when the outputs of one network become inputs of another, creating multiple layers.
In addition to weighted connections, neural networks often feature bias vectors at the output of each layer and typically pass outputs through an activation function that may be used to introduce non-linearities in the response.
Example non-linear activation functions are rectified linear units (ReLUs) \cite{pmlr-v15-glorot11a}.
Neural networks are trained using back-propagation \cite{backprop}. 
This technique computes the error in the output of the network as compared to the known label for the input during the training phase, and employs the chain rule of calculus to propagate that error through the weights in the network.

Deep convolutional neural networks (DCNNs) \cite{lenet} are a type of neural network that employ a structure inspired by the biological structure of the eye, and they are particularly well-suited to computer vision problems.
This family of neural networks is useful for deriving features automatically when the problem under consideration does not have an obvious set of engineered features from data to use.
%DCNNs are a class of algorithms for dealing with data that may be represented in image form.
Their success is rooted in the exploitation of symmetries in image data representation through weight-sharing in convolutional kernels.
These kernels structure the features extracted from data and keep the number of parameters in the network from growing too large, as described below.

In a DCNN, the inputs of each layer, typically an image at the beginning of the network, are mapped into outputs by convolving a kernel with the inputs according to a process discussed below.
By sharing weights across a layer, the network is able to express translational symmetries in the input.
Stacking multiple layers with many kernels per layer allows a DCNN to learn a hierarchically organized set of features for extracting information about the input image without requiring ``by-hand'' feature engineering.
%As an image progresses through the layers of a DCNN, the features extracted grow in semantic complexity \cite{alexnet}.
%In this way, a DCNN learns the important features of a dataset without requiring ``by-hand'' feature engineering.

Convolving kernels with images is a well-established image processing technique.
The kernel is matrix-multiplied in patch-wise fashion across the entire image, with different options for the  step sizes in each direction (``strides'') and multiple techniques for handling the edge of the image (``padding'') available.
The weights in the kernel are kept fixed throughout the convolution process (they are shared across the entire image).
Certain kernels are known to produce or highlight features of images.
For example, the kernel in Equation \ref{eqn:sharpkernel} is known to sharpen images.
%, as seen in Figure \ref{fig:generic-taj-convmatrix-sharpen}:

\begin{equation}
k_{\text{sharpen}} = 
\begin{pmatrix}
0 & -1 & 0 \\
-1 & 5 & -1 \\
0 & -1 & 0 \\
\end{pmatrix}.
\label{eqn:sharpkernel}
\end{equation}
\smallskip 

\noindent However, the kernel in Equation \ref{eqn:embosskernel} produces an embossing effect.
%, as seen in Figure \ref{fig:generic-taj-convmatrix-emboss}:

\begin{equation}
k_{\text{emboss}} = 
\begin{pmatrix}
-2 & -1 & 0 \\
-1 & 1 & 1 \\
0 & 1 & 2 \\
\end{pmatrix}.
\label{eqn:embosskernel}
\end{equation}
\smallskip 

\noindent See, e.g. references \cite{gimp,gimpmanual} for more kernel examples and for visualizations.

% Figure source:
% https://docs.gimp.org/en/plug-in-convmatrix.html
%\begin{figure}[h]
%	\begin{subfigure}{0.5\textwidth}
%    	\centering
%		\includegraphics[width=0.99\linewidth]{generic-taj-convmatrix-sharpen}
%  		\caption{A sharpened image.}
%  		\label{fig:generic-taj-convmatrix-sharpen}
%	\end{subfigure}
%	\begin{subfigure}{0.5\textwidth}
%    	\centering
%		\includegraphics[width=0.99\linewidth]{generic-taj-convmatrix-emboss}
%  		\caption{An embossed image.}
%  		\label{fig:generic-taj-convmatrix-emboss}
%	\end{subfigure}
%	\caption{A demonstration of the operation of convolutional kernels on images \cite{gimpmanual}.}
%	\label{fig:convkerns}
%\end{figure}

The central idea behind DCNNs is to ask the MLA to find kernels that highlight important features for the problem under consideration rather than attempting to hand-engineer them.
A DCNN will begin with random kernels for feature selection, but during training it will find kernels well-suited to the task through the usual optimization methods (minimization of the network loss function via gradient descent and backpropagation).
Generally, DCNNs are comprised of multiple convolutional layers, with the output of one layer becoming the input of the next, and with the number of kernels growing for deeper layers.
This enables the network to build a hierarchical representation.
There is evidence \cite{alexnet} that the lower layers of the network learn to identify feature primitives (e.g., lines, corners, etc.) and deeper layers learn to combine the features into more semantically meaningful compositions.

DCNNs also traditionally utilize pooling, fully-connected, dropout and softmax layers.
Pooling layers subsample their inputs (the most common operation is a \texttt{max()}) - for example, a \texttt{2x2} max-pooling unit reduces four pixels to one, retaining only the maximum value.
Fully-connected layers are defined by connecting every neuron in one layer of the network to every neuron in the next layer.
Dropout layers \cite{Srivastava:2014:DSW:2627435.2670313} randomly drop connections between layers during training to improve network robustness and reduce the possibility of over-fitting.
Softmax layers normalize network outputs so training does not overweight events with higher-amplitude input values, and are discussed further in Section \ref{sec:trainstrat}.
Convolution and pooling layers act as feature extraction and dimensionality reduction layers.
The image is extended into a feature space of depth equal to the number of filters and may shrink spatially depending on how the filters are applied at the boundary of the image.
%A pooling layer also passes a filter over the image, but instead reduces the resolution by the size of the filter. 
%Dropout layers assist with keeping the DCNN well behaved to improve generalization. 
Non-linear activation layers are used so the DCNN may model non-linear features.
When solving classification problems, the final features extracted are fed into a softmax loss layer which compares these features to a ``one-hot'' representation of the target label or value.

In 2014, domain adversarial neural networks (DANNs) were developed to ameliorate the problem of domain adaption \cite{JMLR:v17:15-239}.
The principle behind a DANN allows the user to leverage a labeled dataset to train on an un-labeled dataset by utilizing only features that are common between the two.
In the field of computer vision, this is an important problem due to the limited availability of labeled training sets - it is common to have no labeled training set for a given task, but to have access to labeled data that is ``similar''.
DANNs cast these different datasets as belonging to different domains - a source domain where the user has labels for the data and a target domain where the user has no labels.
The meaning of ``domain'' is flexible - it might be a digit recognition problem with samples of different handwriting or typesetting methods, or an animal classification problem with animals in captivity versus animals in the wild, etc.

In particle physics we have an analogous problem: we have large amounts of labeled simulation for training but need to understand algorithm performance in a different domain - detector data.
In this study we investigate the applicability of DANNs to one of the core problems with using ML in particle physics: that of the trustworthiness of using a supervised network trained on simulation.
Here, we do not seek to derive a universal answer to the question of domain similarity for DANNs - which is to say where the dividing lines are for the applicability of technique in cases where the domains are too similar or too different for the approach to work.
Rather, we seek to test the effectiveness only in this specific context and try to establish whether DANNs are potentially useful tools in ML problems in particle physics. 
Relative to previous work on neutrino vertex finding \cite{ijcnn7966131}, here we extend the number of classes, investigate the impacts of the technique on the physics measurement, and examine the effects of using domain adversarial neural networks. 

There are many frameworks which have been developed to train and apply deep neural networks. 
For this work, we employed Theano \cite{2016arXiv160502688T} along with Lasagne framework \cite{lasagne} for network design and the first round of updates to the physics analysis.
We employed Caffe \cite{jia2014caffe} to study domain adversarial neural nets (see Section \ref{sec:dann}).

\subsection{Overview of the paper}
\label{sec:paperplan}

The direction of the remainder of the paper is as follows. 
First, in Section \ref{sec:background} we discuss details of the \minerva detector and analysis goals.
Next, in Section \ref{sec:sampledetails} we discuss the simulated dataset and features of the training and evaluation samples.
In Section \ref{sec:dnnforvert} we discuss the DCNN architecture and training strategies.
Then, in Section \ref{sec:improvement} we examine the impacts of the new vertex finder on a measurement of nuclear effects in charged current deep inelastic scattering.
In Section \ref{sec:dann} we discuss the application of a domain adversarial neural network (DANN) to our problem.
Finally, in Section \ref{sec:conclusions} we discuss future plans and our conclusions.

\section{Detector and analysis details}
\label{sec:background}

The analysis task here is reconstruction of the neutrino interaction vertex.
In particular, we are not concerned here with the transverse position of the interaction, but only the longitudinal position where (and in which material) the interaction took place.
The analysis is taking place in the context of a deep inelastic scattering measurement, but the details of the physics measurement, and the interaction physics, are not relevant for this work.
Details about deep inelastic scattering and other neutrino interactions may be found elsewhere \cite{McFarland:2008xd,doi:10.1146/annurev-nucl-102010-130255,Mosel:2012kt}.
For our purposes here it is sufficient to think of the reactions of interest as being ``inelastic'' and likely to produce particle cascades, as opposed to reactions that are more ``elastic'' and instead produce a small number of particles likely to be measured as tracks in the detector.

\subsection{Neutrino scattering in \minerva}
\label{sec:neutrinophys}

Measurements of neutrino-nucleus scattering face two crucial challenges: the flux of a neutrino beam is broad and difficult to measure, and the cross sections for neutrino interactions are small.
To compensate for small cross sections, we must use dense neutrino target materials.
This complicates the interpretation of detector response, as these materials increase the re-interaction rate of produced particles both inside and outside of the struck nucleus.

\minerva features passive and active neutrino targets.
The active targets are plastic scintillator (a hydrocarbon); these measure the passage of electrically charged particles produced during the neutrino interaction.
The passive targets are solid layers of carbon, iron, and lead, and tanks of liquid helium and water. 
Interactions within these targets may only be identified and reconstructed based on the particles observed in the surrounding scintillator.
In the analysis considered here, the liquid targets are empty and we are only concerned with the solid passive targets, the layers of fine-grained plastic scintillator between them, and the tracker region made up of additional layers of fine-grained plastic scintillator downstream of the passive targets.

\subsection{\minerva detector}
\label{sec:minerva}

\begin{figure}
  \centering
  \includegraphics[width=1.0\textwidth]{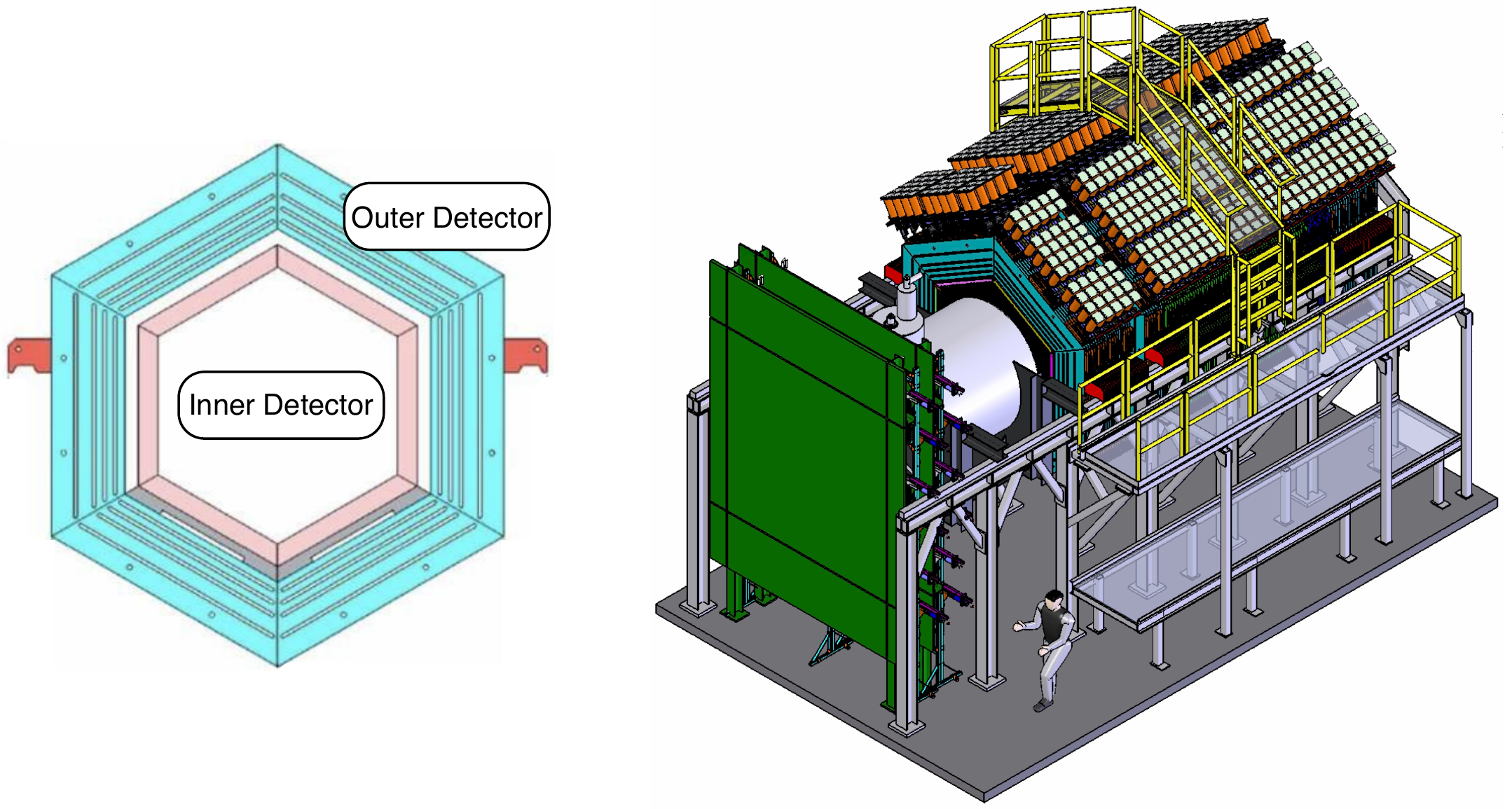}
  \caption{
  The \minerva detector.
  The subfigure on the left shows the relative structure of the inner and outer detectors for a given plane (the outer detector is composed of steel-scintillator calorimetry).
  The inner detector features an outer ring of lead on the surface of the plane to prompt rapid conversion of electromagnetic particles leaving the inner detector. 
  The subfigure on the right shows the relative size and location of the main detector and the helium neutrino target tank and veto wall.
  }
  \label{fig:detector_schem}
\end{figure}

The \minerva detector consists of a passive neutrino target region, a fine-grained tracker which is an active target, downstream and circumferential calorimetry, and a muon spectrometer.
See Figure \ref{fig:detector_schem} for a schematic.
The detector is described in detail elsewhere \cite{Aliaga2014130}.

\begin{figure}
  \centering
  \includegraphics[height=0.42\textheight]{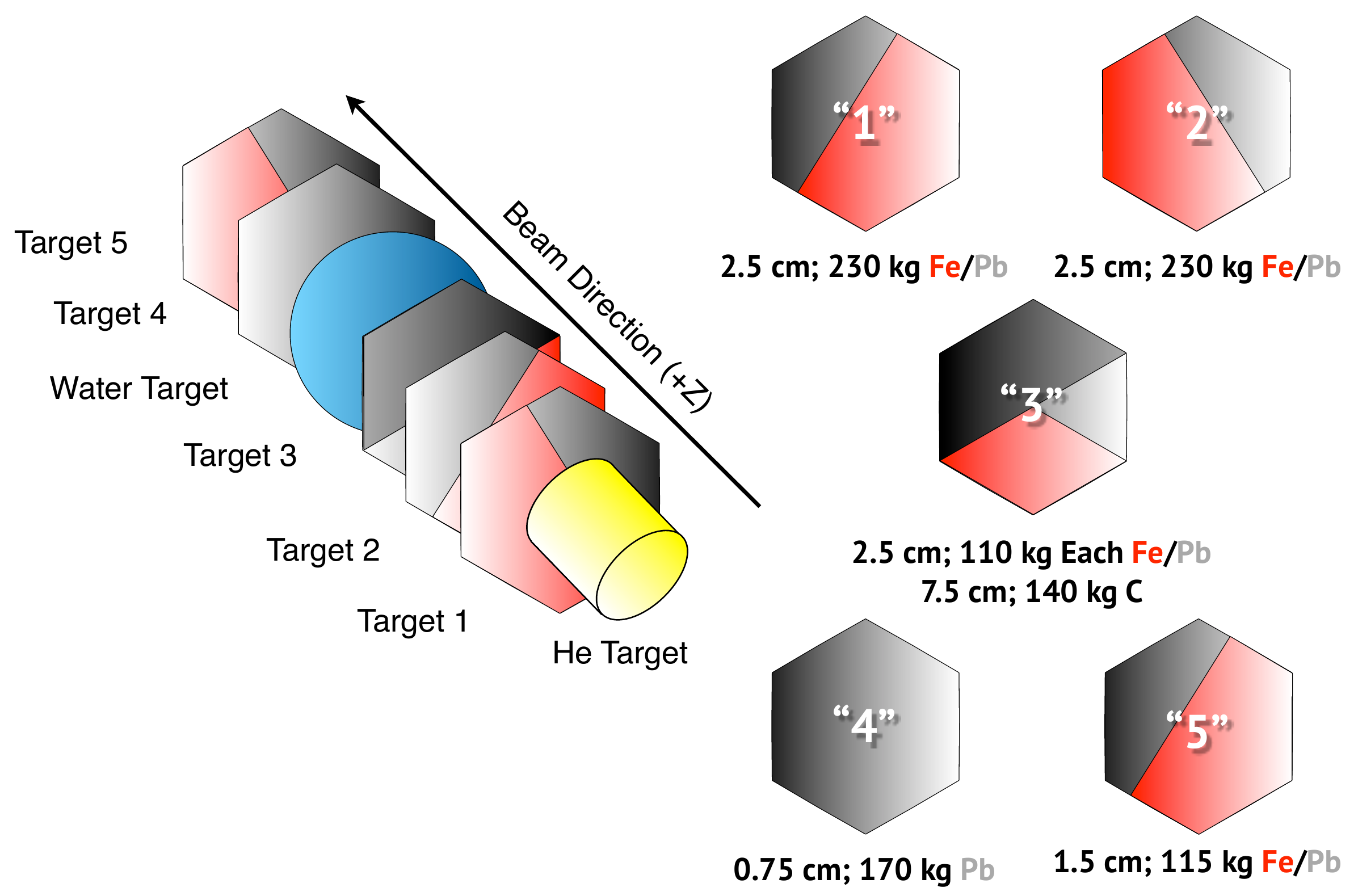}
  \caption{Neutrino target material shapes and layout.
  Note that the targets are composed of segments of pure material - so targets 1, 2, and 5 contain a jointed slab of iron and lead, target 3 contains a jointed slab of iron, lead, and carbon, and target 4 is a slab of pure lead.
  The thicknesses of the different slabs are listed in Table \ref{tbl:tgtthick}.
  }
  \label{fig:targets_sketch}
\end{figure}

The tracker region forms the core of the detector.
Bundles of polystyrene (CH) strips, called planes, are aligned in one of three orientations: X, U or V.
Strips in X planes are oriented vertically and U and V strips are oriented $\pm 60^0$ relative to X.
The three views allow for stereoscopic track reconstruction, and allow for track reconstruction even when two tracks overlap in one of the views.
Planes are housed in modules, with two planes per module.
Each module contains a U or a V plane, followed by an X, such that the pattern is ``UXVXUXVX,'' etc.
The neutrino target region sits upstream of the tracker and consists of an external veto wall and helium target, an internal water target, five passive heavy nuclear targets, and active scintillator between each nuclear target.
Each heavy passive target consists of a module with some combination of lead (Pb), iron (Fe), and graphite (C) and is separated from the other targets by a number of tracker modules.
There are four modules between targets 1 and 2 and between targets 2 and 3.
There are eight modules between targets 3 and 4, with the (empty) water target in the center of that region. 
Finally, there are two modules between targets 4 and 5.
For a sketch of the target region see Figure  \ref{fig:targets_sketch}.

\begin{figure}
  \centering
  \includegraphics[width=1.0\textwidth]{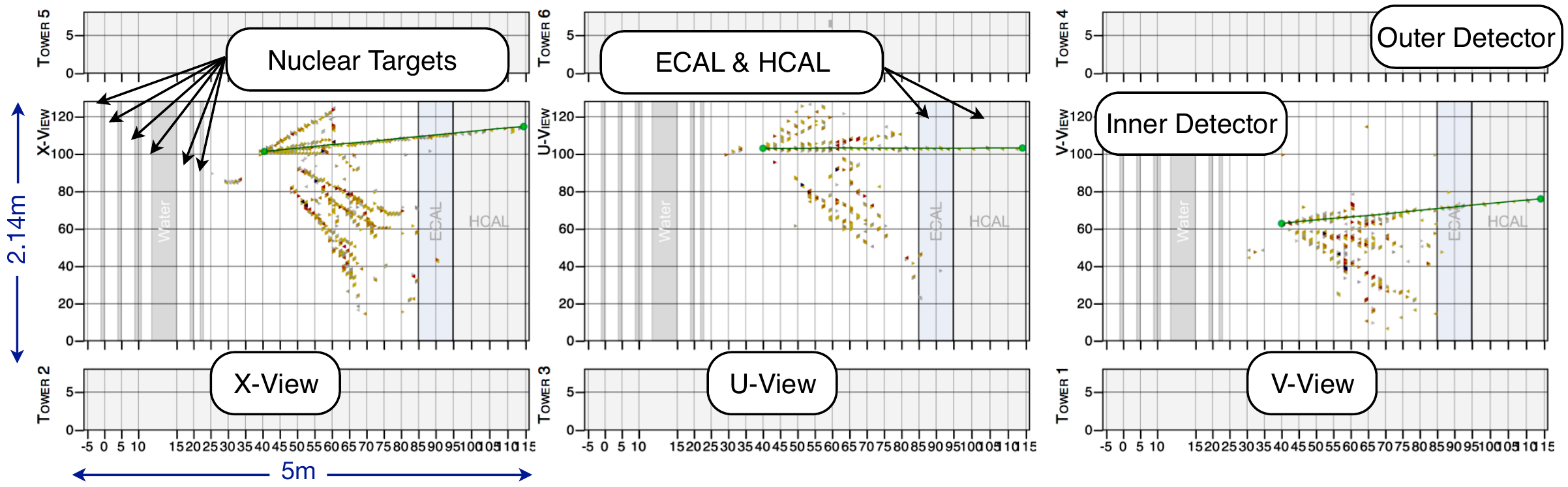}
  \caption{
  A \minerva event display.
  The three columns represent the X, U, and V views respectively, and the three rows show relevant sums of the outer detector, the inner detector, and outer detector sums again when moving from top to bottom through the detector.
  Tick marks on the inner detector show module label (horizontal axis) and strip number (vertical axis).
  The green line shows a reconstructed muon track that was matched to MINOS for sign determination with the green points at either end representing reconstructed start and end points (within the detector volume). 
  }
  \label{fig:eventdisp_labeled}
\end{figure}

Downstream of the neutrino target and tracker regions are the electromagnetic and hadronic calorimeters, with iron and lead absorbers between scintillator planes to induce energy loss.
The \minerva muon spectrometer is the MINOS near detector \cite{refminos}, a magnetized iron-scintillator detector that sits approximately 2 meters downstream of the end of the calorimeters. 
A labeled example event within \minerva can be seen in Figure \ref{fig:eventdisp_labeled}.

\subsection{The \minerva neutrino-interaction simulation}
\label{sec:mnvsim}

%% Description of nuclear effects/interactions (GENIE)

\minerva has a complete simulation, including neutrino beam production, neutrino scattering, and propagation of the interaction products through the detector.
It includes a full simulation of the photosensors and readout electronics.
The simulation has been painstakingly tuned to agree well with the data where we are confident in our modeling of the physics processes.
 
\minerva uses the GENIE Monte Carlo event generator \cite{Andreopoulos:2009rq} to simulate the initial interaction of a neutrino with a nuclear target. 
GENIE is responsible for the primary interaction of the neutrino with nuclei in the detector and for the propagation of the scattering products to the surface of the nucleus.
The code is tuned to reproduce neutrino-nucleon scattering data, but limited neutrino-\emph{nucleus} scattering data means there are deficiencies in the primary physics model when compared to \minerva data.
%, where the nuclear targets include carbon, iron, and lead. 
See references \cite{PhysRevLett.112.231801, PhysRevD.93.071101}. 

The Geant 4 software toolkit \cite{Agostinelli:2002hh} simulates the propagation of all particles exiting the nucleus through the different materials in the detector, and custom software simulates the response of the photosensors and electronics.
Geant 4 has been extensively tested, and the simulation accurately reproduces many aspects of the data. 
The detector response to charged hadrons, for example, is simulated to within about four percent \cite{Aliaga:2015aqe}.

%% Description of Detector Response (although we don?t really adjust this)
% -- this doesn't fit the flow well anymore, although we may want to keep the comment about low energy hadron response somewhere... (although our test beam quantified the problems reasonably well)
%Beyond the surface of the nucleus, the \minerva detector simulation is based on Geant4 \cite{Agostinelli:2002hh}. 
%Geant4 is a complete toolkit for describing a detector geometry and simulating the passage of radiation through matter. 
%Like GENIE though, it is a model with limitations, particularly with respect to the response to low energy hadrons.

\subsection{Traditional vertex reconstruction}
\label{sec:tradreco}

\minerva finds vertices by first reconstructing tracks using a variety of pattern recognition algorithms \cite{Aliaga2014130}. 
Tracking proceeds by first identifying the longest track candidate in an event, the so-called ``anchor track'', and then recursively searching its beginning and ending points for other tracks. 
The origin of the anchor track is the first candidate for the primary vertex. 
The position of the primary vertex may be adjusted by subsequent algorithms. 

Traditional vertex reconstruction methods that rely only on identified tracks can fail when tracks are created by secondary interactions or decays, or when particle cascade activity occludes the vertex region.
The best vertex finding performance is realized for multi-track events with few or no particle cascades.
Multi-track vertices are fit using a Kalman filter \cite{kalmanfilter} that adjusts the vertex position along with track parameters to produce the best overall fit \cite{tice2014thesis}, but tracks from secondary interactions can shift the reconstructed vertex from the true position. 
Single-track events must use algorithms to propagate the anchor muon track back through any activity that may be surrounding the true event vertex.
For analyses based in the active tracker, where missing the primary vertex by a few centimeters imposes essentially no cost on the analysis quality, this is an acceptable situation. 
For analyses of neutrino interactions in the passive neutrino targets though, missing the vertex by a few centimeters is a significant problem because it changes our assignment of the target nucleus.

\subsection{Analysis}
\label{sec:analysis}

We will present two strategies for vertex finding in the inclusive neutrino target ratio analysis.
First, we will perform a large ``segment'' hybrid \locclass exercise (hereafter referred to as ``classification'') and classify events as being in a particular target or anywhere in the scintillator between the targets; second, we will build a ``plane'' classifier based on individual planes in the detector.

\begin{figure}
  \centering
  \includegraphics[height=0.32\textheight]{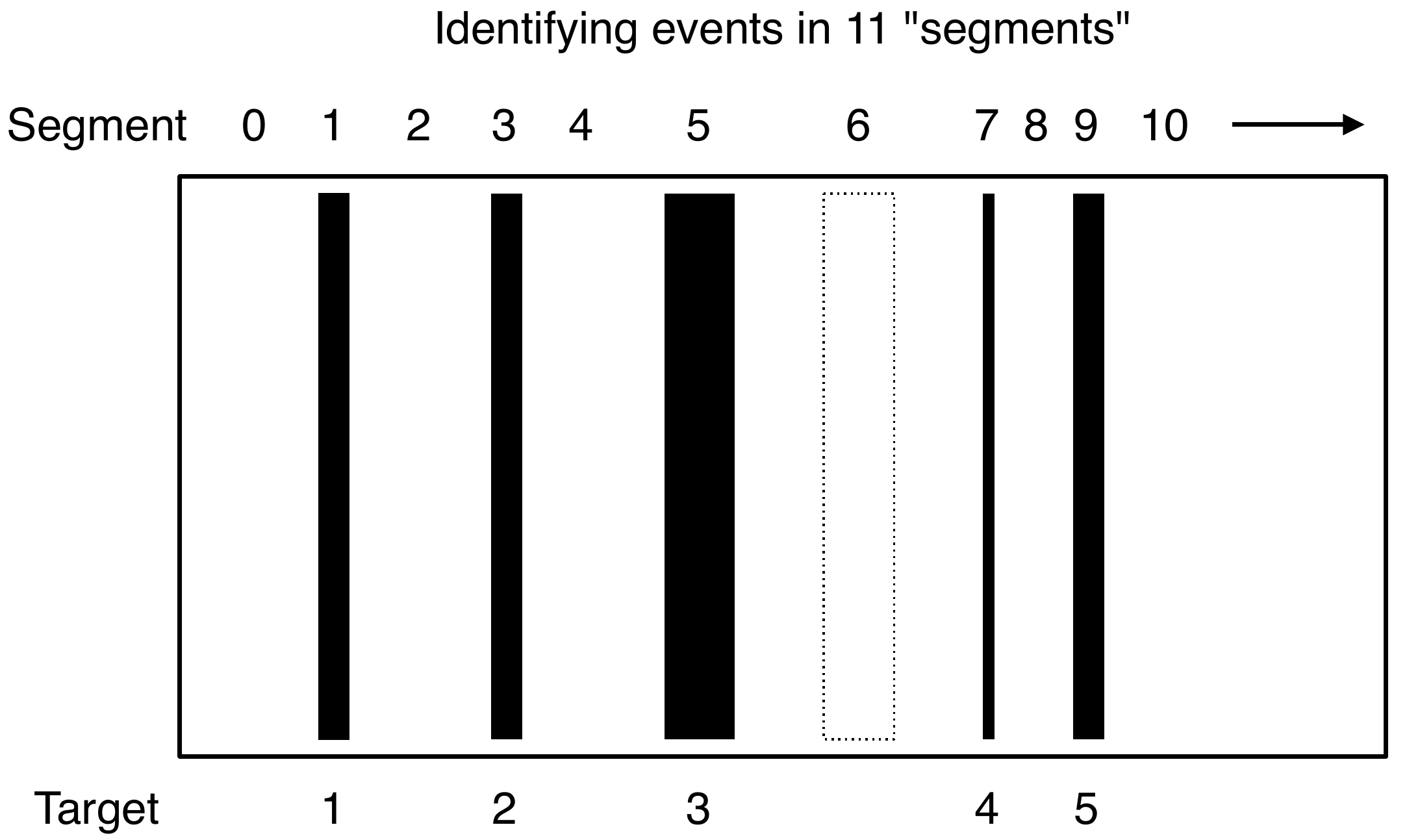}
  \caption{The eleven segments used in vertex segment classification for inclusive nuclear ratio analysis. 
  Note that segment number does not generally correspond to neutrino target number.
  Note also that these regions are only roughly to scale (target 3 is much thicker than the other targets, target 4 much thinner, the gap between targets 4 and 5 is relatively small, etc.).
  The dotted region in the center of segment 6 represents the empty water target.}
  \label{fig:eleven_segments}
\end{figure}

When considering the neutrino target regions used for reconstruction in the inclusive target ratio analysis, there are  eleven ``segments'' - the five targets and the regions upstream and downstream of the targets, as well as in between the targets, as shown in Figure \ref{fig:eleven_segments}.
The regions are non-homogenous. Regions 0, 2, and 4 are all four modules of scintillator, but region 6 is much larger and divided in the middle by the passive water target.
% that is not considered in this analysis.
In the datasets considered  here, the water target is empty so there is little scattering as particles pass through its region. 
Region 8 is smaller with only two modules of scintillator, equal to two planes of scintillator in the X view and one in the U and V view respectively.
Finally, region 10 is quite large.
% (essentially everything downstream of the target region).
The passive targets are of varying thicknesses - see Table \ref{tbl:tgtthick}.

\begin{table}[htb]
\centering
\begin{tabular}{crr}
\toprule
Target & $z$-location (cm) & Thickness (cm) \\
\midrule
1-Fe  & 448.1 & $2.567 \pm 0.006$ \\
1-Pb  & 448.1 & $2.578 \pm 0.012$ \\
% target 1 is at z = 452.5 in the NIM, but we trust the wiki number (448.1) more
% {\color{red}*)I think Wiki has the correct position for target 1}
2-Fe  & 470.2 & $2.563 \pm 0.006$ \\
2-Pb  & 470.2 & $2.581 \pm 0.016$ \\
3-Fe  & 492.3 & $2.573 \pm 0.004$ \\
3-Pb  & 492.3 & $2.563 \pm 0.004$ \\
3-C   & 492.3 & $7.620 \pm 0.005$ \\
Water & 528.4 & $17 - 24$ \\
4-Pb  & 564.5 & $0.795 \pm 0.005$ \\
5-Fe  & 577.8 & $1.289 \pm 0.006$ \\
5-Pb  & 577.8 & $1.317 \pm 0.007$ \\
\bottomrule
\end{tabular}
\caption{Neutrino target thicknesses. \cite{Aliaga2014130}
The water target thickness varies (when filled) because it is built from flexible materials and bows out in the center.
Here we work with samples that featured an empty water target.}
% https://cdcvs.fnal.gov/redmine/projects/minerva-sw/wiki/Data_Run_Periods
% No water for A or B
\label{tbl:tgtthick}
\end{table}

In the segment classifier, we will follow the analysis of \cite{tice2014thesis} and include the neighboring plastic planes (one upstream and two downstream) in the definition of the ``neutrino target'' and subtract the plastic background later using constraints from the \minerva tracker region during the physics analysis.
% (the details of that subtraction are not considered here).
This is the approach followed by track-based methods because it is impossible to distinguish single-track events in a target from those in the neighboring scintillator planes when using only information from a single track to compute the event vertex.

The plane classifier identifies the plane where the interaction occurred.
Because the planes are of varying thicknesses, 
%(recall Table \ref{tbl:tgtthick}), 
we treat the problem as a classification problem rather than a regression problem.
To classify the data with the plane classifier, we use a total of 67 ``planecodes''.
This includes two ``overflow'' classes - code 0 for events upstream of the detector, and code 66 for everything downstream of the last considered plane.

Two potential sources of domain differences for this work are the cross section model and the simulation of the hadronic system created by a neutrino interaction.
Theoretical calculations are more reliable for the leptonic side of the neutrino interaction than the hadronic side; differences between the model and nature will show up only in subtle distortions in the outgoing lepton energy and angle spectra that our DCNN is likely insensitive to in this task.
In contrast, problems in the cross section model will change the relative frequency of occurrence of different reaction channels (e.g., quasi-elastic versus deep inelastic scattering, etc.) and $A$-dependent effects in the cross section will change the class balance for events in different neutrino targets. 
These effects are what \minerva was designed to measure and our current challenges in understanding these effects are cause for caution when training a neural network.

Problems in the simulation of the hadronic side of the reaction will lead to different final state particle spectra, with potentially important differences in particle type, energy, and direction.
%Because the hadronic recoil system is visually the primary determinant of the vertex location, the DCNN vertex finder is potentially sensitive to the way it is modeled.
Because hadronic recoil system is visually the primary determinant of the vertex location, and its existence is why DCNNs are expected to outperform traditional vertex reconstruction, the DCNN vertex finder is potentially sensitive to the way it is modeled.

\section{Simulated data sample details}
\label{sec:sampledetails}

%We used official \minerva simulation for these studies.
% Mont\'e Carlo (MC)
Simulated datasets were drawn from two different flux distributions - the Low Energy (LE) and Medium Energy (ME) configurations of the NuMI beam.
The NuMI beamline \cite{Adamson:2015dkw} features a tunable energy spectrum.
The peak of the beam energy distribution is about twice as large in the ME configuration as the LE. 
These higher energies lead to a larger overall cross section and increased likelihood for inelastic, particle cascade-rich events.
The \minerva flux has been described in detail in  \cite{Aliaga:2016oaz,Soplin:2016qrt}. 
See Section \ref{sec:fluxstudy} for more details on the beam configurations.
The main physics analysis we are pursuing with this study uses the ME flux, so the LE simulated datasets form one of the special physics variations sets we used to evaluate DANNs.
See Table \ref{tbl:numbrevt} for a breakdown of the sample sizes, with additional explanation below.

%Simulated data sets carry labels, e.g. ``1'' or ``1A'' etc. that reflect internal \minerva run period assignments that are not important here.
%We used two separate ``runs'' with effectively identical beam and physics configurations for the ME simulated dataset (ME 1A and 1B) and two versions of a \minerva run from the LE configuration - one with and one without re-scattering in the nucleus activated in the GENIE nuclear model. 
To study the impact of the DCNN on our DIS physics analysis, we divided our ME flux sample into two sets - one larger sample for physics evaluation (6.7 million events) and a smaller sample for ML training and hyperparameter optimization (3.9 million events).
Training was conducted with the second sample, with a small piece reserved for model validation between training epochs.
This sample was used to tune hyperparameters and network topology, with the final (test) performance evaluation taking place on the large (6.7 million events) sample.
We preserved the larger sample in this way to maximize the events available for physics studies unrelated to this work.

To study systematic bias in physics modeling, we use DANNs with out-of-domain samples based on event kinematics, different beam fluxes, and different treatments of the nuclear model.
From the ME training sample of 3.7 million events, we constructed two special kinematic samples.
The kinematic split separated events into one sample at $W \leq 1$ GeV, where $W$ is the invariant mass of the hadronic system produced in the interaction, and the other at $W > 1$ GeV.
These events are a subset of the regular ME training set, but models studying the kinematic split are not used for other purposes - every model sees completely statistically independent events for training as compared to validation or testing.
We used two configurations of the \minerva simulation for the LE running - one with and one without re-scattering in the nucleus activated in the GENIE nuclear model.
We refer to the re-scattering of daughter particles created by the primary neutrino interaction as Final State Interactions (FSI).
We used three-way sample splits for training, validation, and testing.
The validation set was used to monitor for evidence of overfitting, and test data saved until training was finished to provide a final evaluation.

\begin{table}[htb]
\centering
\begin{tabular}{lrrr}
\toprule
Sample                 & Training   & Validation & Test    \\
\midrule
\midrule
Physics analysis improvements  &  &  \\
\midrule
ME Physics Evaluation  & NA         & NA         &  6,661,231 \\
ME DL Training         & 3,700,000  & 230,000    & NA      \\ % 3,942,331 raw events
\midrule
\midrule
Model variation studies & & \\
\midrule
ME DL Study High $W$   &   625,000  & 100,000    & 100,000 \\ % 3,223,927 raw events
ME DL Study Low $W$    &   625,000  & 100,000    & 100,000 \\ %   706,811 raw events
LE                     & 1,900,000  & 150,000    & 150,000 \\
LE w/ No FSI           & 2,000,000  & 140,000    & 140,000 \\
\bottomrule
\end{tabular}
\caption{The sample sizes for the datasets used here in number of events.
For the ME flux, we retained a large sample for physics evaluation that doubled as the ``test'' sample for the deep learning work.
}
\label{tbl:numbrevt}
\end{table}

We represented the data as two-channel images using deposited energy and hit time to train the DCNN.
Each event contains three images - one for each tracker view - that are fed into separate convolutional towers.
We tested a stacked three-channel image that used deposited energies only and padded the U and V views so all three had the same dimensions, but found worse performance with that arrangement.
We believe there are two reasons for this.
First, the longitudinal resolution in the U and V views is different than the X.
This exacerbates the problem of mapping a kernel over a non-linear space (which is already theoretically a problem as the spacing between planes in the detector is not perfectly uniform).
%Second, the relationship between views is different geometrically - it stands to reason one might want differently shaped and separately evolved convolutional filters, etc. 
Second, the views have different relationships to the neutrino target modules and so the distributions of particles traversing them will be locally different, so differently shaped and separately evolved convolutional kernels are appropriate.

\begin{figure}
  \centering
  \includegraphics[width=0.99\textwidth]{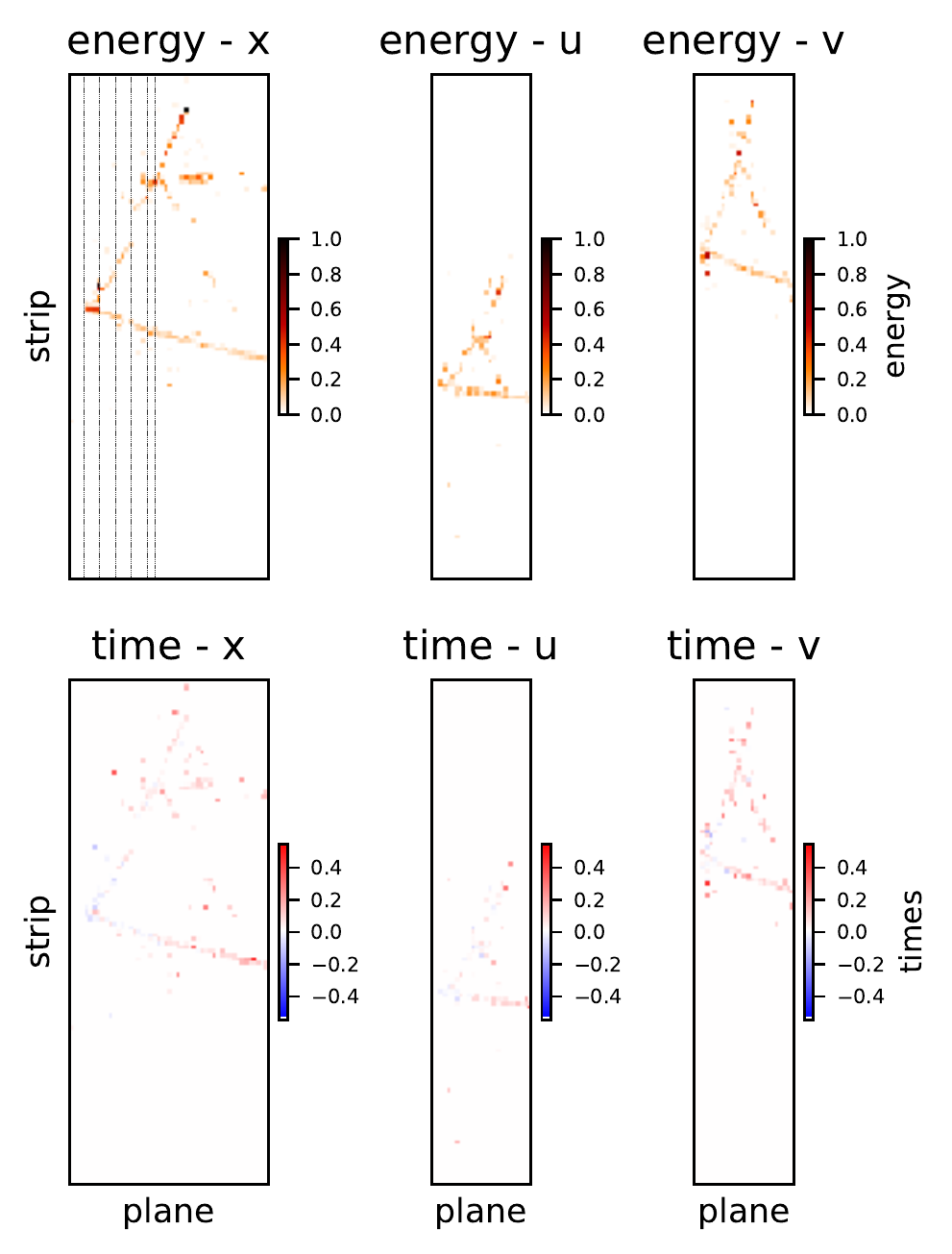}
  \caption{Example event in neutrino target 1.
  Note the discontinuity where the water target sits, reflecting the gap that is present there.
  The images are 127 pixels tall and 50 (25) pixels wide in the X (UV) views.
  }
  \label{fig:evt_1122000001004001_targ1}
\end{figure}

\begin{figure}
  \centering
  \includegraphics[width=0.99\textwidth]{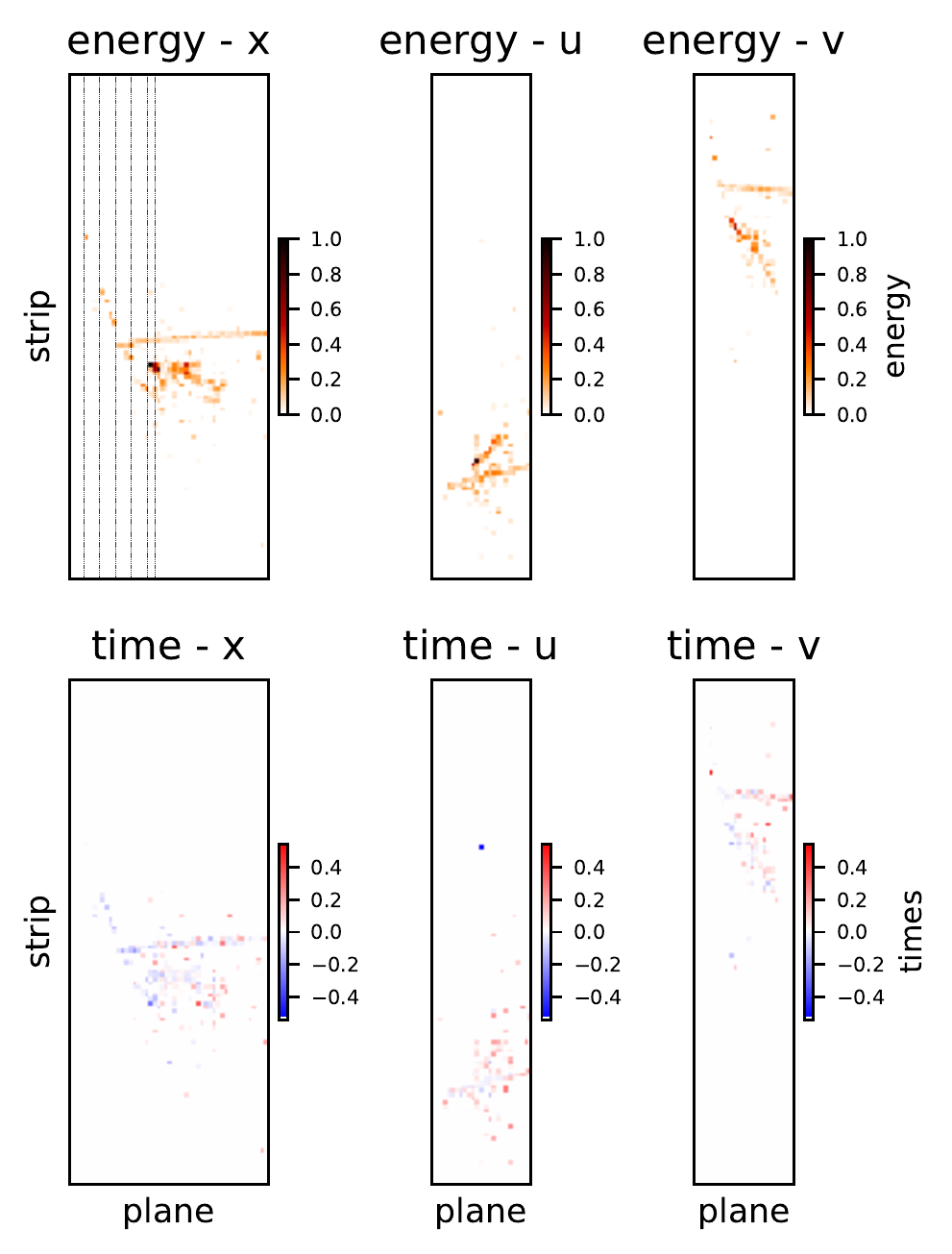}
  \caption{Example event in neutrino target 3.
  Note the particle cascade activity and the backwards going track.
  }
  \label{fig:evt_1122000001010701_targ3}
\end{figure}

\begin{figure}
  \centering
  \includegraphics[width=0.99\textwidth]{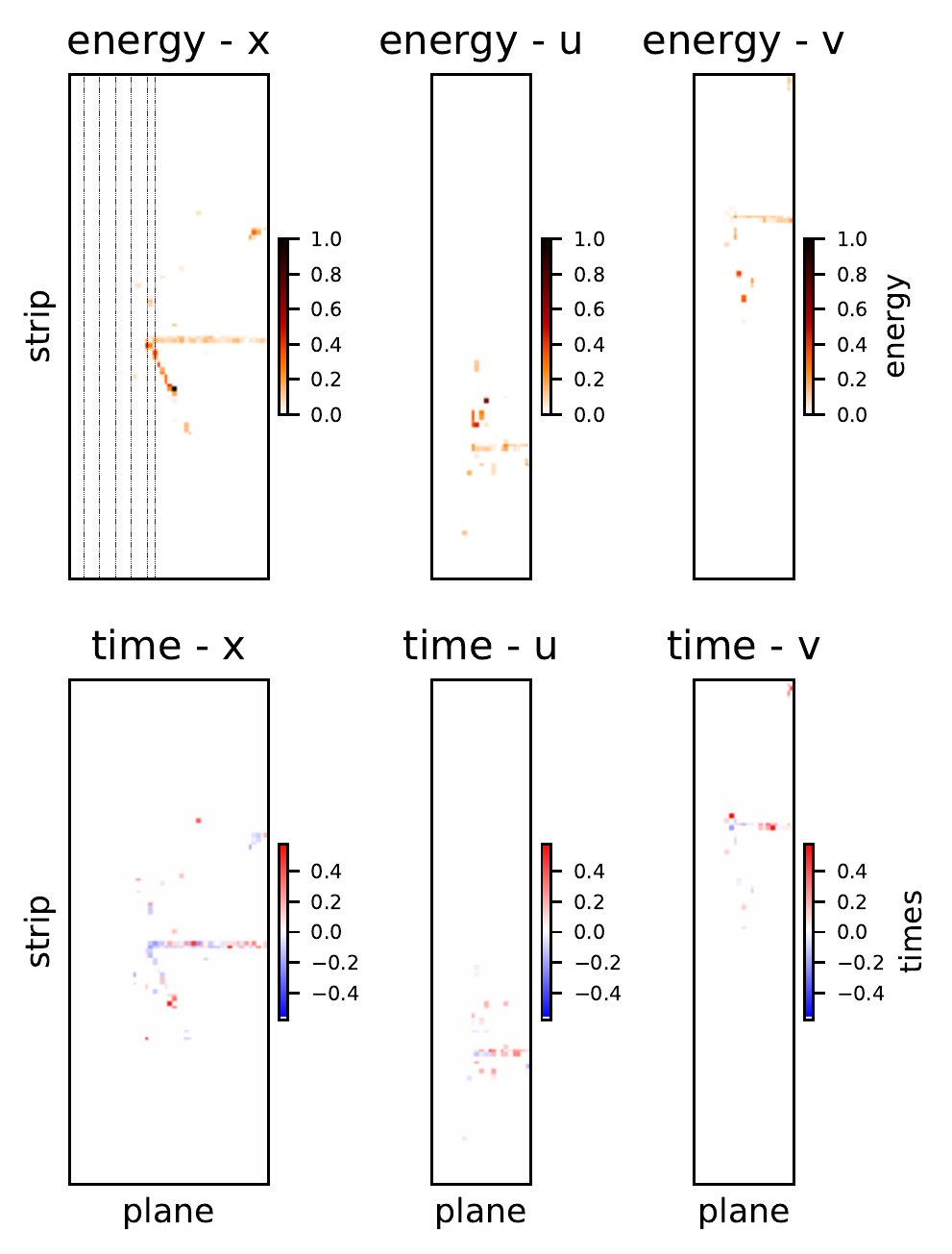}
  \caption{Example event in neutrino target 4.
  This is an example of an elastic-like event dominated by tracks.
  }
  \label{fig:evt_1122000001086001_targ4}
\end{figure}

The NNs consumed four-dimensional tensors of shape:
%\begin{center}
%\begin{verbatim}
%S = (Batch-size, Channel-depth, Image-height, Image-width),
%\end{verbatim}
%\end{center}
\[
\texttt{S} = \texttt{(Batch-size, Channel-depth, Image-height, Image-width),}
\]
where \texttt{Batch-size} describes the size of a mini-batch used during training based on stochastic gradient descent, and the other dimensions describe the shape of an individual image three-tensor.
Images for the NNs were pre-processed by normalizing the sum of the deposited energy values to unity on an event-by-event basis.
Timing values were pre-processed by subtracting the event time extracted by fitting a linear model to the anchor track and scaled by the largest absolute value of all the relative times in the event.
See Figures \ref{fig:evt_1122000001004001_targ1}, \ref{fig:evt_1122000001010701_targ3}, and \ref{fig:evt_1122000001086001_targ4} for a visualization of several example images.
The dashed lines in the X-view energy pane in the Figures go through the plane immediately upstream of, in order from left to right, targets 1, 2, 3, the water target (empty in this analysis), 4, and 5.
These dashed lines were not part of the image during training - only the raw images were used.
In these Figures, we break the event into six panels, with the two-deep tensor components separated into deposited energy and hit-time images.
%The two-deep tensors are broken into two images here for presentation purposes.
Note that daughter particles from the primary scattering event often re-interact in the target region - in Figure \ref{fig:evt_1122000001004001_targ1} we have a re-interaction in target 5, and in Figure \ref{fig:evt_1122000001010701_targ3} we have a re-interaction in target 4.

\begin{figure}[htbp]
  \begin{minipage}{1.0\linewidth}
    \centering
    \includegraphics[height=0.4\textheight]{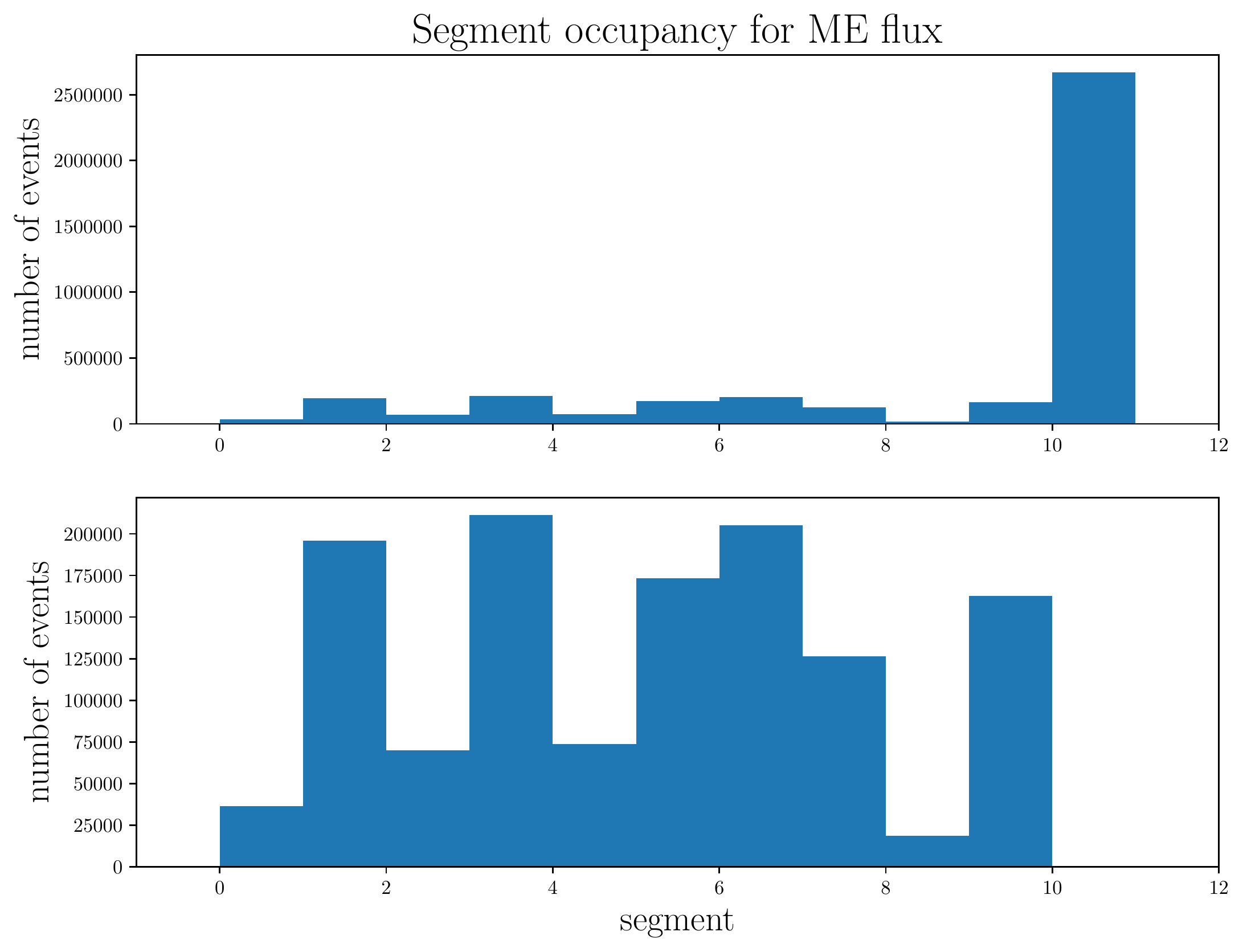}
    \subcaption{Segment occupancy for the ME flux, with the downstream class removed in the lower panel.
    Neutrino targets 1, 2, 3, 4, and 5 correspond to segments 1, 3, 5, 7, and 9.
    }
    \label{fig:segment_occupancy_me1Bmc}\par \medskip \vfill
    \includegraphics[height=0.4\textheight]{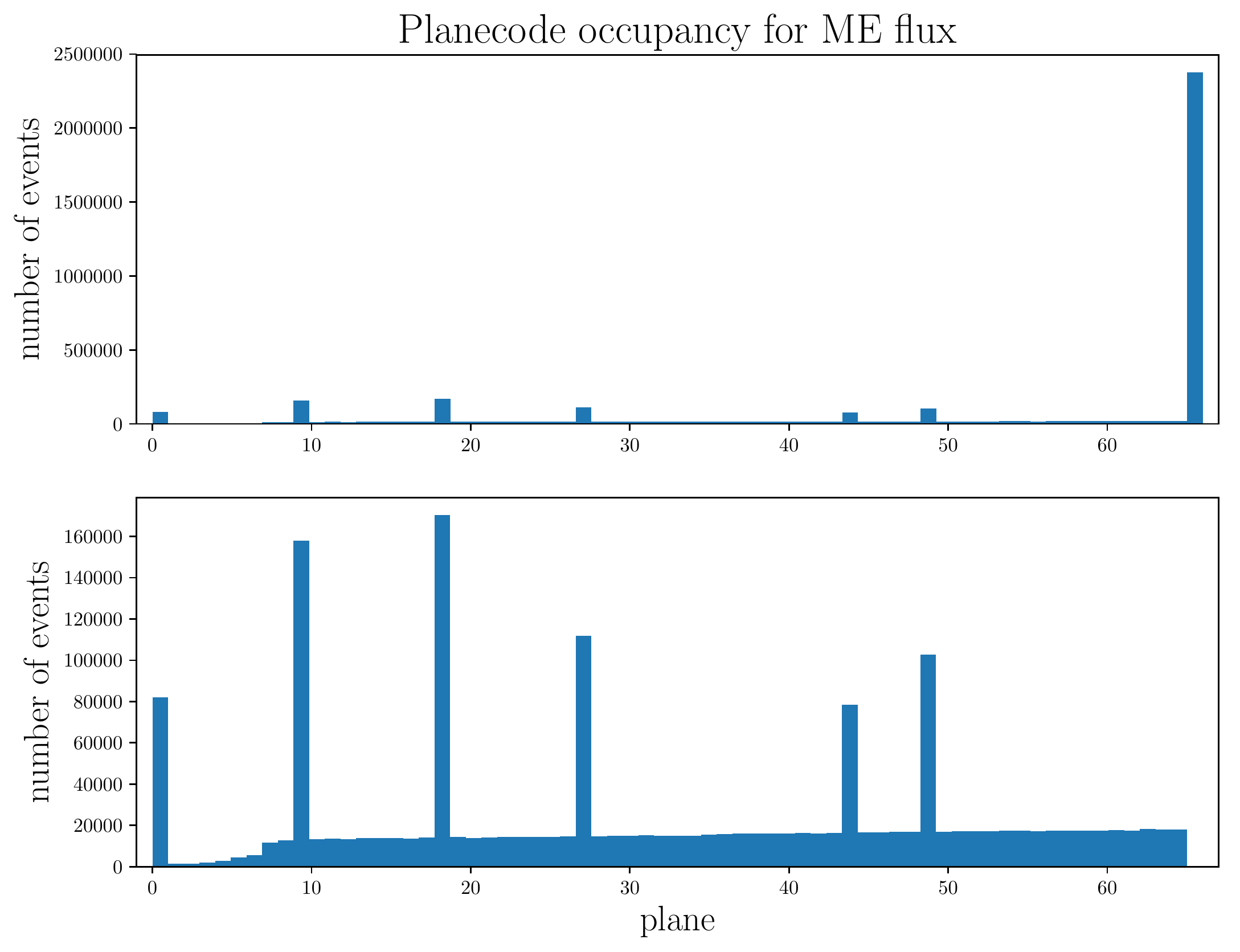}
    \subcaption{Planecode occupancy for the ME flux, with the downstream class removed in the bottom panel.
    The peaks in the bottom panel correspond to residual front-entering events, and then events in targets 1, 2, 3, 4, and 5 respectively.}
    \label{fig:planecode_occupancy_me1Bmc}\par \medskip \vfill
  \end{minipage}
  \caption{Class balance in the ME simulated dataset.
  The LE dataset shows similar trends.
  }
  \label{fig:classbal}
\end{figure}

The simulated data have a significant class imbalance.
See Figure \ref{fig:classbal} for a visualization of the class distribution.
There is a filter that runs prior to the creation of these data samples that removes front-entering tracks from the detector as a background source, and this sculpts the event distributions upstream of the neutrino targets.
Note that the front-entering events were added as a class to the plane classifier (Figure \ref{fig:planecode_occupancy_me1Bmc}) after studying the segment classifier performance.
The targets themselves are denser than the surrounding scintillator and so feature much higher event rates.
The thickness of the regions between targets varies, so the segment and plane classifiers have different event counts.
Downstream planes in the detector have higher geometric acceptance in the muon spectrometer and are therefore more likely to enter the sample. 
Finally, our overflow category - the region downstream of the targets - is much larger than the target region and so there were more events in that region due to detector mass scaling.

Because our final goal was to apply this technique to real detector data, where the vertex location is not known \emph{a priori}, we used all the events in the simulation to train the classifier, and not just events restricted by a cut on the true simulated vertex to the neutrino target region.
For faster execution time we limited the region that is exposed to the neural network - roughly two-thirds of the full span of the \minerva detector is not included in the images provided to the network (the downstream half of the active tracker, the electromagnetic calorimeter, and the hadronic calorimeter are cropped from the images).

\section{Deep neural network for vertex finding}
\label{sec:dnnforvert}

\subsection{Network topology}
\label{sec:networktopology}

DCNNs are not inherently localization algorithms.
In this problem we make some non-standard choices (e.g., non-square kernels) to adapt a classification algorithm for the task of localization.
%The neutrino beam lies nearly along the  $z$-axis of \minerva and the the three views that define the detector geometry all include the  $z$-axis.
A great deal of localization information is contained directly in the energy distribution in $z$.
By deliberately breaking the translational symmetry inherent in a DCNN along the $z$-axis, but preserving the symmetry on the orthogonal axes, we achieve better hybrid localization-classification performance.
During the early stages of classifier design we discovered it was possible to achieve over 80\% localization accuracy in the neutrino targets using the planar sum information only, ignoring transverse segmentation.
This inspired our approach of factorizing the problem into a classification task along the transverse axes that studies the radiation patterns coming from the candidate vertices and into a localization problem along the $z$-axis.

%In the process of network design we found significant advantage in using non-square kernels and non-square pooling units.
In this work, we allowed the images to shrink along the transverse dimension but largely preserved the image size along the $z$-axis.
To achieve this we only perform pooling along the ``X'',  ``U'' and ``V'' axes in the corresponding views and the kernels which extract the features are non-square; they are much larger along the transverse direction than they are along the $z$-direction. 
Additionally, we pooled tensor elements together only along the transverse axis, not along the $z$-axis.
By fully preserving the image along the $z$-axis, we allow the network to simultaneously utilize the gross features with respect to where activity stops, starts, and peaks.
The fully connected layers are able to correlate the transverse patterns with the distribution along $z$ to build an effective localization algorithm.

\begin{figure}
  \centering
  \includegraphics[height=0.95\textheight]{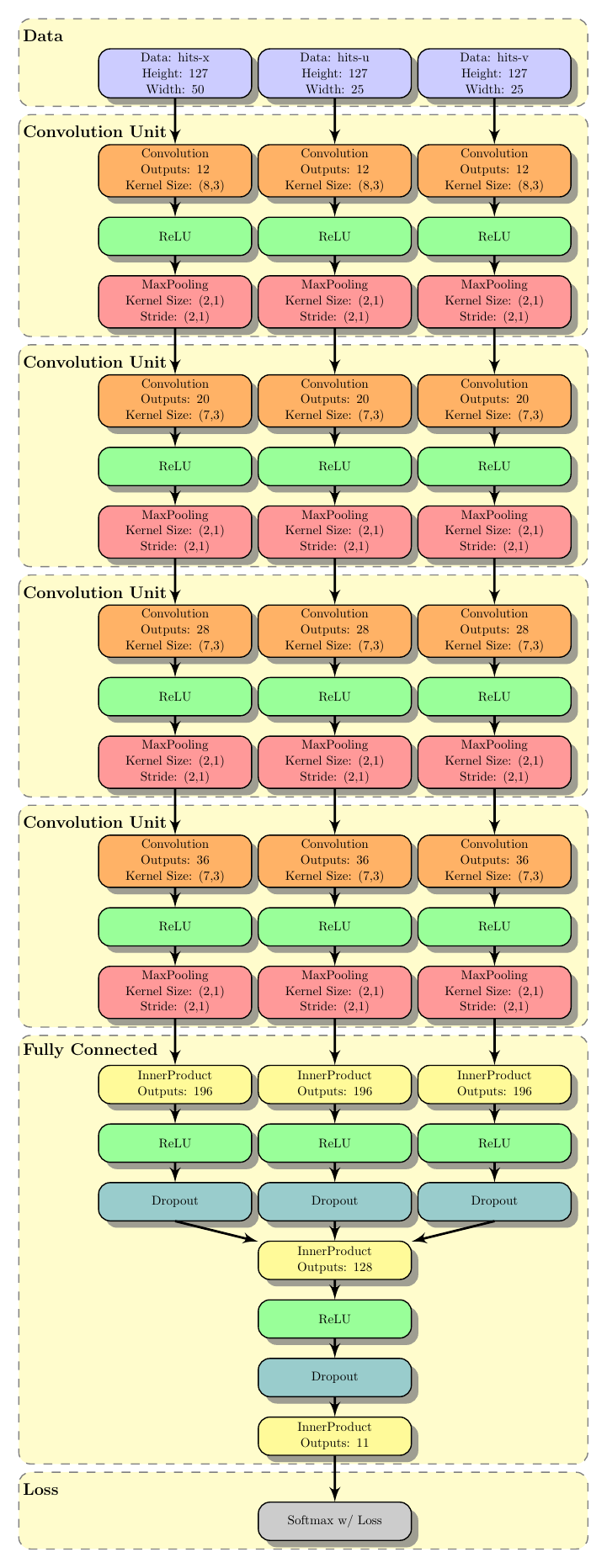}
  \caption{The \minerva vertex finding neural network.}
  \label{fig:minerva_net}
\end{figure}

The network consists of three separate towers for X,  U and V views, as shown in Figure \ref{fig:minerva_net}. 
Each tower consists of four iterations of convolution and max pooling layers with ReLUs acting as the non-linear activations.
Each pooling layer consists of a kernel which decreases the dimension along the transverse axis by one.
After the four iterations of Convolution, ReLU and Pooling, there is a fully connected layer with 196 semantic output features.
The outputs for the three views are concatenated and fed to another fully connected layer with 98 outputs which in turn is input for a final fully connected layer with 11 (67) outputs for the segment (plane) classifier that is input to a softmax layer.
The fully connected layers operate on the semantic representation at the end of the feature discovery layers to associate features to desired outputs.
Finally, we assign the network cost using a cross-entropy function, which is subsequently minimized.
Further details about the network design choices are described elsewhere \cite{ijcnn7966131}.

\subsection{Training strategies}
\label{sec:trainstrat}

We consider two applications in this work: in the first we employ a DANN segment classifier (recall Figure \ref{fig:eleven_segments}) using one-deep tensor inputs (energy only) and in the second we employ a plane classifier using two-deep tensor inputs.
In both cases we use identical network topologies, although the structure of the last fully connected layer is different between the two because of the different number of classes (11 vs 67).
Training of the two types of networks is independent - we do not train one and then export the convolutional feature layers to the other, etc.

Our networks are trained using a softmax cross-entropy loss function.
The output from the final fully-connected layer of the network is passed to a \emph{softmax} function such that the sum of the values in the final layer is one.
The softmax function is defined in Equation \ref{eqn:softmax}:
\begin{equation}
S_i = \frac{e^{x_i}}{\sum_j e^{x_j}}.
\label{eqn:softmax}
\end{equation}
Here the $x_j$ are the input layer values to the softmax, with $j$ running from one to the number of nodes in the layer, and $S_i$ is the $i$th node value in the softmax layer.
The \emph{cross entropy} between two vectors $p$ and $q$ is defined according to Equation \ref{eqn:crossentropy}:
\begin{equation}
H\left(p, q\right) = - \sum_{x_i} p \left(x_i\right) \log q\left(x_i\right) 
\label{eqn:crossentropy}
\end{equation}
For training purposes, one vector, $p$ is a one-hot representation of the truth label and the other vector, $q$, is the softmax of the final output of the network.
A one-hot vector converts a number $i$ in a set of numbers up to $n$ to a vector of length $n$ with the $i^{\text{th}}$ element equal to one and all other elements equal to zero.
During training, parameters are adjusted to minimize this loss function.

% comment on weight decay and L2 reg from: https://bbabenko.github.io/weight-decay/
The training approaches are slightly different between the segment and plane classifiers.
% due to some asynchronicity in the work.
For both applications we use Glorot's prescription \cite{AISTATS2010_GlorotB10} for weight initialization when first training a model.
For the by-plane application we use the ``AdaGrad'' \cite{AdaGrad} algorithm for the learning rate decay schedule with an initial value of 1e-3, employ L2 regularization of the weight tensors with a scale factor of 1e-4, and train with a batch size of 500.
For the by-segment application, we use an inverted exponential weight decay algorithm for the learning rate, we use weight decay with a regularization scale of 1e-4 (roughly equivalent to L2 regularization with a scale factor of 5e-5), and we use a batch size of 50.
The batch size is reduced because the DANN network occupies more space in memory on the GPU owing to the extra fully-connected layers in the domain classifier.
We do not modify the training strategy for the segment classifier when a DANN is active versus when it is not for simplicity of comparison between the methods.
% A different training strategy under the changed conditions is likely optimal and might result in different long-term behaviors.

\section{Results of DCNN study}
\label{sec:improvement}

We consider a DIS cross section measurement in the \minerva passive neutrino targets \cite{annethesis, mayathesis}.
DIS events involve scattering with partons inside a nucleus and are characterized by high momentum transfer squared ($Q^2$) and high invariant mass values of the final state hadronic system ($W$).
These features typically appear in events as more than three final state particles and energetic hadrons that are likely to produce particle cascades in their interactions with the detector material.

The DCNN performs significantly better than our previous track-based method for vertex finding.
We do this comparison using the segment classifier because it defines the ``targets'' in the same way the track-based algorithm does, namely, each target includes one plane of scintillator from the upstream and two from the downstream.
This is further illustrated below.
In the plane classifier, we do not include neighboring scintillator planes in the definition of the ``target.''

\begin{figure}
  \centering
  \includegraphics[width=0.95\textwidth]{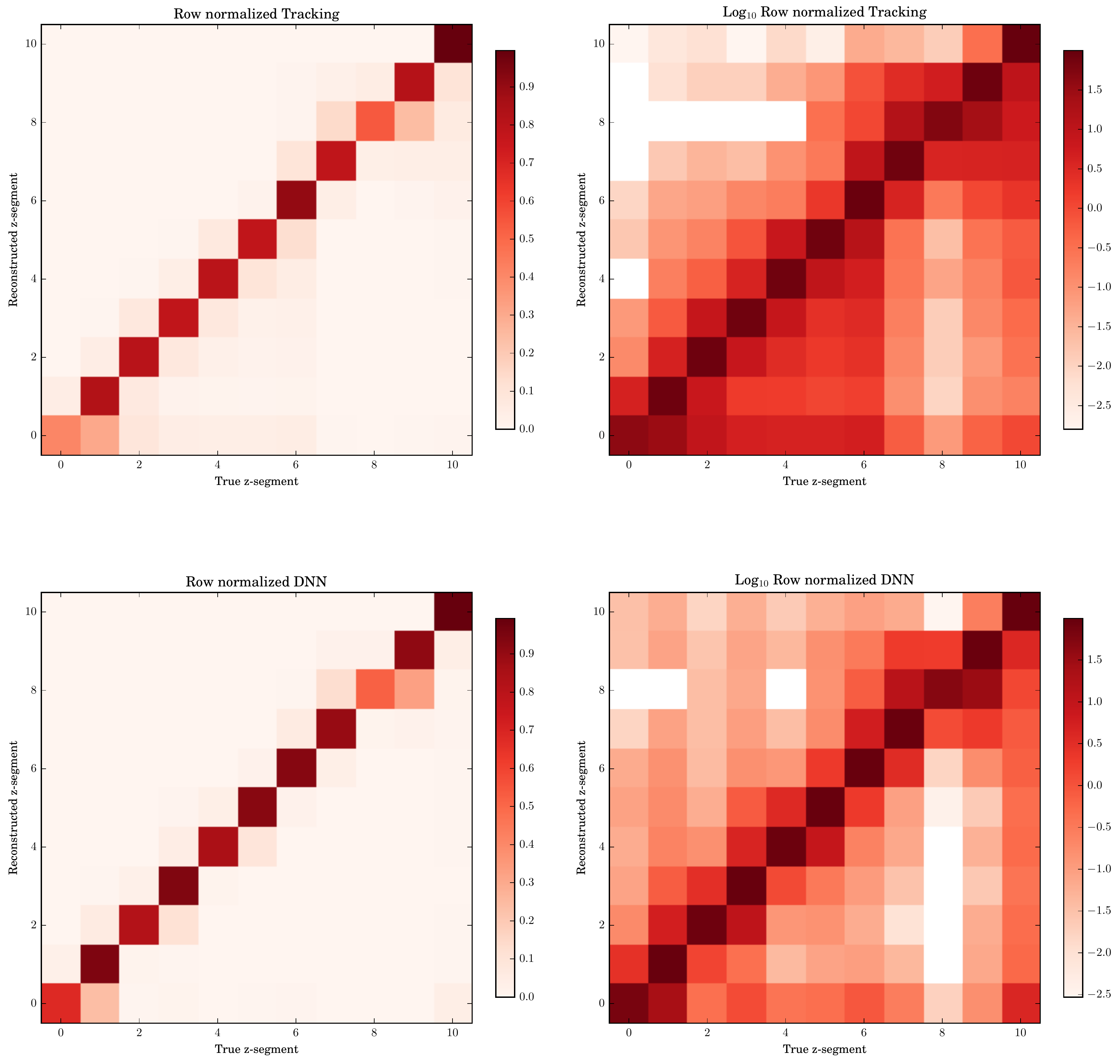}
  \caption{Track-based reconstruction performance (top row) and deep neural network reconstruction performance (bottom row) for the segment classifier in ME simulation.
  Here, we show row-normalized confusion matrices (so the sum of the matrix entries reading along a row are normalized to one).
  Because we row-normalize, the plot should be read by row and interpreted as revealing for a given reconstructed segment (row value), what fraction of the events truly originated in each of the possible 11 segments.
  The plots on the left have a linear color scale, and the plots on the right show $\log_{10}\left(\text{percentages}\right)$. 
  See Table \ref{tbl:epsilonpur} for the numerical values in the plots shown here.
}
  \label{fig:confusion_matrices_trk_v_dnn_loglin_pur_ME_MC+No_DANN_Partner+10700000_iterations}
\end{figure}

Consider first the row-normalized confusion matrix for the segment classifier presented in Figure \ref{fig:confusion_matrices_trk_v_dnn_loglin_pur_ME_MC+No_DANN_Partner+10700000_iterations}.
A confusion matrix presents the correlation between a reconstructed value and a true value.
Here, we chose to row-normalize to account for the class imbalances and to ease interpretation.
The DCNN-based reconstruction method presents a clearly more diagonal confusion matrix, indicating that it is more likely to correctly reconstruct a vertex.
See Table \ref{tbl:epsilonpur} for a numerical accounting of the difference in performance.

The performance in the neutrino targets improves to a larger degree than the performance in the regions between the targets.
We believe this is because the targets are made of much denser material and, as the cross section scales roughly with the number of nucleons in the target, the network has learned to preferentially draw ``borderline'' events into the targets.
This truthfully reflects the relative cross sections in the target materials versus the interstitial scintillator as modeled in the simulation, but naturally raises concerns that we have introduced a dependency on the cross section model in our training sample.
This is investigated further in Section \ref{sec:dann}.

%% time python Lasagne/minerva_triamese_epsilon.py -t -d ./minosmatch_nukecczdefs_127x50x25_xuv_me1Bmc.hdf5 -s models/prediction1461944084_epsilon_v2r0.npz -v
\begin{table}[htb]
\centering
\begin{tabular}{ccrrr}
\toprule
Segment & Neutrino Target & Track-based & DCNN & Change\\
        &                 &     (\%)    & (\%) &  (\%) \\
\midrule
 0 &   & 41.1 $\pm$ 0.95 & 68.1 $\pm$ 0.60 & 27.0  $\pm$ 1.14 \\
 1 & 1 & 82.6 $\pm$ 0.26 & 94.4 $\pm$ 0.13 & 11.7  $\pm$ 0.30 \\
 2 &   & 80.8 $\pm$ 0.46 & 82.1 $\pm$ 0.37 &  1.30 $\pm$ 0.60 \\
 3 & 2 & 77.9 $\pm$ 0.27 & 94.0 $\pm$ 0.13 & 16.1  $\pm$ 0.30 \\
 4 &   & 80.1 $\pm$ 0.46 & 84.8 $\pm$ 0.34 &  4.70 $\pm$ 0.60 \\
 5 & 3 & 78.0 $\pm$ 0.30 & 92.4 $\pm$ 0.16 & 14.4  $\pm$ 0.34 \\
 6 &   & 90.5 $\pm$ 0.20 & 93.0 $\pm$ 0.14 &  2.50 $\pm$ 0.25 \\
 7 & 4 & 78.3 $\pm$ 0.35 & 89.6 $\pm$ 0.22 & 11.3  $\pm$ 0.42 \\
 8 &   & 54.3 $\pm$ 1.12 & 51.6 $\pm$ 0.95 & -2.70 $\pm$ 0.15 \\
 9 & 5 & 81.6 $\pm$ 0.30 & 91.2 $\pm$ 0.18 &  9.6  $\pm$ 0.34 \\
10 &   & 99.6 $\pm$ 0.01 & 99.3 $\pm$ 0.13 & -0.30 $\pm$ 0.02 \\
\bottomrule
\end{tabular}
\caption{Vertex reconstruction purity by segment and technique. Here ``purity'' is defined by row normalizing the confusion matrix (see Figure \ref{fig:confusion_matrices_trk_v_dnn_loglin_pur_ME_MC+No_DANN_Partner+10700000_iterations}) and is interpreted as the probability that a given reconstructed value was equal to the underlying true generated value.
The same test sample of 600,000 events was used for track and DCNN-based evaluation.
Uncertainties quoted here are purely statistical.
}
\label{tbl:epsilonpur}
\end{table}

\begin{figure}
	\begin{subfigure}{0.5\textwidth}
    	\centering
		\includegraphics[width=0.99\linewidth]{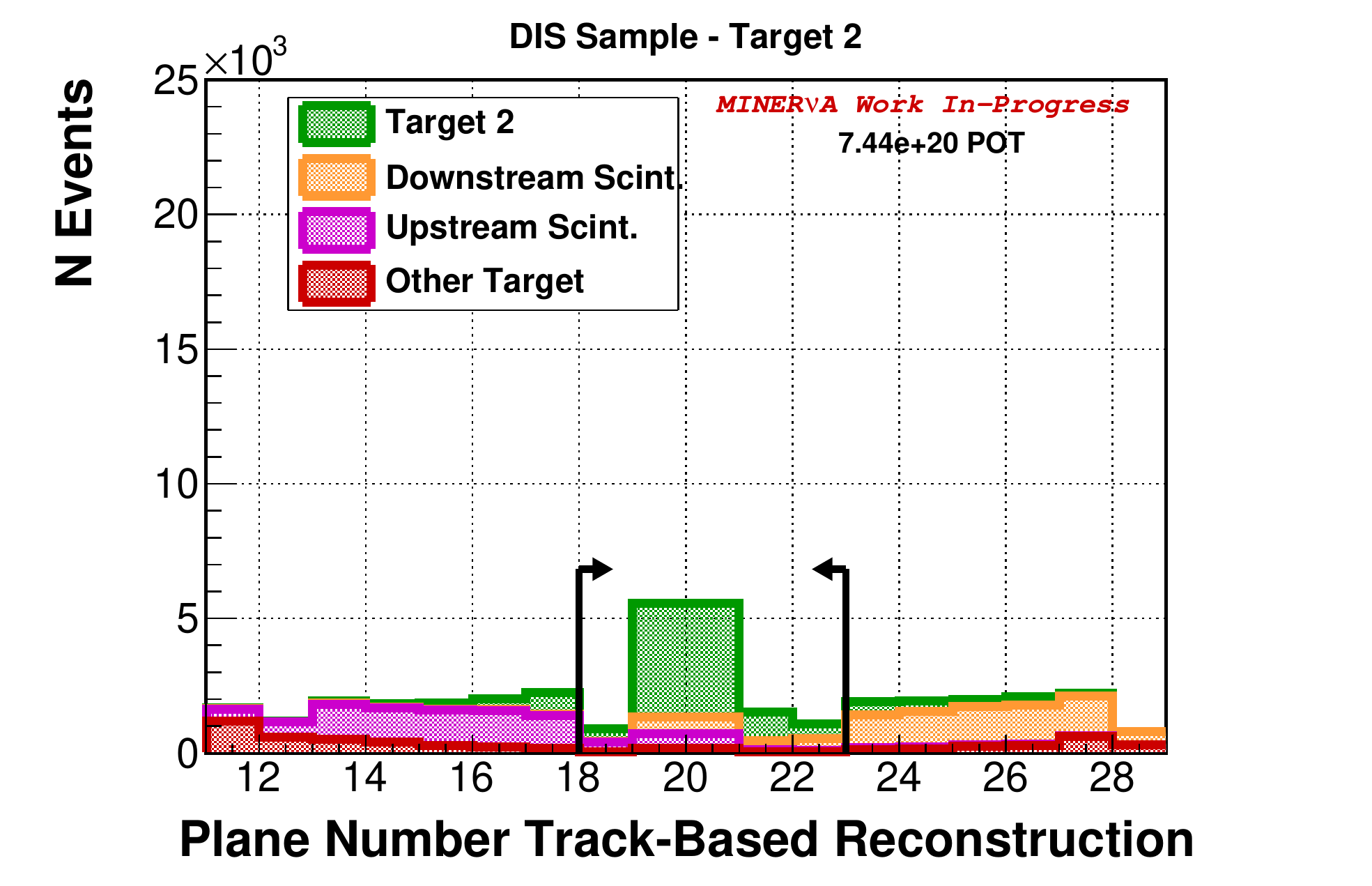}
  		\caption{Track-based vertex finding.}
  		\label{fig:RegionPosStackPlot_plane_t02_z82_minervame1A_vTrackBasedVertex}
	\end{subfigure}
	\begin{subfigure}{0.5\textwidth}
    	\centering
		\includegraphics[width=0.99\linewidth]{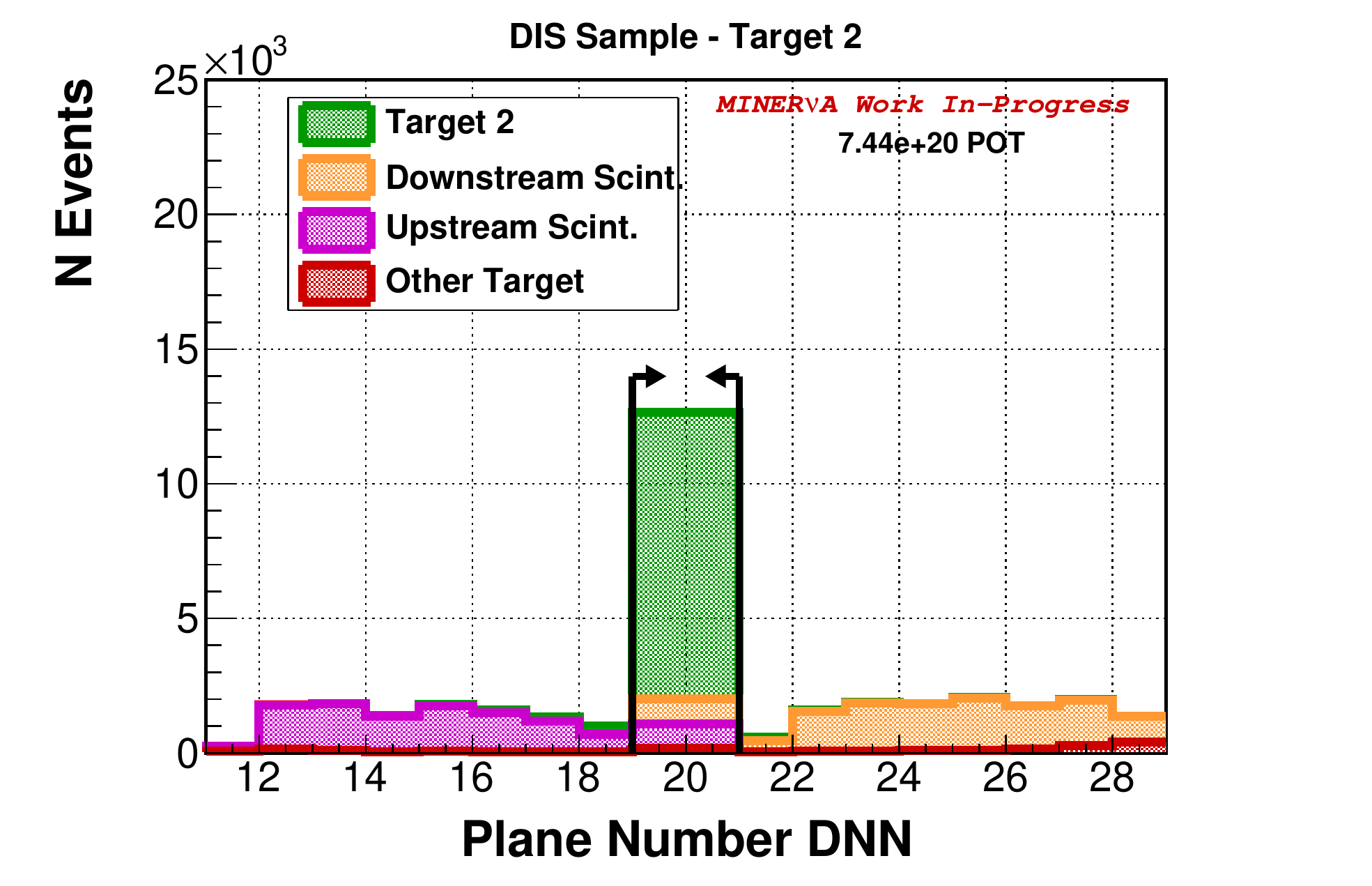}
  		\caption{DCNN-based vertex finding.}
  		\label{fig:RegionPosStackPlot_plane_t02_z82_minervame1A_vDNNVertex}
	\end{subfigure}
	\caption{Source interaction material by reconstructed vertex location near neutrino target 2 (here, planes 19 and 20).
	The color corresponds to the true source material.
	All the scintillator planes between targets 1 and 2 are shown on the left side of the plots, and all the scintillator planes between targets 2 and 3 are shown on the right side of the plots.}
	\label{fig:RegionPosStackPlot_plane_t02_z82_minervame1A}
\end{figure}

\begin{figure}
	\begin{subfigure}{0.5\textwidth}
    	\centering
		\includegraphics[width=0.99\linewidth]{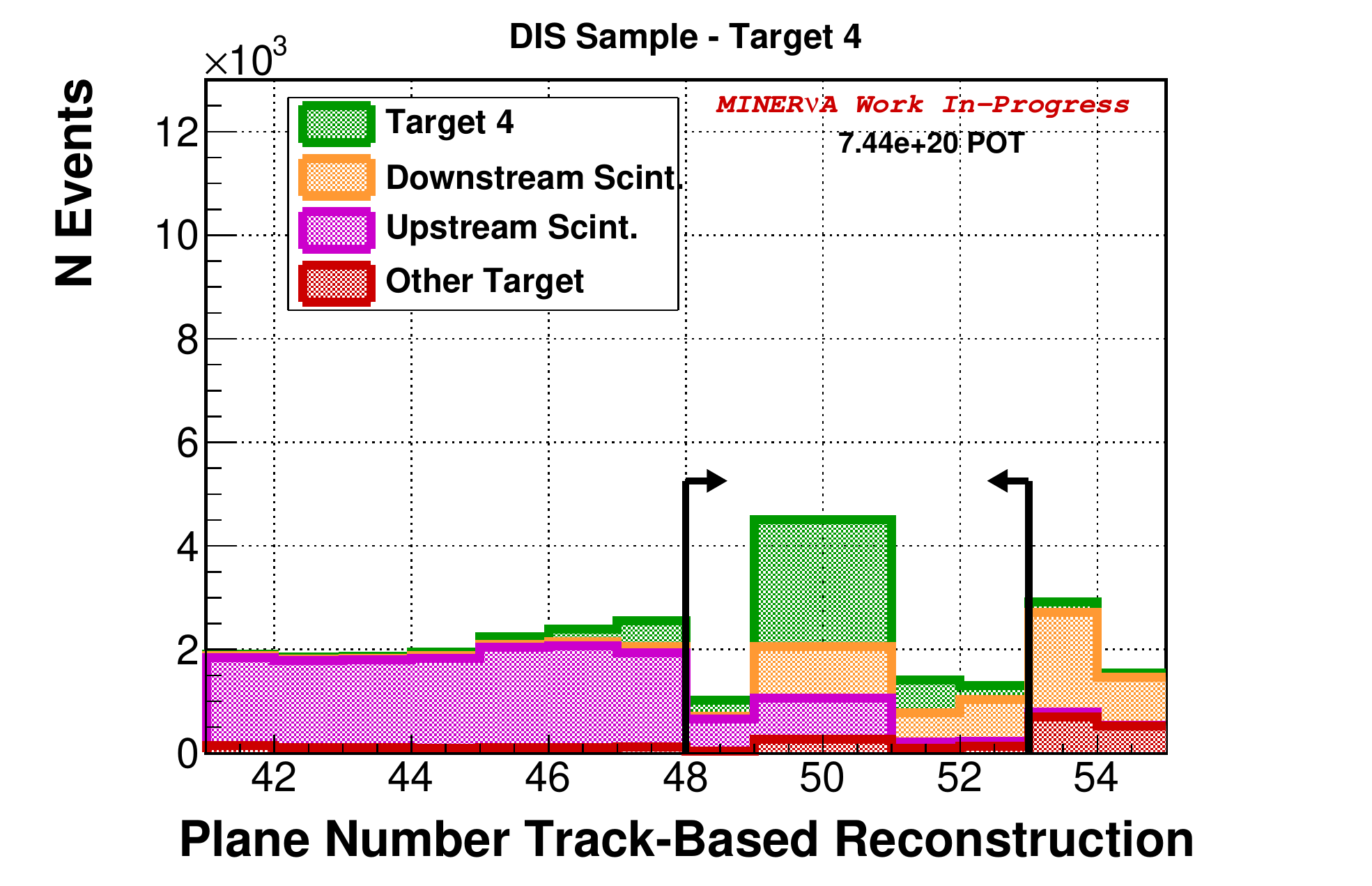}
  		\caption{Track-based vertex finding.}
  		\label{fig:RegionPosStackPlot_plane_t04_z82_minervame1A_vTrackBasedVertex}
	\end{subfigure}
	\begin{subfigure}{0.5\textwidth}
    	\centering
		\includegraphics[width=0.99\linewidth]{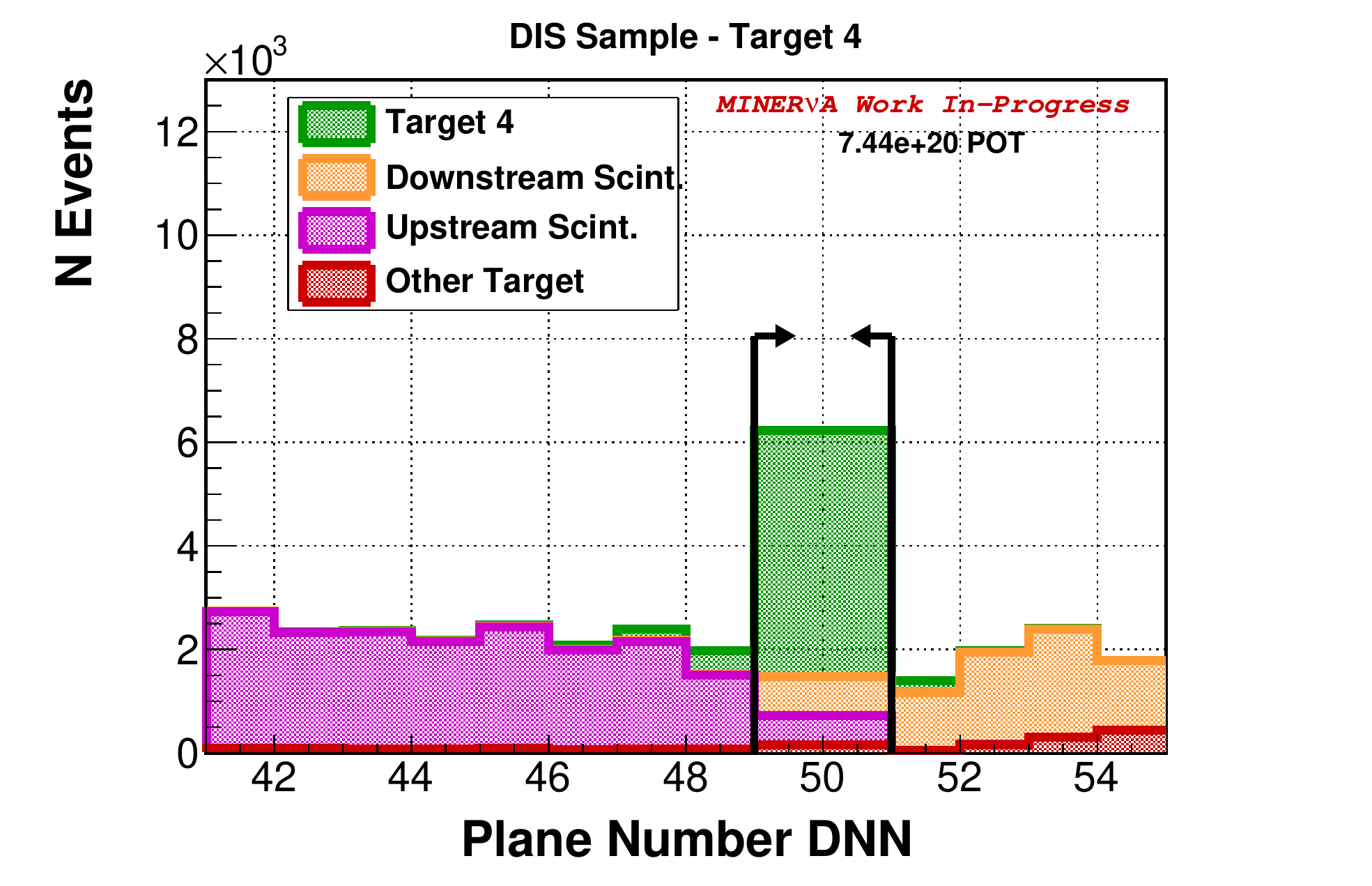}
  		\caption{DCNN-based vertex finding.}
  		\label{fig:RegionPosStackPlot_plane_t04_z82_minervame1A_vDNNVertex}
	\end{subfigure}
	\caption{Source interaction material by reconstructed vertex location near neutrino target 4 (here, planes 49 and 50).
	The color corresponds to the true source material.
	All the scintillator planes between targets 3 and 4 are shown on the left side of the plots, and all the scintillator planes between targets 4 and 5 are shown on the right side of the plots.
	Note that the gap between targets 4 and 5 is smaller than any of the other gaps between targets.}
	\label{fig:RegionPosStackPlot_plane_t04_z82_minervame1A}
\end{figure}

\begin{figure}
	\begin{subfigure}{0.5\textwidth}
    	\centering
		\includegraphics[width=0.99\linewidth]{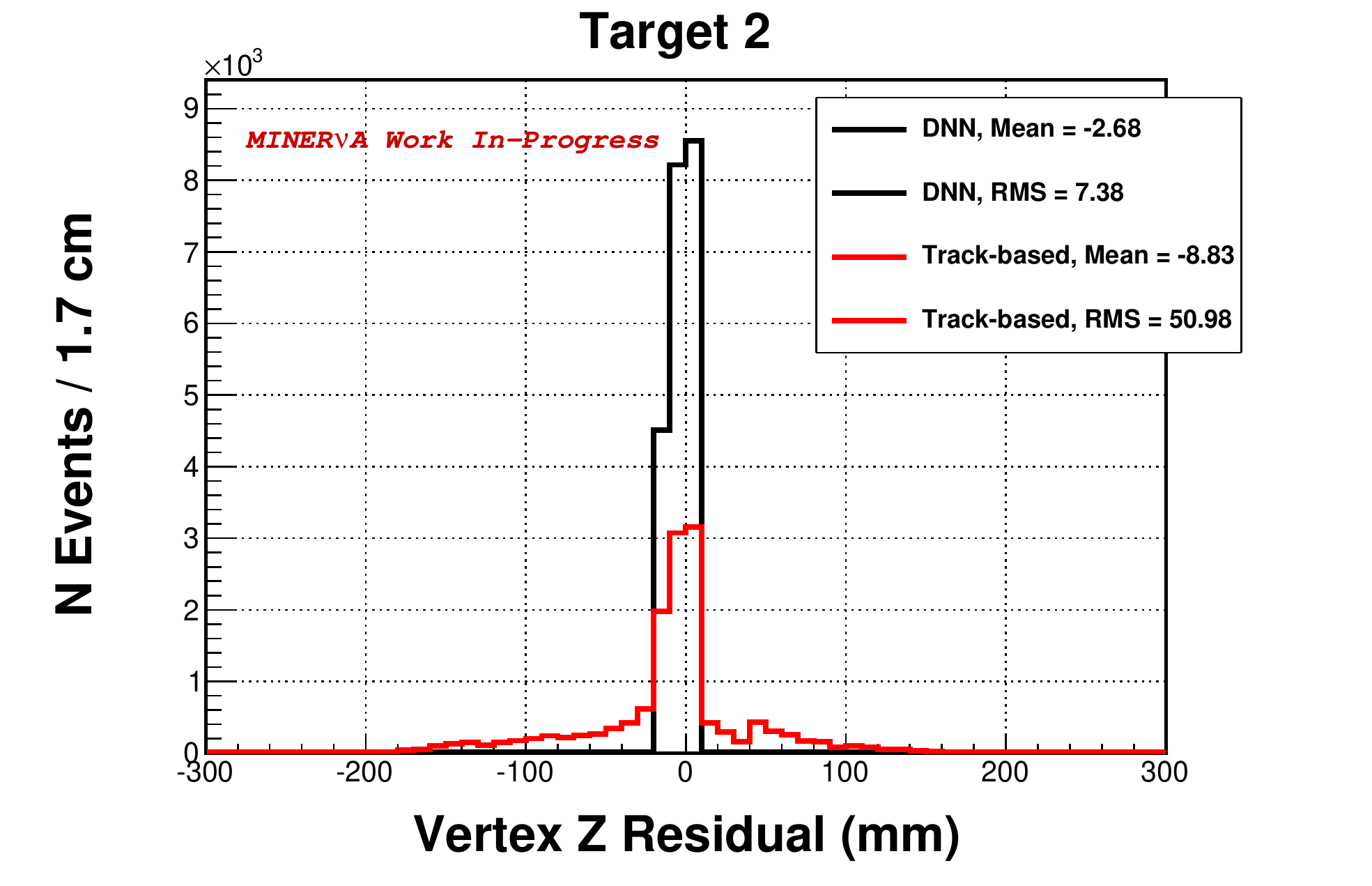}
  		\caption{Neutrino target 2 residuals.}
  		\label{fig:ZResiduals_t02_z82_minervame1A_vDNNVertex}
	\end{subfigure}
	\begin{subfigure}{0.5\textwidth}
    	\centering
		\includegraphics[width=0.99\linewidth]{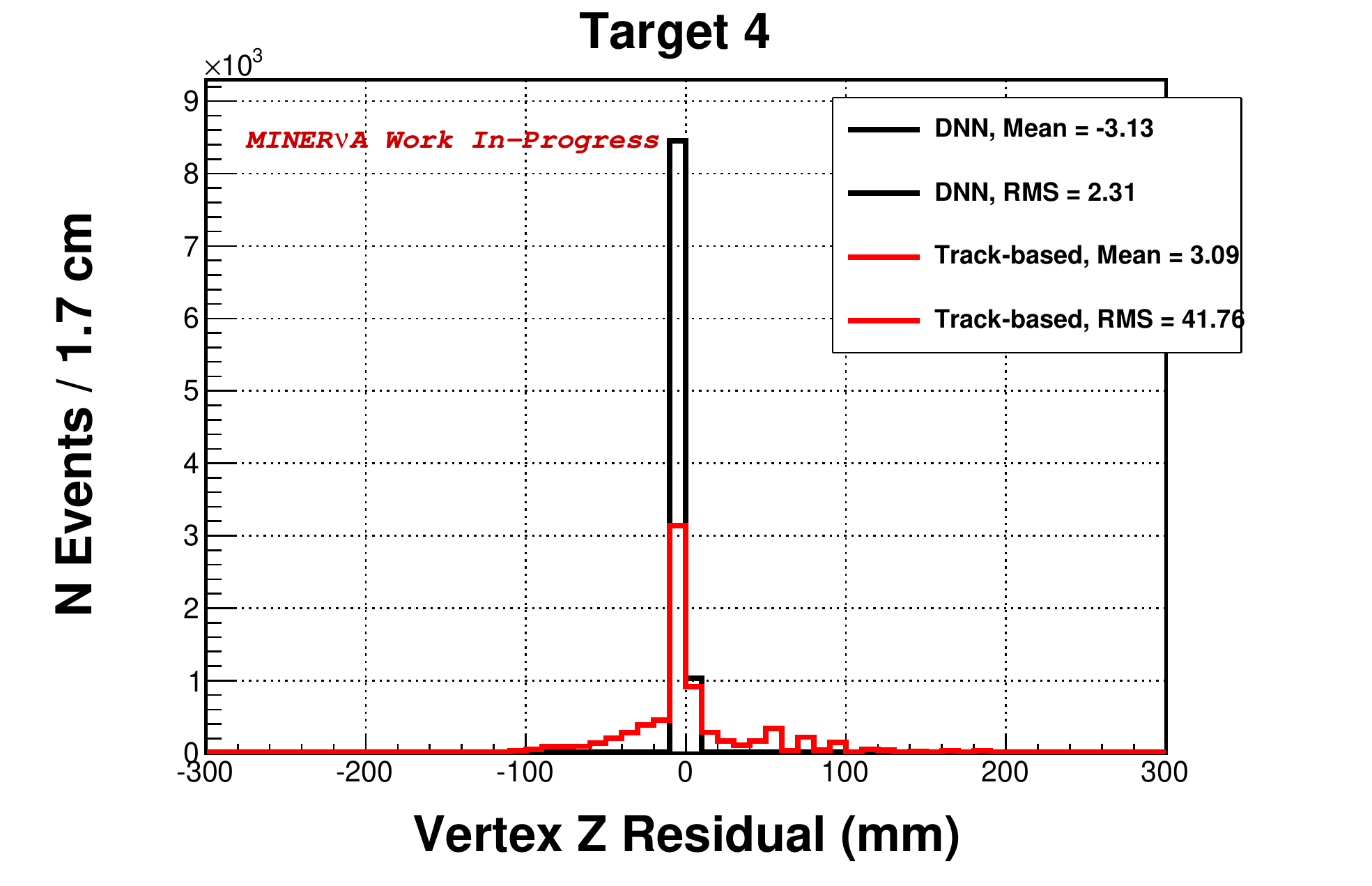}
  		\caption{Neutrino target 4 residuals.}
  		\label{fig:ZResiduals_t04_z82_minervame1A_vDNNVertex}
	\end{subfigure}
	\caption{Reconstructed vertex position residuals for neutrino targets 2 and 4.
		Here, we present integer differences based on the reconstructed plane directly for the DCNN and by computing which plane the floating point $z$ vertex found by the track-based method corresponds to by look-up table and then computing the difference in the number of planes.}
	\label{fig:ZResiduals_z82_minervame1A_vDNNVertex}
\end{figure}

In the plane classifier, we choose 60 scintillator planes in the neutrino targets region and the five targets, and attempt to localize events directly into a specific plane or target.
We additionally include an underflow and overflow class.
Figures \ref{fig:RegionPosStackPlot_plane_t02_z82_minervame1A} and \ref{fig:RegionPosStackPlot_plane_t04_z82_minervame1A} show the difference in the performance in the inclusive nuclear ratio analysis in the vicinity of targets 2 and 4 using the plane classifier algorithm.
The left side of the figures shows the performance of the track-based algorithm and explains why we included additional layers of scintillator in the definition of a ``target'', namely a large amount of signal bleeds out of the target and into the surrounding scintillator, so it is ultimately advantageous to include those regions and perform more stringent background subtraction to drive sufficient statistical power in the measurement.
Figure \ref{fig:ZResiduals_z82_minervame1A_vDNNVertex} compares the vertex plane residuals for a subset of the events when using the DCNN vs the track-based method.

% Figure source:
% https://minerva-docdb.fnal.gov/cgi-bin/private/ShowDocument?docid=14537

% Figure source:
% https://minerva-docdb.fnal.gov/cgi-bin/private/ShowDocument?docid=14537

% Figure source:
% https://minerva-docdb.fnal.gov/cgi-bin/private/ShowDocument?docid=14537

\begin{figure}
	\begin{subfigure}{0.5\textwidth}
    	\centering
		\includegraphics[width=0.99\linewidth]{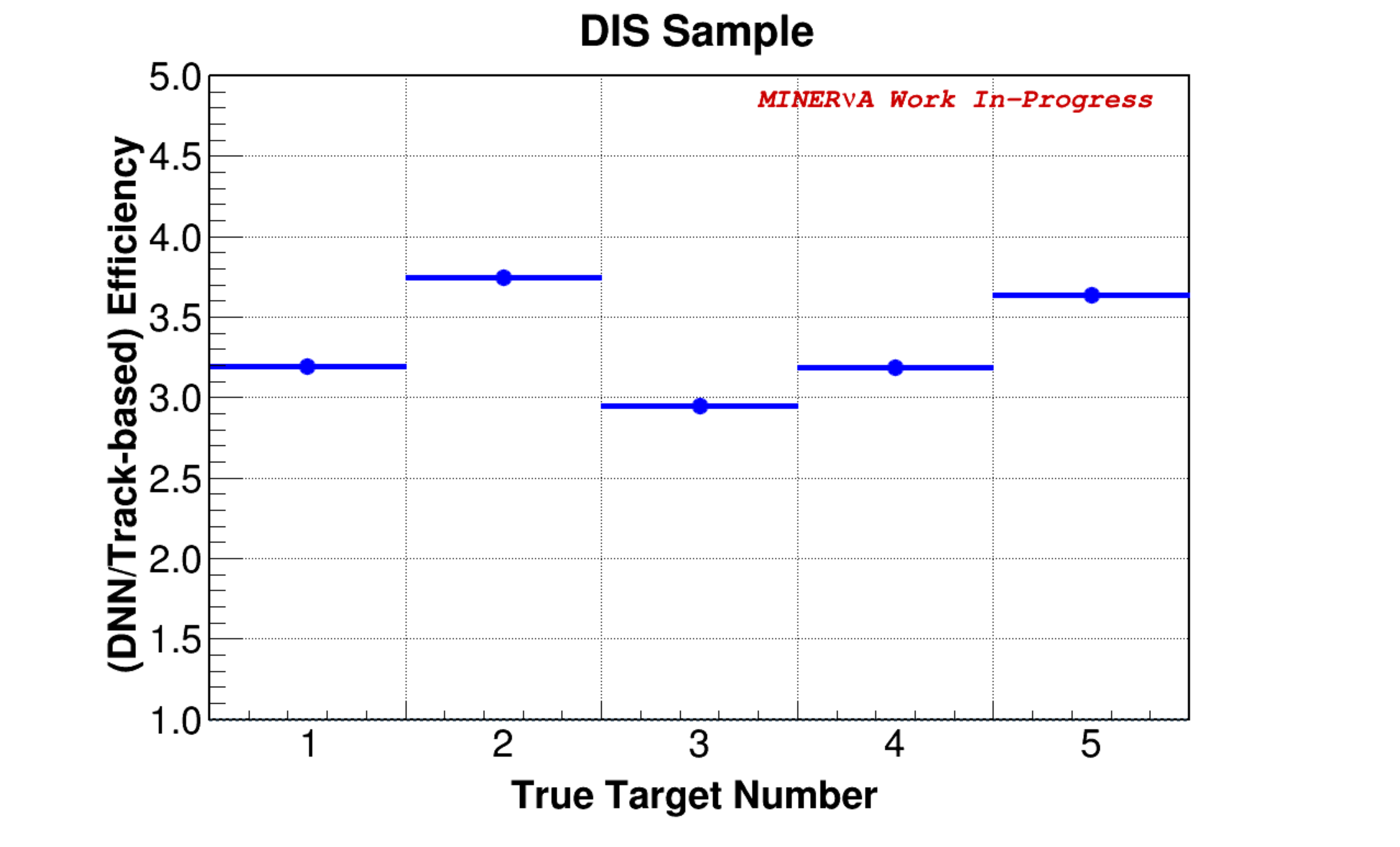}
  		\caption{Efficiency improvement factors}
  		\label{fig:EffPurRatio_DISEff}
	\end{subfigure}
	\begin{subfigure}{0.5\textwidth}
    	\centering
		\includegraphics[width=0.99\linewidth]{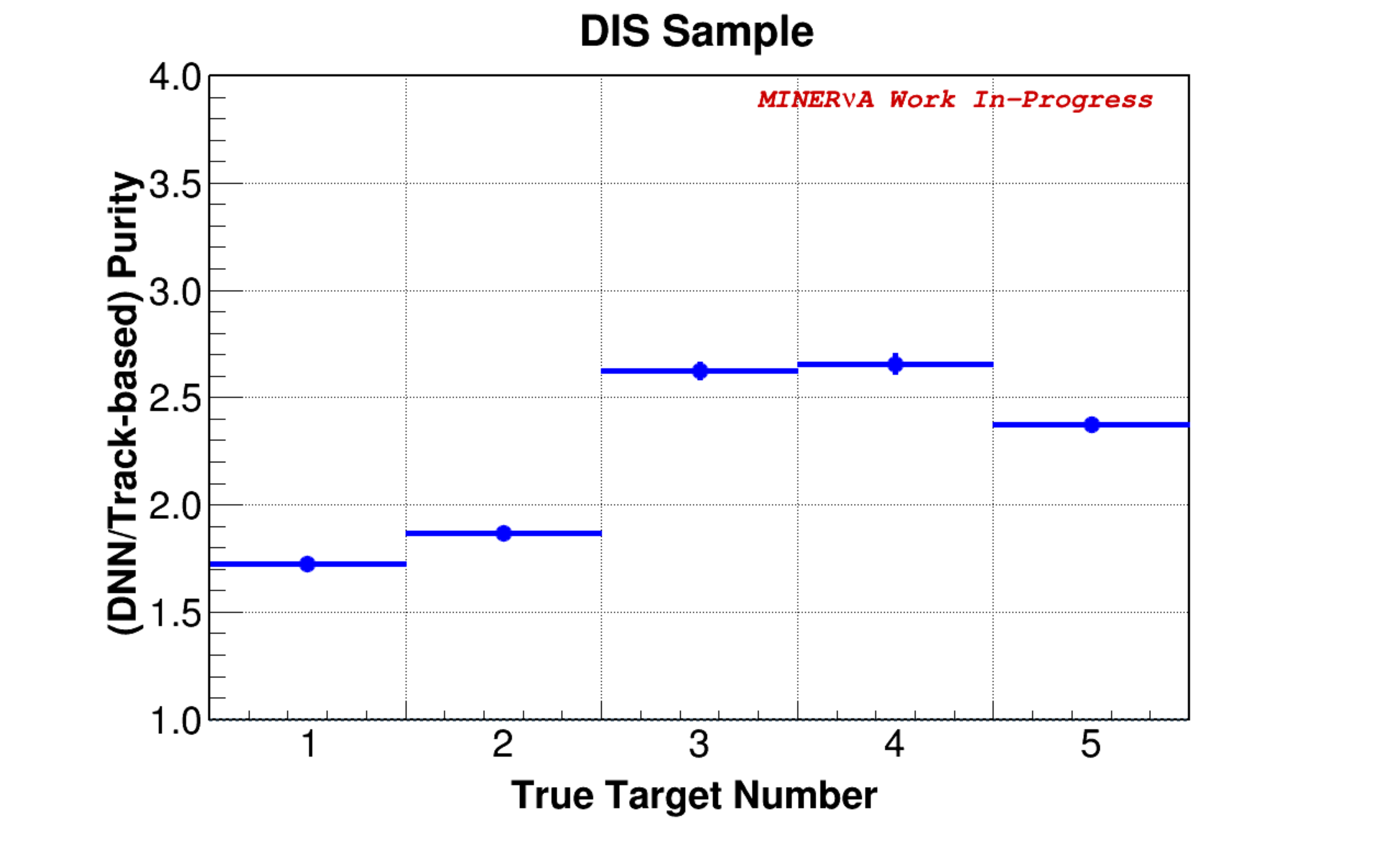}
  		\caption{Purity improvement factors}
  		\label{fig:EffPurRatio_DISPur}
	\end{subfigure}
	\caption{Efficiency and purity gains as a multiplicative factor (i.e., DCNN divided by tracking) for the Deep Inelastic Scattering (DIS) analysis.
	}
	\label{fig:EffPurRatio_DIS}
\end{figure}

\begin{figure}
	\begin{subfigure}{0.5\textwidth}
    	\centering
		\includegraphics[width=0.99\linewidth]{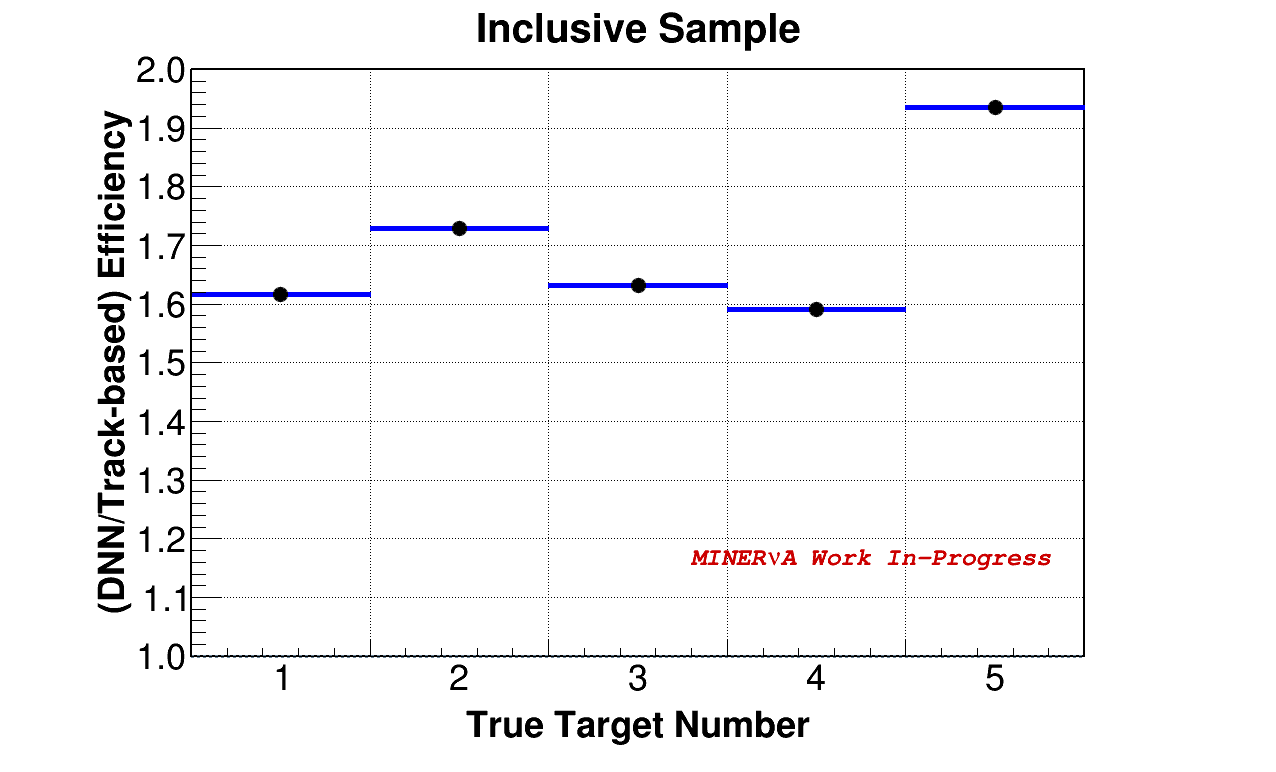}
  		\caption{Efficiency improvement factors}
  		\label{fig:EffPurRatio_InclEff}
	\end{subfigure}
	\begin{subfigure}{0.5\textwidth}
    	\centering
		\includegraphics[width=0.99\linewidth]{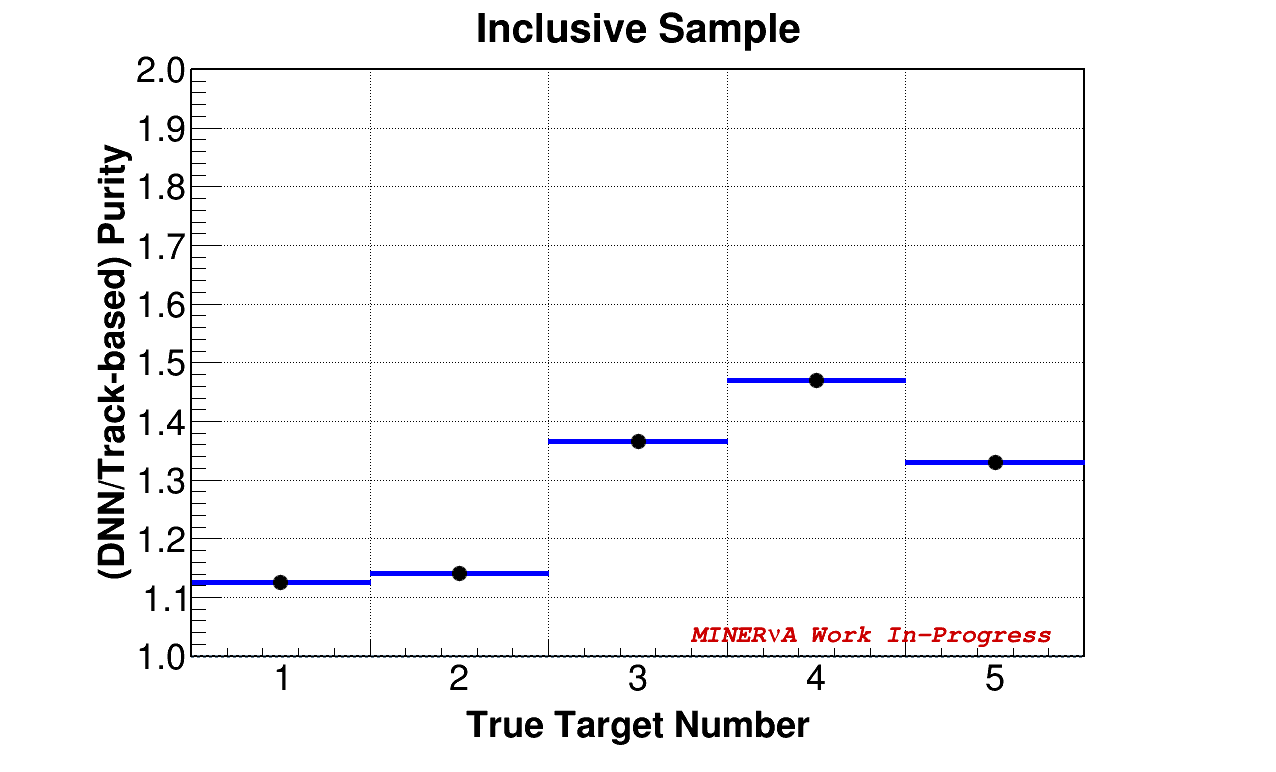}
  		\caption{Purity improvement factors}
  		\label{fig:EffPurRatio_InclPur}
	\end{subfigure}
	\caption{Efficiency and purity gains as a multiplicative factor (i.e., DCNN divided by tracking) for the inclusive analysis.
	Note the different vertical scales as compared to Figure \ref{fig:EffPurRatio_DIS}.}
	\label{fig:EffPurRatio_Incl}
\end{figure}

The final impact on the signal efficiency and purity for the DIS analysis is shown in Figure \ref{fig:EffPurRatio_DIS}, where efficiency is fraction of true events reconstructed, and purity is the fraction of events where the reconstructed value was equal to the underlying true generated value.
The efficiency gains over track-based vertex finding methods are striking.
%The gains are largely coming from better handling of events with particle cascade-rich topologies.
The efficiency is improved largely as a result of better vertex finding in events with cascades.
%The DIS events are a high momentum transfer ($Q^2$), high final state hadronic invariant mass ($W$) subset of the inclusive charged current analysis.
%Because these events tend to be contain more activity (especially particle cascades), the DCNN vertex reconstruction has a more important relative advantage. 
DIS events tend to be cascade-rich, and we obtain a better performance improvement from the introduction of the DCNN than in the set of inclusive events.
The gains in the inclusive sample, Figure \ref{fig:EffPurRatio_Incl}, are smaller because a larger fraction of that sample is composed of low produced particle count final states where the track-based methods are successful.
The track-based vertex finding method employed in \minerva was originally designed to handle low multiplicity elastic and resonance events with only two or three particles in the final state.
The track-based vertex finding performs well for these events; indeed, it is very difficult to improve the fit resolution of a very clean multi-track event.
% - and so the relative overall improvement is much smaller.

% Figure source:
% Maya Slack

% Figure source:
% Maya Slack

\section{Domain adversarial neural networks (DANNs)}
\label{sec:dann}

A known weakness of machine learning in discovery science is the use of an imperfect simulation model for training.
Classification may rely on features that are systematically different between the simulation and detector data.
Traditionally, this is managed by running the simulation many times with different parameters to study domain-sensitive differences.
However, this is computationally expensive, particularly so for machine learning based approaches where re-training is often required when the physics model underlying the simulation is varied.
What is needed is an efficient way to control for domain differences between the simulation and detector data.

Here we report a study of the use of DANNs \cite{JMLR:v17:15-239} for domain adaption in the \minerva vertex finding problem.
DANNs are designed to correct a supervised learner trained in a source domain where truth labels are available but applied in a target domain where they are not.
Here, the network produces two outputs - one is a regular feature classifier (e.g., vertex location) while the other is domain discriminator that attempts to distinguish events from the source and target domains. 
%The DANN strategy is to jointly minimize the loss from the feature classifier and maximize the loss from the domain discriminator. 
%DANNs use the same training methods as DCNNs in the source domain, when the truth information is available, and a second loss function is introduced to distinguish source from target domain events.  The DANN strategy is to minimize the first loss function, called the feature classifier, while maximizing the second loss function, called the domain classifier.

\begin{figure}
  \centering
  \includegraphics[height=0.95\textheight]{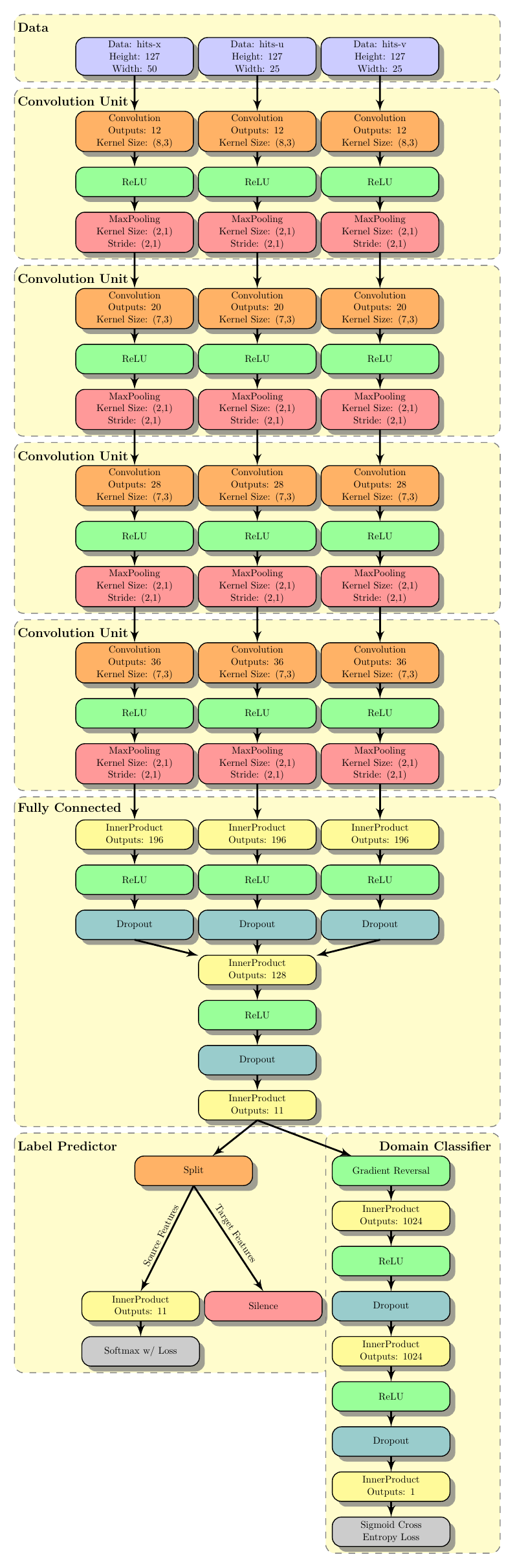}
  \caption{The \minerva vertex finding neural network with the DANN.
  The network used for this diagram used named input layers to distinguish sample sources.}
  \label{fig:minerva_net_dann}
\end{figure}

During training, batches are constructed with half the samples coming from the source domain and half from the target domain.
Only images from the source domain are used to compute the loss for the features classifier, while images from the source and target domains are both used to compute the loss for the domain discriminator.
By jointly minimizing the loss on the features classifier and maximizing the loss on the domain discriminator in the feature generating layers, the features classifier is trained to use only features that are present in both the source and target domains.
%The final outputs of the network are a feature prediction (e.g., a vertex position) and a domain prediction.
At evaluation time, the feature prediction is meant to solve the physics problem under investigation, while the domain prediction may be monitored for bias.
If the network domain discrimination performance is different from random guessing, that is a sign of a failure in the domain adaption.
A diagram of the DANN can be seen in Figure \ref{fig:minerva_net_dann}.

During physics analysis, simulation is the source domain and detector data is the target domain.
We would like to compare the performance on detector data of a classifier trained only on simulation to a classifier trained on simulation with a DANN ``partner'' from the detector data, but this evaluation is difficult because the detector data lacks truth labels.
Therefore, to validate the method, we will compare multiple simulation-only datasets that feature different, unrealistically large perturbations to the underlying physics model and see if the DANN allows the networks to recover proper in-domain performance.
In essence, we are masking the labels on our partner sample and using only the raw images and the domain information to augment training.
But by using simulation, we can test the level of success for the approach.

An additional application of a DANN (not explored here) would be to examine the results of the features classifier for both simulation samples, but continue to enforce loss maximization on the domain classifier.
This technique would possibly enable the use of two simulations of different quality (for example, a fast MC that is cheap to produce and a fully rendered MC that is expensive to produce) to boost sample statistics for training while saving computation time.

%% -- Overfitting study --
%%
%\input{sec_danns_subsec_overfit.tex}

\subsection{Technical implementation details}
\label{sec:techdetails}

Our DANN is implemented in Caffe and we now describe some of the technical details of that implementation making reference to class names in the code.
The implementation we used was based on work \cite{GaninDANN} from the original DANN paper \cite{JMLR:v17:15-239}.
The DANN version of the network includes special data layers at the bottom of the network.
These layers add an adversarial image into the network and also provide labels for the domain (1 for target and 0 for source). 
The target image is concatenated with the source image and the domain labels are also concatenated. 
The network then follows the design as above until the final \texttt{InnerProduct} (fully connected) layer in the non-adversarial network is reached.
After this layer is a \texttt{Slice} layer which divides the features into source features or target features and then discards the target features (or the features from the adversarial domain) with a \texttt{Silence} layer during the training of the network and only the source features are used during the testing/validation of the network.
These final features are used as an input to a final \texttt{InnerProduct} layer with the 11 (or 67) outputs corresponding to target region.
A \texttt{GradientScalar} layer also uses the final \texttt{InnerProduct} layer from the non-adversarial network as an input and then outputs to a series of two \texttt{InnerProduct}, \texttt{ReLU} and \texttt{Dropout} layers. 
This finally is input into an \texttt{InnerProduct} layer which, with the domain labels, is the input to a \texttt{SigmoidCrossEntropyLoss}.
This final layer is what selects features which exist in both domains.

\subsection{Simulation studies}
\label{sec:simstudies}

%The simulation used in these studies include simulation with FSI turned on and off and simulation with the LE flux and preliminary ME flux. 
%We would use simulation corresponding to one run period for training and testing (a small portion would be used for testing and not used for training) while simulation from another run period would be used for validation or to test the robustness and behavior of the network. 
%When we used data as an adversarial partner, we also divided it into the run periods and treated it in an equivalent manner to simulation.

We consider three variations in the simulation physics model, motivated by likely problems in the neutrino analysis.
We change (a) the flux model, (b) the FSI nuclear model, and (c) we split the sample by event kinematics into low and high $W$ (hadronic invariant mass) regions to emulate a mis-modeling of the process of hadron formation.
To understand the robustness of the results in these three studies, we also perform a study of the sensitivity of the results to the network initialization scheme in the context of the kinematic split analysis.

Neutrino flux is notoriously difficult to model and it is the dominant uncertainty in most cross section measurements.
Because we actually measure a convolution of the neutrino flux with the cross section, changing the flux model is strongly correlated to changing the measured cross section.
Both effects modulate the observed rate in the detector as a function of neutrino energy.

The nuclear model is a significant uncertainty and one of the most problematic pieces is the modeling of the propagation of hadrons produced at the initial interaction point through the dense nuclear medium after creation (FSI).
Modeling of this propagation in different nuclei could produce differences between simulation and data.
FSI produces low energy charged hadrons which obscure the region near the interaction point.
They are also a prominent cause of backward going hadrons, which could confuse the longitudinal  position assessment (and for which transverse information processed by the neural networks can help distinguish the true position).
This could lead to biased reconstruction in some neutrino targets over others when switching from the simulation domain to the data domain.

The performance of a machine learning system for the measurement of DIS is also sensitive to the model for hadronization, the process by which partons become hadrons in the nucleus.
By splitting the sample into low and high $W$ samples we create two samples with very different compositions of tracks and particle cascades (the low $W$ sample is dominated by tracks, and the high $W$ sample is dominated by particle cascades).
Hadronization models are strong functions of $W$, so by splitting the sample we can approximate a move to extremes in the modeling of the hadronization process.

\subsubsection{Flux model}
\label{sec:fluxstudy}

\begin{figure}
\centering
\begin{subfigure}{.5\textwidth}
  \centering
  \includegraphics[width=1.0\linewidth]{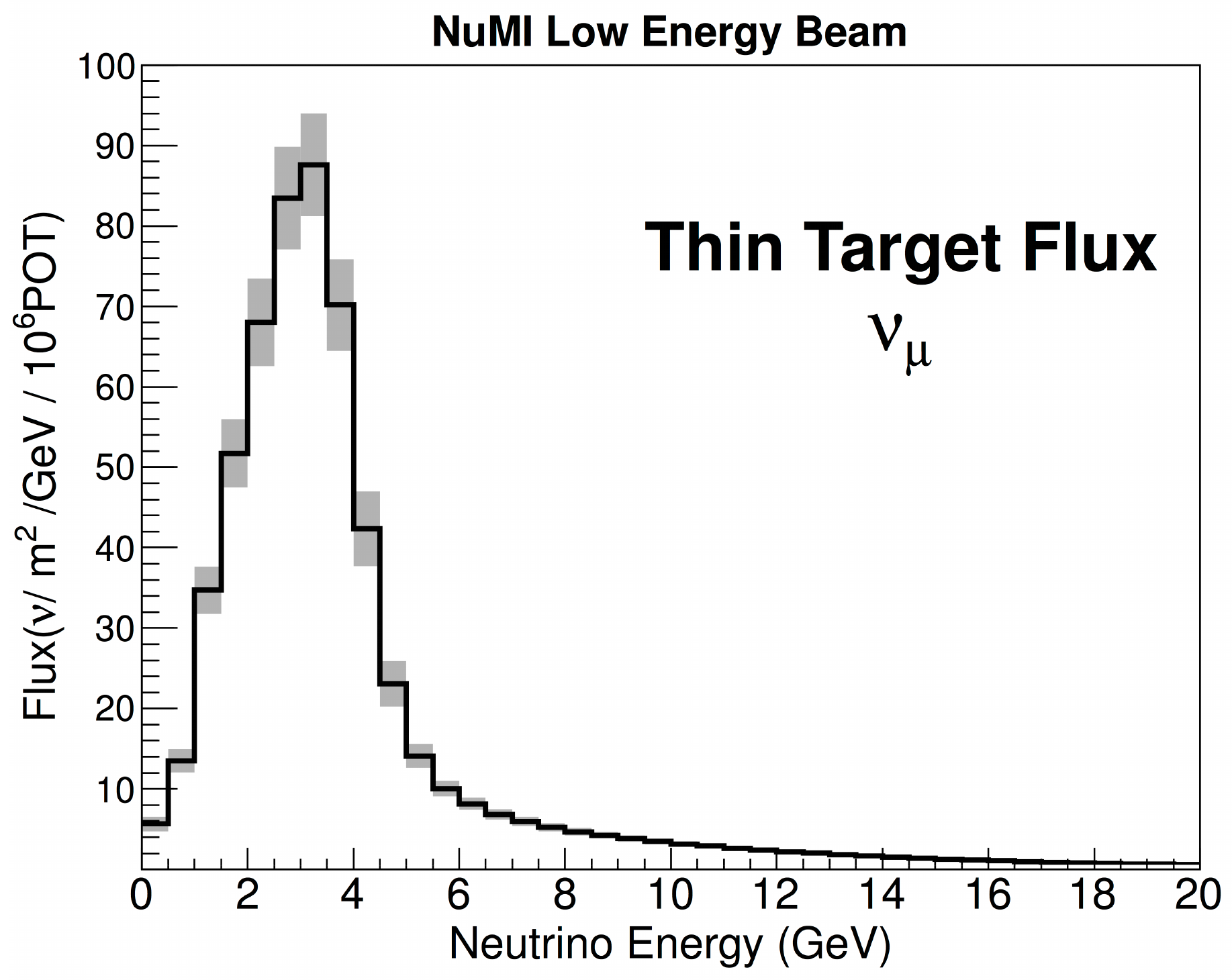}
  \caption{}
  \label{fig:spec_rat_minerva13_Gen2thin_numu}
\end{subfigure}%
\begin{subfigure}{.5\textwidth}
  \centering
  \includegraphics[width=1.0\linewidth]{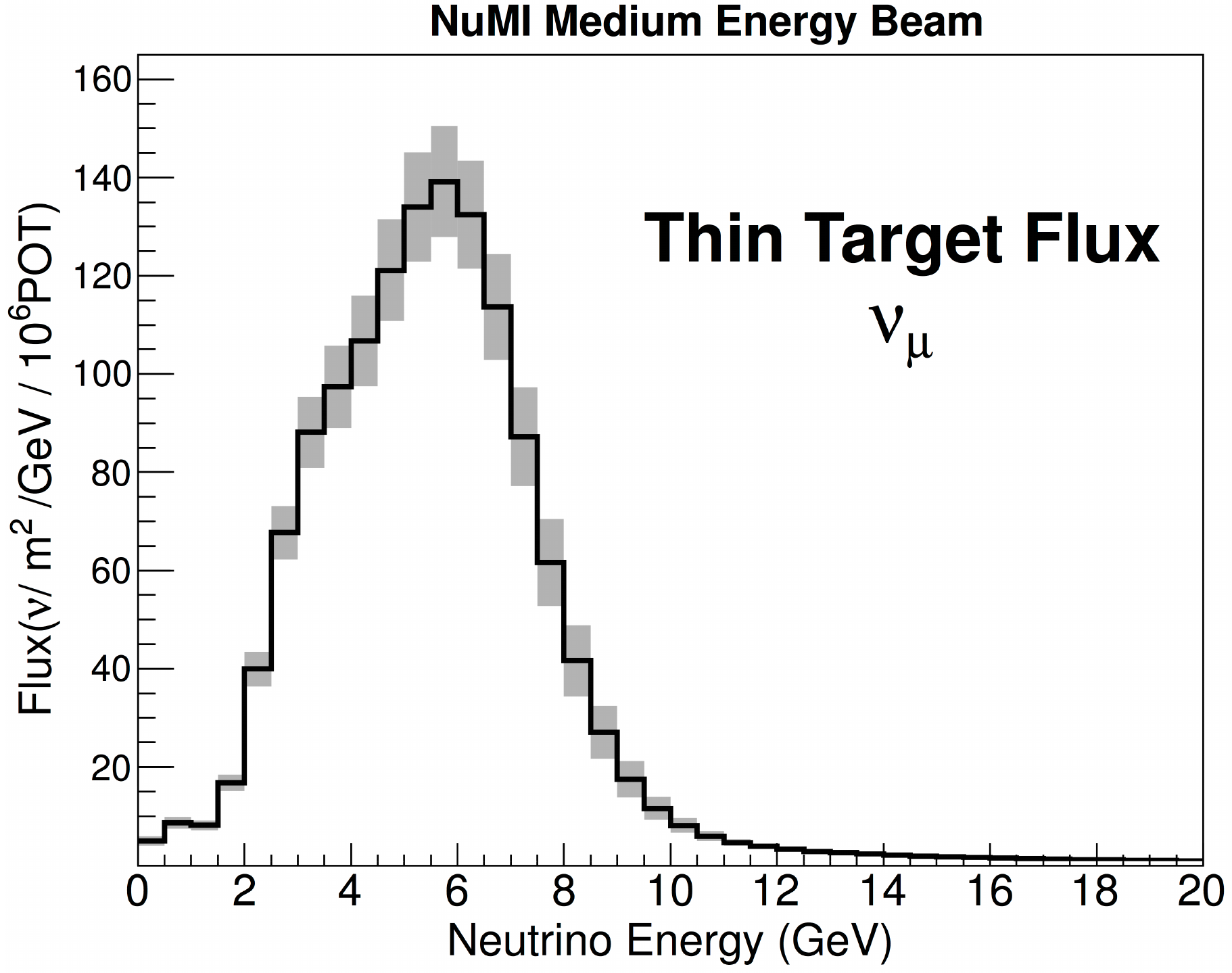}
  \caption{}
  \label{fig:spec_rat_minervaME_numu.eps}
\end{subfigure}
\caption{Muon-neutrino flux in (a) the ``Low-Energy'' (LE) configuration and (b) the ``Medium Energy'' (ME) configuration \cite{Aliaga:2016oaz,Soplin:2016qrt}.
%The ratios shown in the lower part of the figures show corrections made to the flux simulation to produce agreement with external hadron production measurements.
%These event weights are considered during physics analysis but are not used anywhere in the machine learning algorithms presented here.
}
\label{fig:spec_rat_minerva}
\end{figure}

Because events created by neutrinos of higher energy tend to be cascade-rich, and the neutrino energy distribution is not available to the DCNN, it is important to validate the robustness of the DCNN we built for vertex finding with respect to changes in the flux.  To test this, we used events simulated with the LE beam flux as the source domain and events simulated with the ME flux as a target domain.  
Figure \ref{fig:spec_rat_minerva} shows the two different neutrino fluxes discussed in this paper.

%Testing performance for models trained with one flux and evaluated with the other provides a convenient way to test the sensitivity of the performance of the DCNN to the modeling of the flux.
%We will consider the LE beam as a proxy for an extreme perturbation in the flux model - our source domain is the LE flux and our target domain is the ME flux. 
%The neural networks do not have direct information about the neutrino flux, but they observe event topologies.
%Because various scattering channels have different cross sections as a function of energy, changing the flux changes the balance of event topologies.
%For example, the relative mix of low multiplicity elastic events to particle cascade-rich inelastic events will be higher in the LE than the ME.

%% Don't know if this actually helps...
%
%\begin{table}[htb]
%\centering
%\begin{tabular}{ccc}
%\toprule
%Source domain & Target domain & DANN partner \\
%\midrule
%LE & ME & ME \\
%\bottomrule
%\end{tabular}
%\caption{Domain assignment for Figures \ref{fig:segments35_trainLE_testME} and \ref{fig:segment4_trainLE_testME}.
%Therefore, the blue curve in those Figures corresponds to training and evaluating using the ME flux, the black curve corresponds to training with the LE flux and evaluating with the ME flux, and the red curve corresponds to training with the LE flux using ME samples as a DANN-partner, and evaluating in the ME flux.
%}
%\label{tbl:fluxdomains}
%\end{table}

%% flux train LE/ME test ME

\begin{figure}[htbp]
  \begin{minipage}{1.0\linewidth}
    \centering
    \includegraphics[height=0.425\textheight]{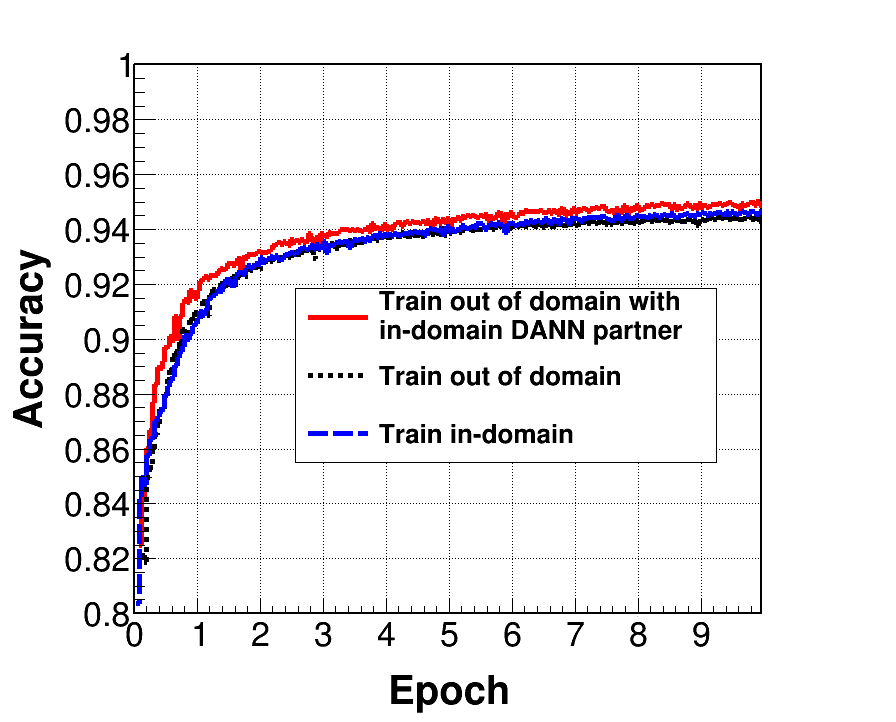}
    \subcaption{Evaluation sample accuracy scores.}
    \label{fig:Flux_model_comparison_accuracy}\par \medskip \vfill
    \includegraphics[height=0.425\textheight]{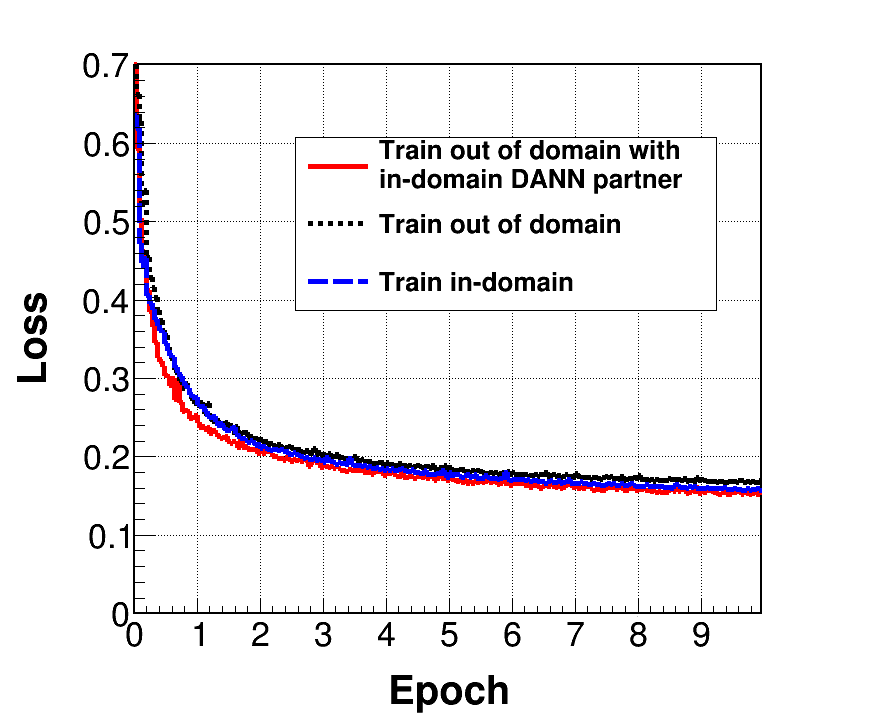}
    \subcaption{Evaluation sample loss scores.}
    \label{fig:Flux_model_comparison_loss}\par \medskip \vfill
  \end{minipage}
  \caption{
 	DANN vs. No-DANN performance as a function of training epoch.
	Here, the target domain sample uses the ME flux.
	See Table \ref{tbl:fluxsampledecoder} for sample-label correspondence.
%	Here we compare models trained on samples with LE flux (``out of domain'')  with and without a DANN-partner that included ME flux samples, and a model trained on the ME flux only (``in domain'').
%	In all cases the target domain uses the ME flux.  
  }
  \label{fig:Flux_model_comparison}
\end{figure}

Figure \ref{fig:Flux_model_comparison} shows a comparison of the accuracy and loss for the vertex finder under different flux conditions.
Throughout this section, ``accuracy'' is defined as the fraction of correct classifications made out of the total number of classifications computed from the multi-class output of the network.
Here, the target domain is the ME flux, so out of domain training means using the LE flux as the source domain.
%The DANN partner uses samples from the ME flux and ignores the vertex label to simulate the role of unlabeled detector data.
The domains are illustrated in Table \ref{tbl:fluxsampledecoder}.
In this case, we see very little differences in model performance regardless of the training and/or evaluation domains, suggesting that the performance of the network is insensitive to the flux model.
This is likely because both fluxes contain a similar set of available event topologies - the only difference is the balance between them.
With large samples, the networks are able to access a sufficient number of examples of each sort of event and so even large changes in the balance between topologies does not degrade the ability of the networks to localize interactions.
The DANN-partner sample outperforms the other domain strategies, a feature we will discuss in more depth in Section \ref{sec:genimprove}.

%%% FIGURE SOURCE: https://minerva-docdb.fnal.gov/cgi-bin/private/ShowDocument?docid=16716

\begin{table}[htb]
\centering
\begin{tabular}{llll}
\toprule
Curve                  & Train         & DANN partner        & Evaluation sample \\
                       & (labeled)     & (unlabeled)         &            \\
\midrule
\midrule
Black                  & Source domain & None                & Target domain \\
(Train out-of-domain)  & LE Flux       &                     & ME Flux       \\
\midrule
Blue                   & Target domain & None                & Target domain \\
(Train in-domain)      & ME Flux       &                     & ME Flux       \\
\midrule
Red                    & Source domain & Target domain       & Target domain \\
(Train out-of-domain   & LE Flux       & ME Flux             & ME Flux       \\
with in-domain         &               &                     &               \\
DANN partner)          &               &                     &               \\
\bottomrule
\end{tabular}
\caption{Sample decoder for Figure \ref{fig:Flux_model_comparison}.
}
\label{tbl:fluxsampledecoder}
\end{table}

\subsubsection{FSI model}
\label{sec:fsistudy}

In \minerva, uncertainties in the model that describes final state interactions (FSI) are often dominant systematic uncertainties.
In some cases the measurement itself is of the FSI effects.
Signal topologies are usually defined by the observed particles in the final state, so processes that change that content directly impact the signal definition.
%For example, a signal might be defined by the presence of at least one charged pion in the final state.
%It is possible for a reaction to produce a charged pion at the hard scattering vertex in the nucleus but for the pion undergo charge exchange while exiting the nucleus, shifting the event from the signal category to background.

In order to investigate the robustness of our network to FSI, we used two domains: simulation with FSI turned on and off. 
In this section, we treat the simulation with FSI turned on as the source domain and simulation with FSI turned off as the target domain.
Only simulation using the LE flux was available with FSI turned off at the time of the study.

\begin{figure}[htbp]
  \begin{minipage}{1.0\linewidth}
    \centering
    \includegraphics[height=0.425\textheight]{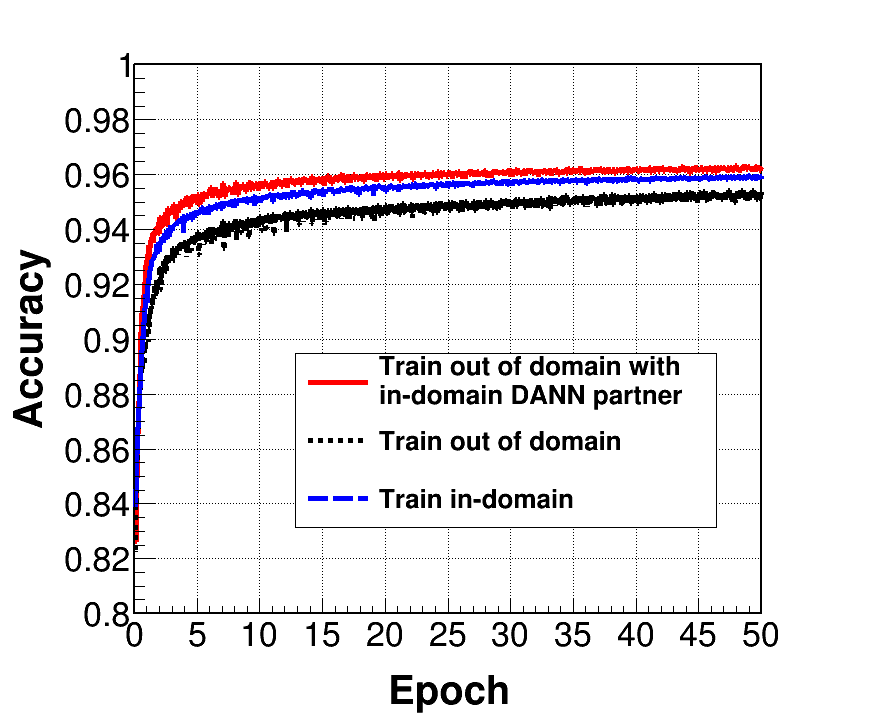}
    \subcaption{Evaluation sample accuracy scores.}
    \label{fig:FSI_model_accuracy_trio}\par \medskip \vfill
    \includegraphics[height=0.425\textheight]{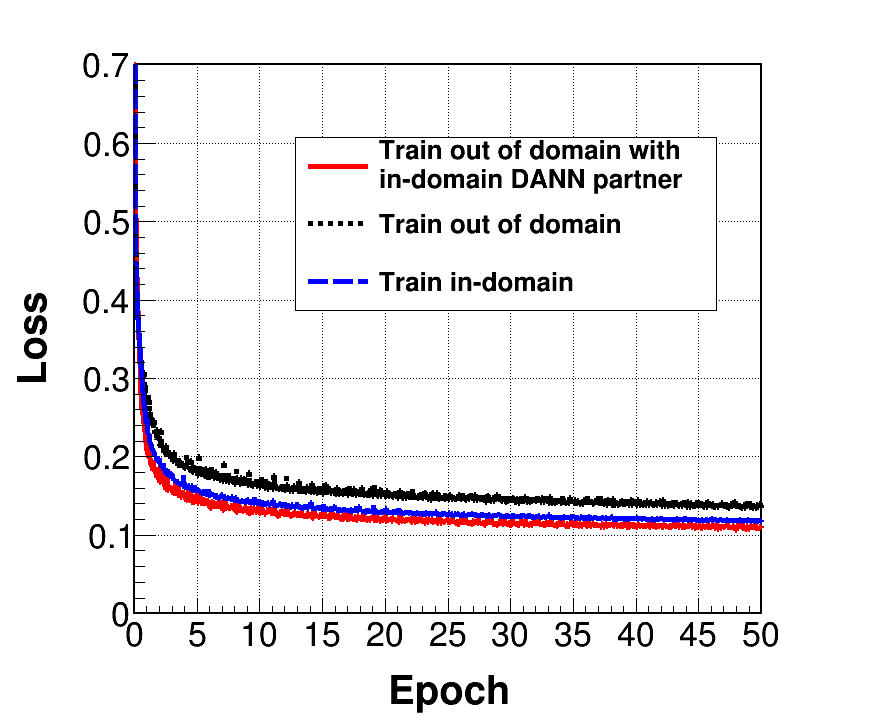}
    \subcaption{Evaluation sample loss scores.}
    \label{fig:FSI_model_loss_trio}\par \medskip \vfill
  \end{minipage}
  \caption{DANN vs. No-DANN performance as a function of training epoch.
	Here, the target domain sample is FSI-active events.
	See Table \ref{tbl:fsisampledecoder} for sample-label correspondence.
  }
  \label{fig:FSI_model_comparison}
\end{figure}

See Figure \ref{fig:FSI_model_comparison} for an example of the impact of the DANN on the total classification accuracy.
The domains are illustrated in Table \ref{tbl:fsisampledecoder}.
We see that if inference is run in a domain with FSI active, that a model trained in the same domain (with FSI active) is more accurate than a model trained with an out-of domain physics model (with FSI inactive).
However, by adding a DANN partner to the model trained in the wrong domain we are able to recover the performance of the model natively trained in the correct domain.
Indeed, for the same set of hyper-parameters we actually achieve better in-domain performance than without a DANN but still with proper in-domain training.

%%% FIGURE SOURCE: https://minerva-docdb.fnal.gov/cgi-bin/private/ShowDocument?docid=16716

\begin{table}[htb]
\centering
\begin{tabular}{llll}
\toprule
Curve                  & Train         & DANN partner        & Evaluation sample \\
                       & (labeled)     & (unlabeled)         &            \\
\midrule
\midrule
Black                  & Source domain & None                & Target domain \\
(Train out-of-domain)  & FSI-off       &                     & FSI-on        \\
\midrule
Blue                   & Target domain & None                & Target domain \\
(Train in-domain)      & FSI-on        &                     & FSI-on        \\
\midrule
Red                    & Source domain & Target domain       & Target domain \\
(Train out-of-domain   & FSI-off       & FSI-on              & FSI-on        \\
with in-domain         &               &                     &               \\
DANN partner)          &               &                     &               \\
\bottomrule
\end{tabular}
\caption{Sample decoder for Figure \ref{fig:FSI_model_comparison}.
}
\label{tbl:fsisampledecoder}
\end{table}

\subsubsection{Kinematic split}
\label{sec:kinsplit}

\begin{figure}[htbp]
  \begin{minipage}{1.0\linewidth}
    \centering
    \includegraphics[height=0.425\textheight]{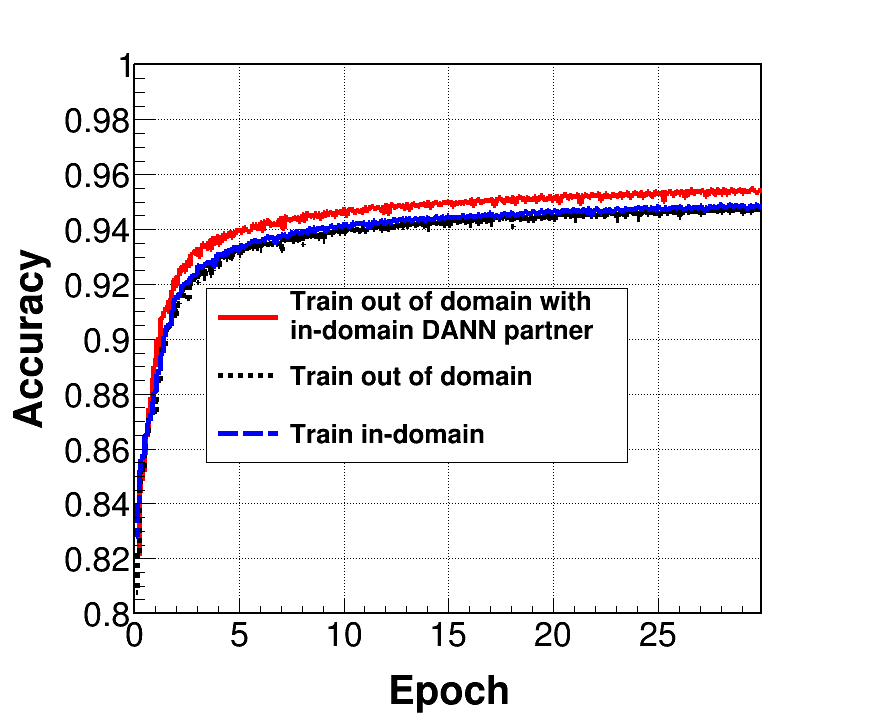}
    \subcaption{Evaluation sample accuracy scores.}
    \label{fig:wsplit_accuracy_trio}\par \medskip \vfill
    \includegraphics[height=0.425\textheight]{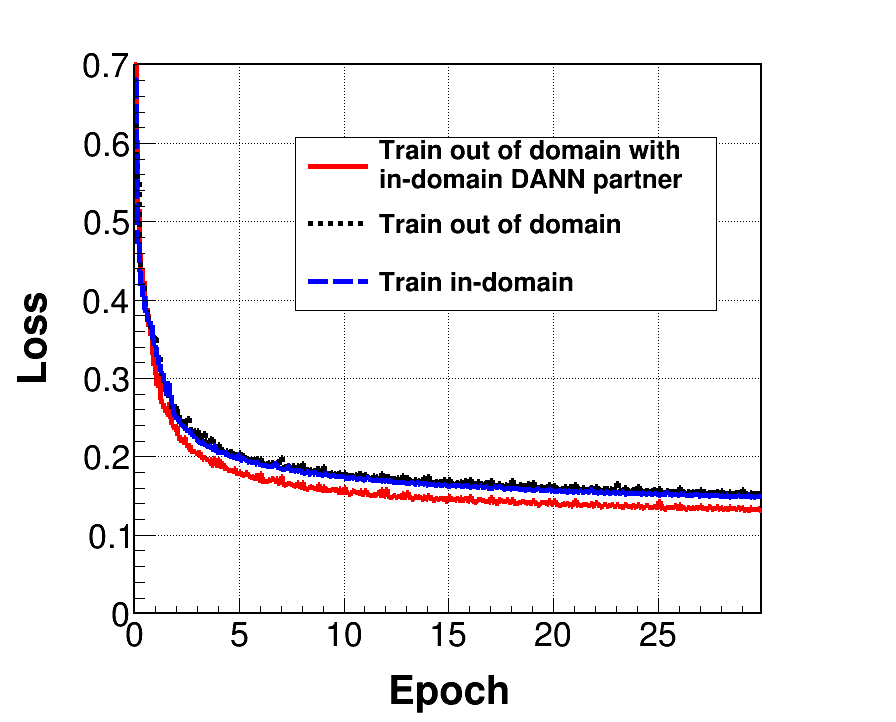}
    \subcaption{Evaluation sample loss scores.}
    \label{fig:wsplit_loss_trio}\par \medskip \vfill
  \end{minipage}
  \caption{DANN vs. No-DANN performance as a function of training epoch.
	Here, the target domain sample is high-$W$ events.
	See Table \ref{tbl:wsampledecoder} for sample-label correspondence.
  }
  \label{fig:wsplit_comparison}
\end{figure}

By splitting the simulation sample by invariant mass into samples with $W < 1$ GeV and $W \geq 1$ GeV (recall Table \ref{tbl:numbrevt}) we can directly test the robustness of the network topology when presented with a set which has topologies which are different than that which it was trained on. 
The $W < 1$ GeV sample features events that are particle cascade poor while the $W \geq 1$ GeV sample contains a range of topologies weighted towards more events that are particle cascade rich. 
We also trained a network with the $W \geq 1$ GeV sample used as the adversarial partner.
In Figure \ref{fig:wsplit_comparison} we see a surprisingly similar agreement in validation accuracy and loss on high-$W$ events regardless of whether the training sample was in or out of domain.
The domains are illustrated in Table \ref{tbl:wsampledecoder}.
Including a DANN partner from the in-domain sample (so training on a low-$W$ sample but with a high-$W$ DANN partner) is the best performing option.
We will discuss this outcome further in Section \ref{sec:genimprove}.

\begin{table}[htb]
\centering
\begin{tabular}{llll}
\toprule
Curve                  & Train         & DANN partner        & Evaluation sample \\
                       & (labeled)     & (unlabeled)         &            \\
\midrule
\midrule
Black                  & Source domain & None                & Target domain \\
(Train out-of-domain)  & Low-W         &                     & High-W        \\
\midrule
Blue                   & Target domain & None                & Target domain \\
(Train in-domain)      & High-W        &                     & High-W        \\
\midrule
Red                    & Source domain & Target domain       & Target domain \\
(Train out-of-domain   & Low-W         & High-W              & High-W        \\
with in-domain         &               &                     &               \\
DANN partner)          &               &                     &               \\
\bottomrule
\end{tabular}
\caption{Sample decoder for Figure \ref{fig:wsplit_comparison}.
}
\label{tbl:wsampledecoder}
\end{table}

\subsection{Initialization and training impacts}
\label{sec:initimpact}

\begin{figure}[htbp]
  \begin{minipage}{1.0\linewidth}
    \centering
	\begin{subfigure}{.5\textwidth}
	  \centering
	  \captionsetup{width=5cm}
	  \includegraphics[width=0.99\linewidth]{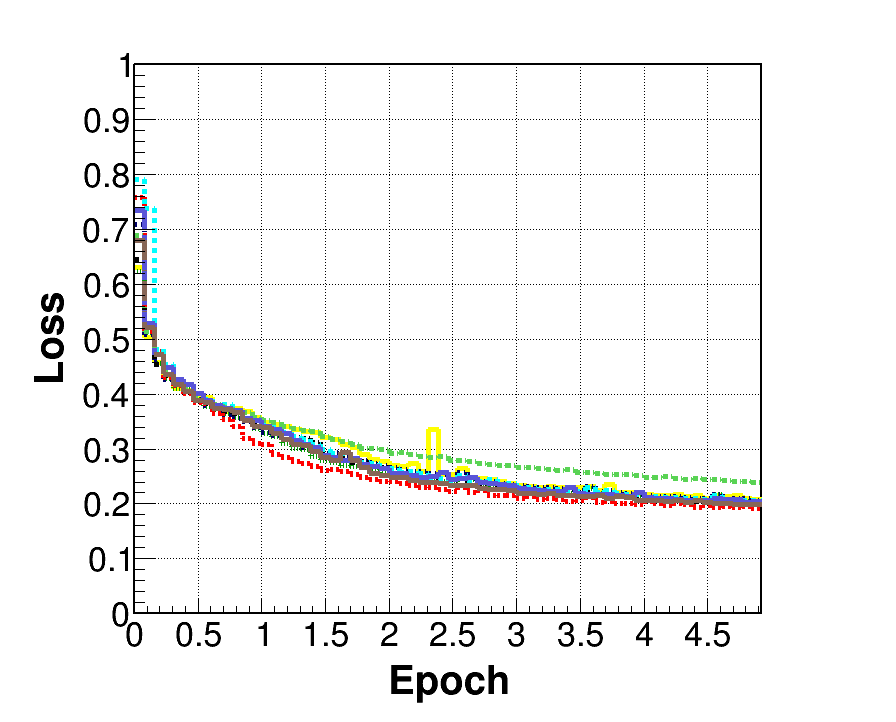}
	  \caption{Evaluation sample losses for 10 runs trained using the high-$W$ sample.}
	  \label{fig:wsplit_model_NODANN_trainhighW_testhighW_multi_initialization}
	\end{subfigure}%
	\begin{subfigure}{.5\textwidth}
	  \centering
	  \captionsetup{width=5cm}
	  \includegraphics[width=0.99\linewidth]{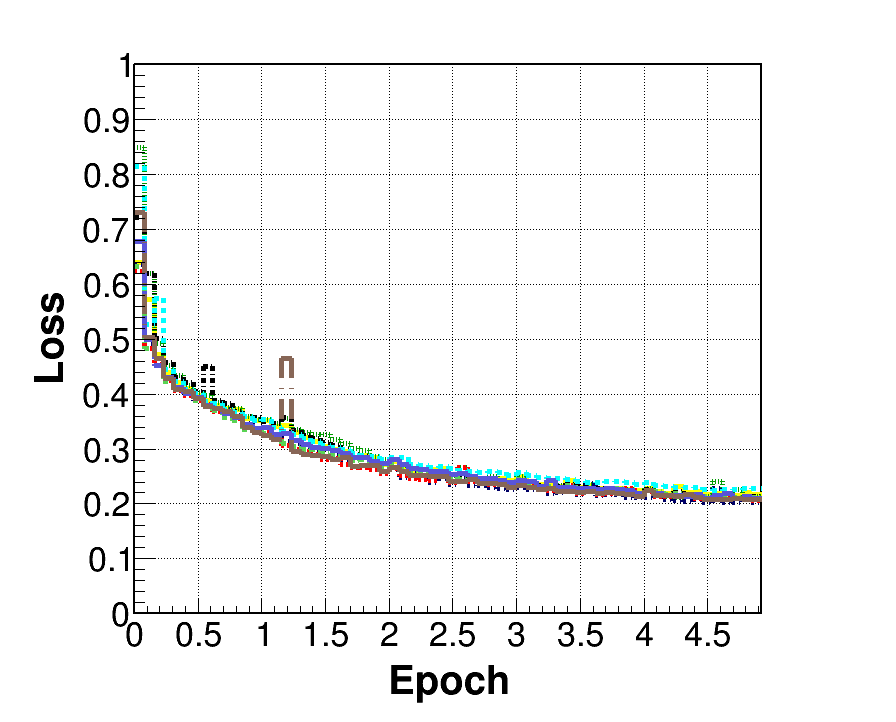}
	  \caption{Evaluation sample losses for 10 runs trained using the low-$W$ sample.}
	  \label{fig:wsplit_model_NODANN_trainlowW_testhighW_multi_initialization}
	\end{subfigure}    
	\begin{subfigure}{.5\textwidth}
	  \centering
	  \captionsetup{width=5cm}
	  \includegraphics[width=0.99\linewidth]{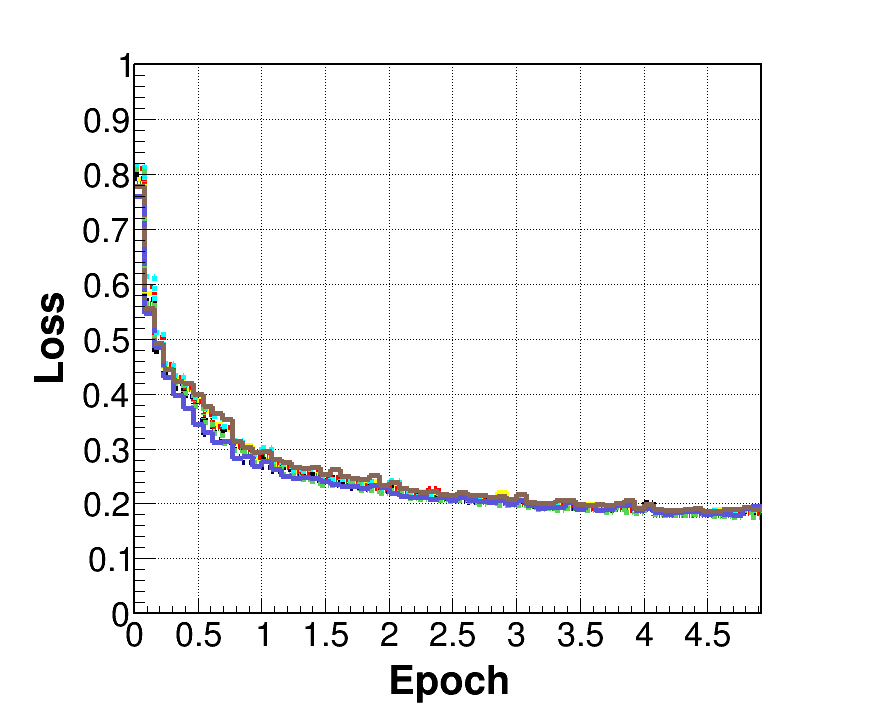}
	  \caption{Evaluation sample losses for 10 runs trained using the low-$W$ sample with a high-$W$ DANN partner.}
	  \label{fig:wsplit_model_DANN_trainlowW_testhighW_multi_initialization}
	\end{subfigure}%
	\begin{subfigure}{.5\textwidth}
	  \centering
	  \captionsetup{width=5cm}
	  \includegraphics[width=0.99\linewidth]{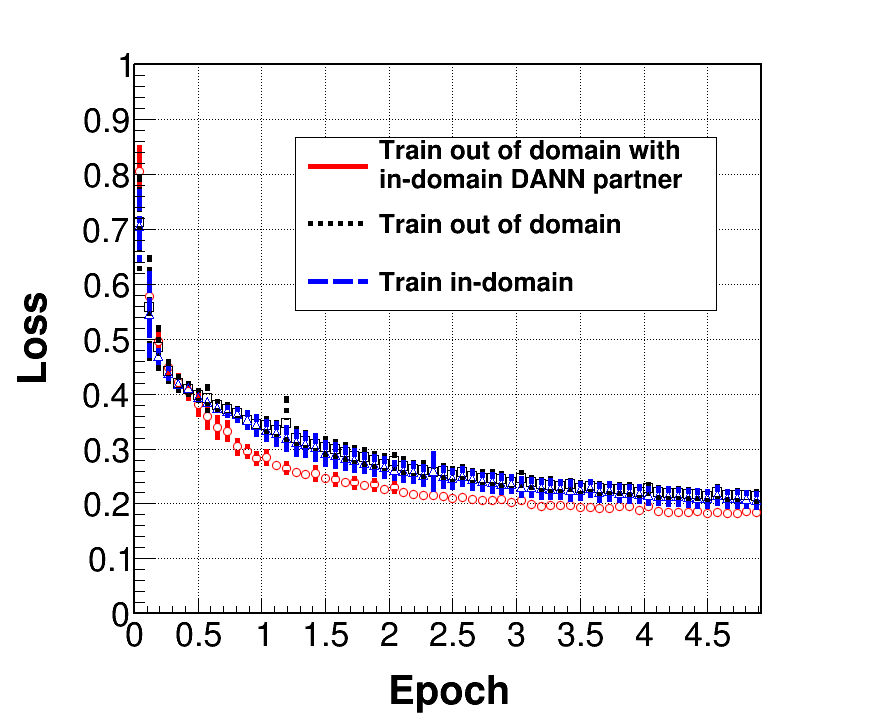}
	  \caption{Training cases (a), (b), and (c), with points equal to the means of the training runs and errors equal to the RMS.}
	  \label{fig:wsplit_mean_trio_multi_initialization}
	\end{subfigure}
  \end{minipage}
  \caption{Many initializations comparison - here using the kinematic split sample.
  The evaluation sample is the high-$W$ region in all cases.
  The results show that the evolution of the training curves is generally quite stable, and not particularly sensitive to the specific random initialization of the network.
  }
  \label{fig:wsplit_multi_init}
\end{figure}

We studied the impact of random initialization on training.
See Figure \ref{fig:wsplit_multi_init}.
The domains considered here correspond to splits in the total sample by event kinematics.
The split will be further explained in Section \ref{sec:kinsplit}
Our conclusion was that the training results were fairly stable even at low epoch count, and that it was not necessary to repeat all of our long training runs many times to account for the initialization variation.

\subsection{Common features of DANN performance tests}
\label{sec:genimprove}

\begin{figure}[htbp]
  \begin{minipage}{1.0\linewidth}
    \centering
    \includegraphics[height=0.425\textheight]{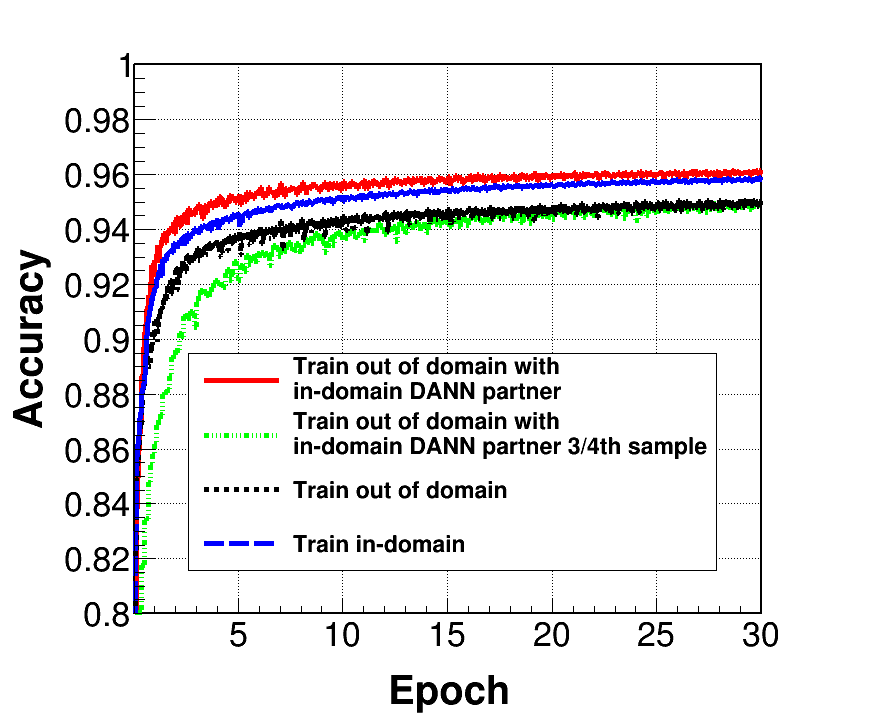}
    \subcaption{Evaluation sample accuracy scores.}
    \label{fig:FSI_model_accuracy_quatro}\par \medskip \vfill
    \includegraphics[height=0.425\textheight]{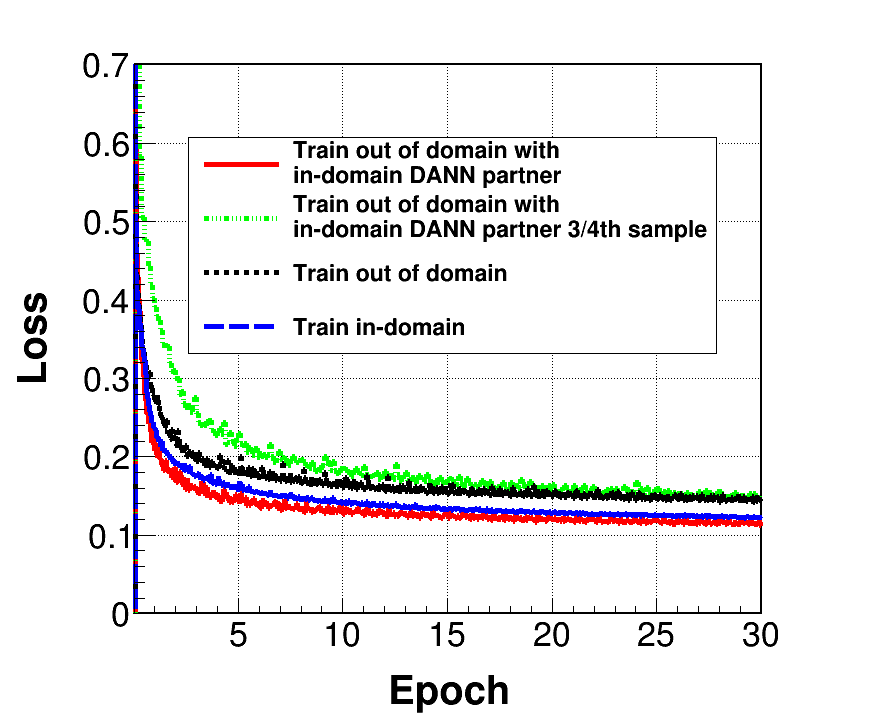}
    \subcaption{Evaluation sample loss scores.}
    \label{fig:FSI_model_loss_quatro}\par \medskip \vfill
  \end{minipage}
  \caption{DANN vs. No-DANN performance as a function of training epoch.
	Here, the target domain sample is FSI-active events.
	The green curve is trained using one quarter of the labeled training set and three quarters of the available DANN partner events.
	See Table \ref{tbl:dannhalfsamplestudy}.
  }
  \label{fig:compare_dann_quarterdann}
\end{figure}

\begin{figure}[htbp]
  \begin{minipage}{1.0\linewidth}
    \centering
    \includegraphics[height=0.425\textheight]{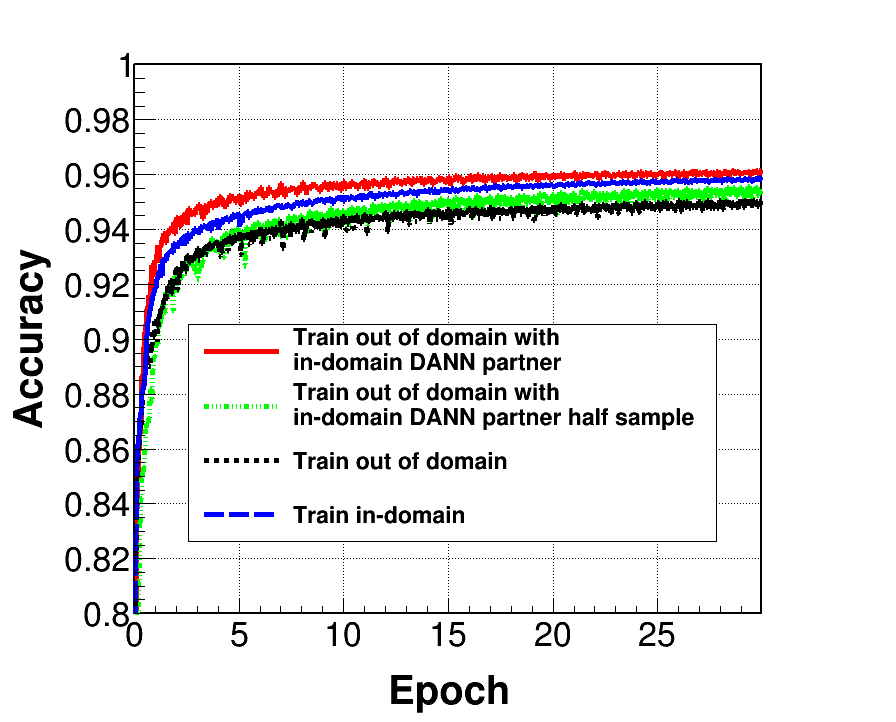}
    \subcaption{Evaluation sample accuracy scores.}
    \label{fig:FSI_model_accuracy_quatro}\par \medskip \vfill
    \includegraphics[height=0.425\textheight]{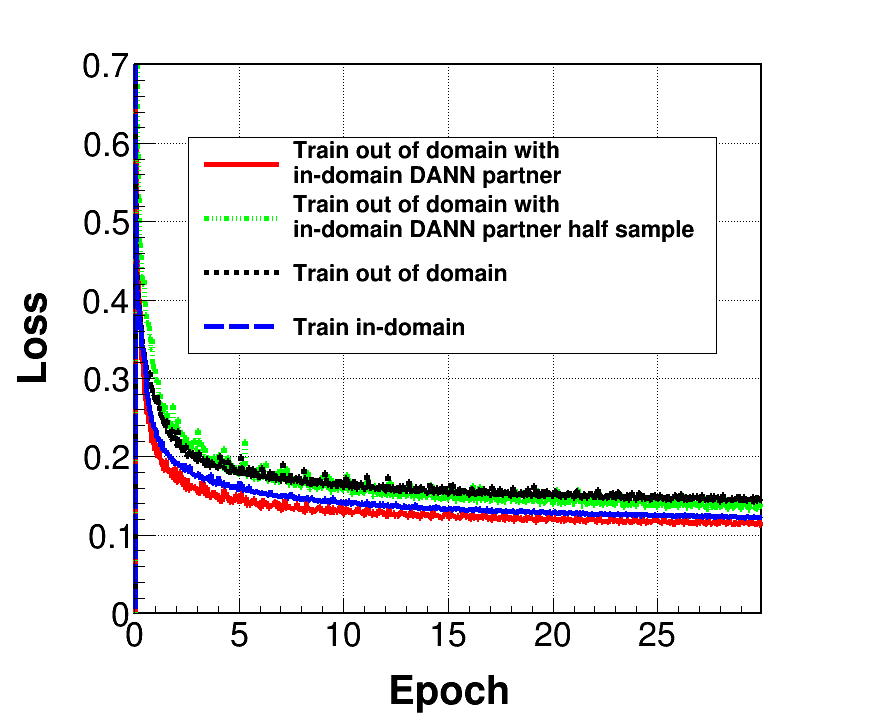}
    \subcaption{Evaluation sample loss scores.}
    \label{fig:FSI_model_loss_quatro}\par \medskip \vfill
  \end{minipage}
  \caption{DANN vs. No-DANN performance as a function of training epoch.
	Here, the target domain sample is FSI-active events.
%	The blue curve corresponds to training and evaluating with FSI active in the simulation.
%	The black curve corresponds to training with FSI deactivated in the simulation, and evaluating using events with FSI active.
%	The red curve corresponds to training with FSI deactivated in the simulation using a DANN partner with FSI activated, and evaluated with FSI active.
%	The combined training plus DANN sample size is twice the size available to the blue or black curves.
%	The green curve is similar to the red curve, and corresponds to training with FSI deactivated in the simulation using a DANN partner with FSI activated, and evaluated with FSI active.
%	However, for the green curve, the training and DANN partner sizes were both divided in half, making the total number of samples between the two used in training to be of the same size as what was available to the blue and black curves.
	The green curve is trained using half of the labeled training set and half of the available DANN partner events.
	See Table \ref{tbl:dannhalfsamplestudy}.
  }
  \label{fig:compare_dann_halfdann}
\end{figure}

\begin{figure}[htbp]
  \begin{minipage}{1.0\linewidth}
    \centering
    \includegraphics[height=0.425\textheight]{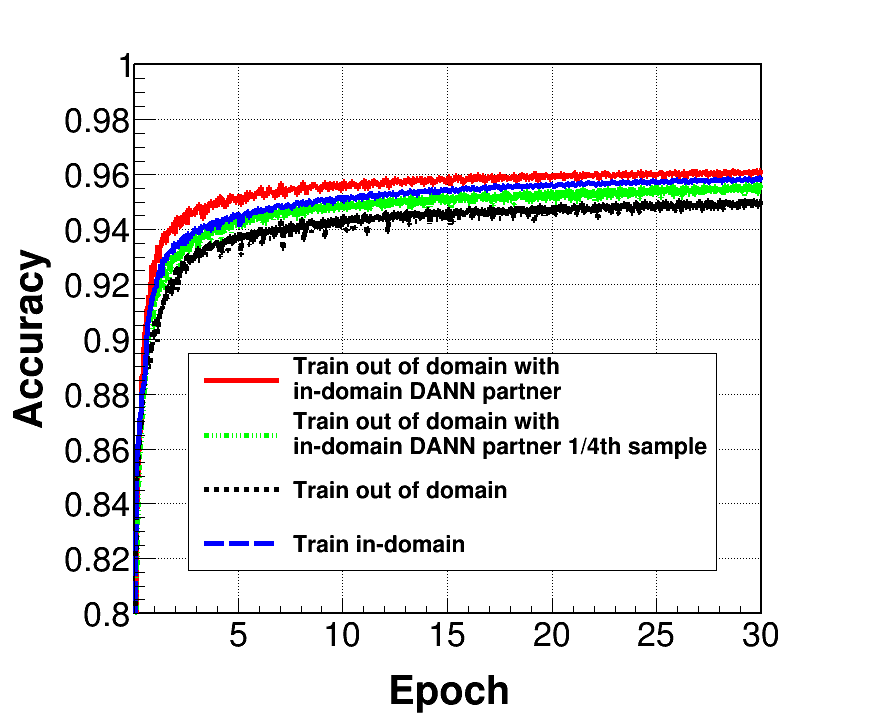}
    \subcaption{Evaluation sample accuracy scores.}
    \label{fig:FSI_model_accuracy_quatro}\par \medskip \vfill
    \includegraphics[height=0.425\textheight]{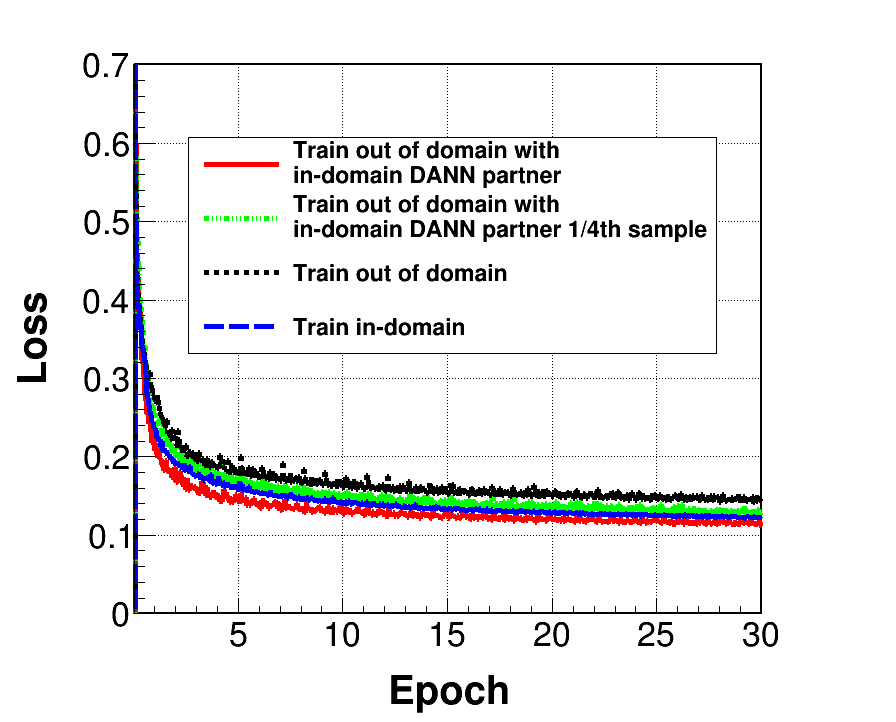}
    \subcaption{Evaluation sample loss scores.}
    \label{fig:FSI_model_loss_quatro}\par \medskip \vfill
  \end{minipage}
  \caption{DANN vs. No-DANN performance as a function of training epoch.
	Here, the target domain sample is FSI-active events.
	The green curve is trained using three quarters of the labeled training set and one quarter of the available DANN partner events.
	See Table \ref{tbl:dannhalfsamplestudy}.
  }
  \label{fig:compare_dann_threequarterdann}
\end{figure}

\begin{figure}[htbp]
  \begin{minipage}{1.0\linewidth}
    \centering
    \includegraphics[height=0.425\textheight]{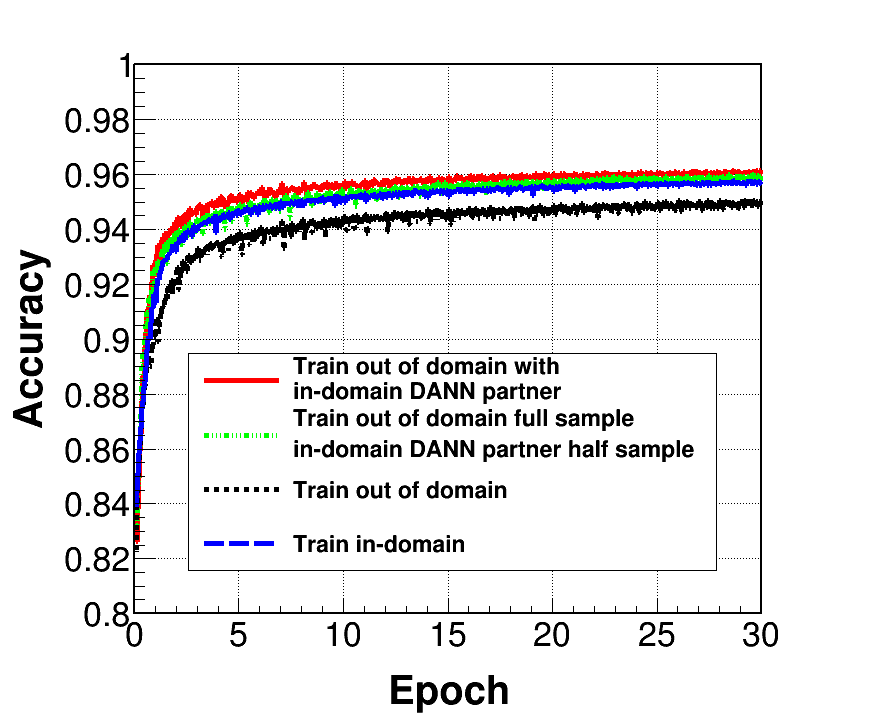}
    \subcaption{Evaluation sample accuracy scores.}
    \label{fig:FSI_model_accuracy_quatro}\par \medskip \vfill
    \includegraphics[height=0.425\textheight]{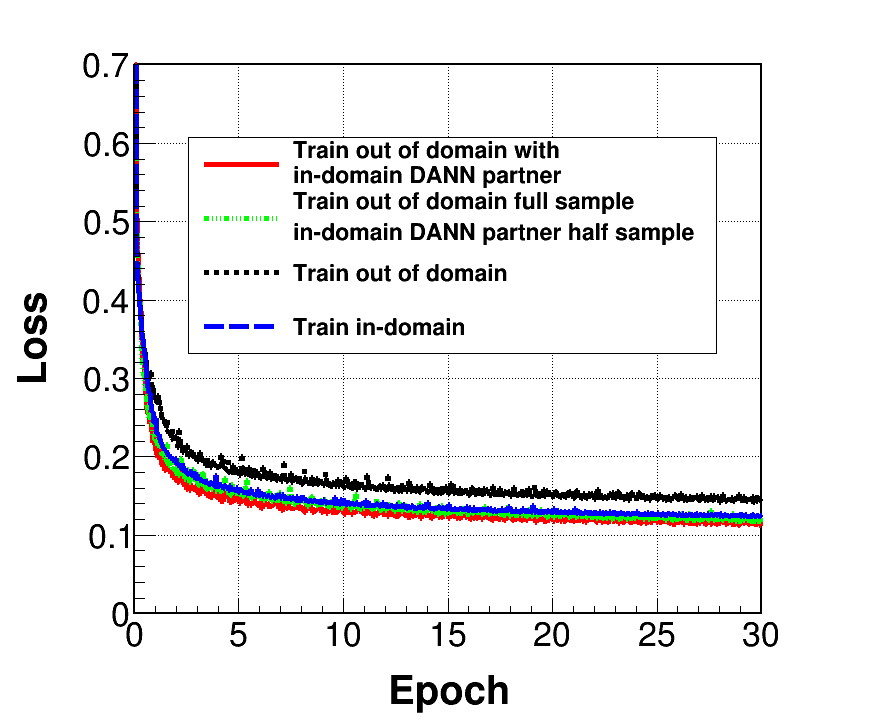}
    \subcaption{Evaluation sample loss scores.}
    \label{fig:FSI_model_loss_quatro}\par \medskip \vfill
  \end{minipage}
  \caption{DANN vs. No-DANN performance as a function of training epoch.
	Here, the target domain sample is FSI-active events.
	The green curve is trained using the full labeled training set and one half of the available DANN partner events.
	See Table \ref{tbl:dannhalfsamplestudy}.
  }
  \label{fig:compare_dann_halfhalfdann}
\end{figure}

The studies in Sections \ref{sec:fluxstudy} to \ref{sec:kinsplit} showed that using a in-domain DANN improved the performance of a network for both the source and target domains.
A surprising result was that networks trained on out-of-domain data with DANN partners were able to outperform networks trained on in-domain data. 
We attempt to explain this effect in Figures \ref{fig:compare_dann_quarterdann}, \ref{fig:compare_dann_halfdann}, \ref{fig:compare_dann_threequarterdann}, and \ref{fig:compare_dann_halfhalfdann} which are the same as Figure \ref{fig:FSI_model_comparison}, but each with an additional curve.
Benchmark performance is established by the blue curve - this is training and evaluating in a domain with FSI active. 
The training sample contains 1.2 million events.
The out-of-domain sample, the black curve, shows degraded performance - this is training on a 1.2 million event sample with FSI deactivated, and evaluating on a sample with FSI activated.
The red curve corresponds to training in a domain with FSI deactivated, but now using a DANN partner sample with FSI activated.
Both samples are 1.2 million events.
We recover the performance of the blue sample and more, although in the asymptotic limit of long training times, the curves are slowly converging.

The question is - why does the red curve outperform the blue?
The answer is supplied by the green curves.
For Figures \ref{fig:compare_dann_quarterdann}, \ref{fig:compare_dann_halfdann}, and \ref{fig:compare_dann_threequarterdann}, we use the same training strategy as for the red curve, but with both samples reduced in size by a factor chosen to keep the total number of events available fixed at 1.2 million, but with varying fractions labeled.
Then in Figure \ref{fig:compare_dann_halfhalfdann} we use all the available labeled data, but only half of the possible DANN partner events.
See Table \ref{tbl:dannhalfsamplestudy} for a breakdown of the event samples for each figure.

\begin{table}[htb]
\centering
\begin{tabular}{cccc}
\toprule
Curve                  & Source domain sample & DANN partner sample & Figures \\
\midrule
\midrule
Blue                   & FSI activated -      & NA                  & \ref{fig:FSI_model_comparison},
                                                                      \ref{fig:compare_dann_quarterdann}, 
                                                                      \ref{fig:compare_dann_halfdann}, 
                                                                      \ref{fig:compare_dann_threequarterdann},
                                                                      \ref{fig:compare_dann_halfhalfdann} \\
                       & 1.2 million events   &                     & \\
\midrule
Black                  & FSI deactivated -    & NA                  & \ref{fig:FSI_model_comparison},
                                                                      \ref{fig:compare_dann_quarterdann}, 
                                                                      \ref{fig:compare_dann_halfdann}, 
                                                                      \ref{fig:compare_dann_threequarterdann},
                                                                      \ref{fig:compare_dann_halfhalfdann} \\
                       & 1.2 million events   &                     & \\
\midrule
Red                    & FSI deactivated -    & FSI activated -     & \ref{fig:FSI_model_comparison},
                                                                      \ref{fig:compare_dann_quarterdann}, 
                                                                      \ref{fig:compare_dann_halfdann}, 
                                                                      \ref{fig:compare_dann_threequarterdann},
                                                                      \ref{fig:compare_dann_halfhalfdann} \\
                       & 1.2 million events   & 1.2 million events  & \\
\midrule
Green                  & FSI deactivated -    & FSI activated -    & \ref{fig:compare_dann_quarterdann} \\
3/4-DANN               & 0.3 million events   & 0.9 million events & \\
\midrule
Green                  & FSI deactivated -    & FSI activated -    & \ref{fig:compare_dann_halfdann} \\
half-DANN              & 0.6 million events   & 0.6 million events & \\
\midrule
Green                  & FSI deactivated -    & FSI activated -    & \ref{fig:compare_dann_threequarterdann} \\
1/4-DANN               & 0.9 million events   & 0.3 million events & \\
\midrule
Green                  & FSI deactivated -    & FSI activated -    & \ref{fig:compare_dann_halfhalfdann} \\
full-source, half-DANN & 1.2 million events   & 0.6 million events & \\
\bottomrule
\end{tabular}
\caption{Training and DANN sample types and sizes for the curves displayed in Figures \ref{fig:FSI_model_comparison}, \ref{fig:compare_dann_quarterdann}, \ref{fig:compare_dann_halfdann}, \ref{fig:compare_dann_threequarterdann}, and \ref{fig:compare_dann_halfhalfdann}.
Note that Figure \ref{fig:FSI_model_comparison} spans a longer training period (50 vs 30 epochs).
}
\label{tbl:dannhalfsamplestudy}
\end{table}

In Figure \ref{fig:compare_dann_quarterdann}, the green curve is utilizing a one-quarter size labeled training set (and three quarters of the available DANN partners) and the observed performance is degraded, although it asymptotically approaches the full sample out of domain model.
Note that because the labeled data controls the size of the epoch that the number of adjustments to the weights per epoch is lower in this case than for the black curve (although the number of events in each computation of the gradients is higher owing to the DANN sample).
Therefore, the slower rise of the green curve is at least partially due to this effect.
Then, in Figure \ref{fig:compare_dann_halfdann}, we see performance recovery towards the blue curve, but we do not observe performance matching at high epoch count.
What this shows is that the domain information from the DANN improves the ability of the network to match in-domain performance, but that the reduced event sample size in the supervised dataset also suppresses performance.
In Figure \ref{fig:compare_dann_threequarterdann}, we see further recovery towards the blue curve, but the overall improvement is limited by the number of events available for training.

Finally, in Figure \ref{fig:compare_dann_halfhalfdann} we utilize all the available labeled data and half of the available DANN sample.
We see a clear move from the black curve towards the red in this case, indicating that the domain information available is able to improve the classifier. 
The fact that we move beyond the blue curve in this case suggests that the DANN sample is large enough to match the domain information that would be available to an in-domain sample and that the additional events available for training, even in semi-supervised form, positively impacts performance.

In other words, the red curve is able to surpass the blue curve and the various green curves because it is able to both utilize an increased effective training sample size due to the larger semi-supervised dataset, and it is able to use domain information from the DANN partner.
This suggests that as long as the additional training time required to train with a DANN can be withstood, and the network architecture is capable of effectively leveraging the additional data (i.e., the network capacity is large enough), it is valuable to supplement training with a DANN partner in a semi-supervised way provided the domains are reasonably similar.
The results of this analysis are not sufficient to draw the dividing line between domains that are similar enough to be used in this way versus those that are too different to add values in general, but in the specific case of the \minerva detector simulation we are able to improve the trained model in this way.

% We should have made the three green curves based on using the full unlabeled set with varying sizes of DANN partner sets, but didn't realize this until after we had made the current set of plots...

\section{Conclusions and discussion}
\label{sec:conclusions}

\subsection{Future work}
\label{sec:futurework}

We will first configure a DANN-based network to produce classifications by plane rather than segment and optimize a new training strategy for the DIS analysis.
Additionally we plan on testing and applying DANNs in other measurements that may be more sensitive to underlying features of the physics model, for example in classifying the number of charged hadrons produced in an interaction or in per-track particle identification, etc.
We also plan on studying the impact of DANNs on optimal network architecture as optimized via an evolutionary algorithm \cite{Young:2017:EDN:3146347.3146355}.

%% Other considerations / Pile-up

Mischaracterized events form a small fraction of the total event count.
Generally, mischaracterized events are very likely to come from the neighboring plane or segment, as seen in Figure \ref{fig:confusion_matrices_trk_v_dnn_loglin_pur_ME_MC+No_DANN_Partner+10700000_iterations}.
Inspection of these events by eye offer little clue about possible pathologies - likely they represent an irreducible reconstruction error background rooted in the fact we must employ passive neutrino targets.
The far off-diagonal elements in the confusion matrix are largely due to pile-up - which is the phenomenon of multiple physics interactions occurring very close in time.
We plan to do a careful investigation of sources for event mischaracterization.

%% DANNs good, DCNNs robust, problem is independent of most simulation issues

%% Conclusions

\subsection{Conclusions}
\label{sec:finalconclude}

We have seen that DCNNs are productive in extracting the relevant features for a task such as finding the vertex in a neutrino charged current event and can do this better than that of traditional feature extraction.
We have also seen that the performance of this network is largely independent of to some of the main sources of uncertainty; networks trained with different hadronization and flux domains had very similar cross-domain performance.
When we have simulation with radically different FSI behavior, we do see cross-domain performance degradation, but by using DANNs to restrict the feature extraction only to features in both domains we can train a domain-invariant classifier.
%While other sources of uncertainty in our simulation, such as geometry and event pile-up, were not directly investigated, we posit that by using a DANN with a data partner to the simulation that our network is restricted to extracting and utilizing features which exist in both data and simulation and thus the network is not sensitive to geometry, hadronization and pile-up.
We also find that DANN partners enable the use of unlabeled data for semi-supervised training in such a fashion as to productively expand the training set for a model.
This has numerous applications in HEP and other problem domains.

%%\begin{acknowledgments}

% Techpubs claims we need to include this paragraph... I (gnp) extended it slightly
% See http://techpubs.fnal.gov for info on what we must include
This document was prepared by the \minerva collaboration using the resources of the Fermi National Accelerator Laboratory (Fermilab), a U.S. Department of Energy, Office of Science, HEP User Facility. 
Fermilab is managed by Fermi Research Alliance, LLC (FRA), acting under Contract No. DE-AC02-07CH11359, which included the \minerva construction project.
The research here was aslo sponsored by the Laboratory Directed Research and Development Program of Oak Ridge National Laboratory, managed by UT-Battelle, LLC, for the U. S. Department of Energy.
This research used resources of the Oak Ridge Leadership Computing Facility at the Oak Ridge National Laboratory, which is supported by the Office of Science of the U.S. Department of Energy under Contract No. DE-AC05-00OR22725. 
Construction support also was granted by the United States National Science Foundation under Award PHY-0619727 and by the University of Rochester. 
Additional support for participating scientists was provided by NSF and DOE (USA) by CAPES and CNPq (Brazil), by CoNaCyT (Mexico), by Proyecto Basal FB 0821, CONICYT PIA ACT1413, Fondecyt 3170845 and 11130133 (Chile), by PIIC (DGIP-UTFSM), by CONCYTEC, DGI-PUCP and IDI/IGI-UNI (Peru), by Latin American Center for Physics (CLAF), by RAS and the Russian Ministry of Education and Science (Russia), and by the National Science Centre of Poland, grant number DEC-2017/01/X/ST2/00128.
We thank the MINOS Collaboration for use of its near detector data.
Finally, we thank the staff of Fermilab for support of the beamline and the detector.

%Notice: This manuscript has been authored by Fermi Research Alliance, LLC (FRA), under contract DE-AC02-07CH11359 and UT-Battelle, LLC, under contract DE-AC05-00OR22725 with the US Department of Energy (DOE).

The US government retains and the publisher, by accepting the article for publication, acknowledges that the US government retains a nonexclusive, paid-up, irrevocable, worldwide license to publish or reproduce the published form of this manuscript, or allow others to do so, for US government purposes. 
The DOE will provide public access to these results of federally sponsored research in accordance with the DOE Public Access Plan (\url{http://energy.gov/downloads/doe-public-access-plan}).

%%\end{acknowledgments}

%\section*{References}
%\label{sec:references}

%%\bibliographystyle{elsarticle-num} 
\bibliographystyle{JHEP} 
\bibliography{Neutrino}


%\begin{thebibliography}{00}


%\bibitem{theano}
%Theano: A Python framework for fast computation of mathematical expressions, 
%http://arxiv.org/abs/1605.02688

%\end{thebibliography}


\end{document}